\documentclass[preprint]{emulateapj}

\newcommand{\etal}{et al.}
\def\kms{\ifmmode{\rm km\,s^{-1}}\else\hbox{$\rm km\,s^{-1}$}\fi}
\def\ergs{\ifmmode{\rm erg\,s^{-1}}\else\hbox{$\rm erg\,s^{-1}$}\fi}
\def\farcs{\hbox{$.\!\!^{\prime\prime}$}}

\usepackage[english]{babel}

\makeatletter
\renewcommand{\paragraph}{\@startsection{paragraph}{4}{0ex}%
   {-3.25ex plus -1ex minus -0.2ex}%
   {1.5ex plus 0.2ex}%
   {\normalfont\normalsize\bfseries}}
\makeatother

\stepcounter{secnumdepth}
\stepcounter{tocdepth}
\slugcomment{To be submitted to Astrophysical Journal}

\shorttitle{Dust and SN~2004dj}
\shortauthors{Meikle et al.}

\begin{document}
\title{Dust and the Type~II-Plateau Supernova 2004dj}

\author{W. P. S. Meikle,\altaffilmark{1}
R. Kotak,\altaffilmark{2}
D. Farrah,\altaffilmark{3} 
S. Mattila,\altaffilmark{4,5} 
S. D. Van Dyk,\altaffilmark{6}
A. C. Andersen,\altaffilmark{7}
R. Fesen\altaffilmark{8},\\
A. V. Filippenko,\altaffilmark{9}
R. J. Foley,\altaffilmark{9,10,11}  
C. Fransson\altaffilmark{5},
C. L. Gerardy,\altaffilmark{12} 
P. A. H\"{o}flich\altaffilmark{12},
P. Lundqvist\altaffilmark{5},
M. Pozzo\altaffilmark{13},
J. Sollerman\altaffilmark{5,7},
and
J. C. Wheeler\altaffilmark{14}
}

\altaffiltext{1}{Astrophysics Group, Blackett Laboratory, Imperial
College London, Prince Consort Road, London SW7 2AZ, United Kingdom;
p.meikle@imperial.ac.uk.}
\altaffiltext{2}{Astrophysics Research Centre, School of Mathematics and
Physics, Queen's University Belfast, BT7 1NN, United Kingdom.}
\altaffiltext{3}{Astronomy Centre, Department of Physics and
Astronomy, University of Sussex, Brighton BN1 9QJ, United Kingdom.}
\altaffiltext{4}{Tuorla Observatory, Department of Physics and
Astronomy, University of Turku, V\"ais\"al\"antie 20, FI-21500 Piikki\"o,
Finland.}
\altaffiltext{5}{Department of Astronomy, Stockholm University, 
AlbaNova, SE-10691 Stockholm, Sweden.}
\altaffiltext{6}{Spitzer Science Center, 220-6 Caltech, Pasadena, CA 91125.}
\altaffiltext{7}{Dark Cosmology Centre, Niels Bohr Institute, University 
of Copenhagen, Juliane Maries Vej 30, 2100 Copenhagen {\O}, Denmark.}
\altaffiltext{8}{Department of Physics and Astronomy, 6127 Wilder Lab., 
Dartmouth College, Hanover, NH 03755.}
\altaffiltext{9}{Department of Astronomy, University of California,
Berkeley, CA 94720-3411.}
\altaffiltext{10}{Harvard/Smithsonian Center for Astrophysics, 60
Garden Street, Cambridge, MA 02138.}
\altaffiltext{11}{Clay Fellow.}
\altaffiltext{12}{Department of Physics, Florida State University, 
315 Keen Building, Tallahassee, FL 32306-4350.}
\altaffiltext{13}{Department of Earth Sciences, University College London, 
London WC1E 6BT, United Kingdom.} 
\altaffiltext{14}{Astronomy Department, University of Texas, 
Austin, TX 78712.}



\begin{abstract}
We present mid-infrared (MIR) spectroscopy of a Type~II-plateau
supernova, SN~2004dj, obtained with the {\it Spitzer Space Telescope},
spanning 106--1393~d after explosion.  MIR photometry plus
optical/near-IR observations are also reported.  An early-time MIR
excess is attributed to emission from non-silicate dust formed within
a cool dense shell (CDS).  Most of the CDS dust condensed between 50~d
and 165~d, reaching a mass of $0.3\times10^{-5}$~M$_{\odot}$.
Throughout the observations much of the longer wavelength ($>10~\mu$m)
part of the continuum is explained as an IR echo from interstellar
dust.  The MIR excess strengthened at later times. We show that this
was due to thermal emission from warm, non-silicate dust formed in the
ejecta.  Using optical/near-IR line-profiles and the MIR
continua, we show that the dust was distributed as a disk whose
radius appeared to be shrinking slowly.  The disk radius may
correspond to a grain destruction zone caused by a reverse shock which
also heated the dust.  The dust-disk lay nearly face-on,
had high opacities in the optical/near-IR regions, but remained
optically thin in the MIR over much of the period studied.  Assuming a
uniform dust density, the ejecta dust mass by 996~d was
$(0.5\pm0.1)\times10^{-4}$~M$_{\odot}$, and exceeded
$10^{-4}$~M$_{\odot}$ by 1393~d.  For a dust density rising toward the
center the limit is higher.  Nevertheless, this study suggests
that the amount of freshly-synthesized dust in the SN~2004dj ejecta
is consistent with that found from previous studies, and adds further
weight to the claim that such events could not have been major
contributors to the cosmic dust budget.
\end{abstract}





\section{Introduction}
\label{sec:intro}
\setcounter{footnote}{0}
\setcounter{section}{1}
\setcounter{subsection}{0}

Massive stars explode via core collapse and ejection of their
surrounding layers \citep[e.g.][and references therein]{arn89}.  The
extent to which core-collapse supernovae (CCSNe) are, or have been, a
major source of dust in the Universe is of great interest.  Of
particular concern is the evidence of enormous amounts of dust
($\gtrsim 10^{8}\,{\rm M}_\odot$) in galaxies at high redshifts
($z\gtrsim 5$).  This comes from a variety of observations such as
sub-mm and near-infrared (NIR) studies of the most distant quasars
\citep{ber03,maio04}, obscuration by dust of quasars in damped
Ly-$\alpha$ systems \citep{pei91}, and measurements of metal
abundances in these systems \citep{pet97}.  Until recently, the
scenario of dust from AGB stars tended to be rejected since it was
thought that their progenitors would not yet have evolved off the main
sequence.  However, \citet{val09} and \citet{dwe11} have argued that,
under certain circumstances, asymptotic giant branch (AGB) stars may
make some contribution to the dust budget at high redshifts. Both
studies nevertheless cannot rule out a supernova contribution.  In
this paper, we examine the supernova option through observations of a
nearby core-collapse event.\\

CCSNe arising from short-lived Population III stars might seem to be a
viable alternative.  It is estimated that each supernova (SN) must
produce 0.1--1\,${\rm M}_\odot$ of dust to account for the
high-redshift observations \citep{dwe07,mei07}. Such masses have been
predicted in models of dust formation in CCSNe \citep{tod01,noz03},
although more recent calculations by \citet{che10} revise such
estimates downward by a factor of $\sim$5.  Perhaps even more
problematic is the fact that actual dust-mass measurements in CCSNe
and SN remnants yield values not exceeding, respectively,
$10^{-3}\,{\rm M}_\odot$ and $10^{-2}\,{\rm M}_\odot$, although only a
handful of such measurements exist. (For a summary of this topic see,
for example, \citealt{kot09}, \S1.)  The {\it Spitzer Space Telescope}
\citep[hereafter, {\it Spitzer};][]{wer04} provided an excellent
opportunity for us to test the ubiquity of dust condensation in a
larger number of CCSNe.\\

Newly-condensed dust in CCSNe can be detected by its attenuating
effects on optical/NIR light and/or via thermal emission from the
grains in the ejecta.  Prior to {\it Spitzer}, the only evidence of
dust condensation in {\it typical} CCSNe was in the Type II-plateau
(IIP) SN~1988H \citep{tur93} and SN~1999em \citep{elm03}.  However the
light curve data used to type SN~1988H was sparse.  In the case of
SN~1999em, \citeauthor{elm03} used optical line suppression to infer a
dust mass lower limit of about $10^{-4}$~M$_{\odot}$.  With the launch
of {\it Spitzer}, we were at last provided with a facility for
high-sensitivity spectroscopy and imaging of nearby CCSNe over the
mid-infrared (MIR) range, covering the likely peak of the dust thermal
emission spectrum.  This can provide a superior measure of the total
flux, temperature, and possibly dust emissivity than can be achieved
at shorter wavelengths.  Moreover, the longer-wavelength coverage of
{\it Spitzer} allow us to detect cooler grains and see more deeply
into dust clumps than was previously possible for typical nearby
CCSNe.  In this paper, we present our late-time {\it Spitzer}
observations of the Type~IIP SN~2004dj.  We use these observations to
study the dust production in this supernova.\\

The paper is arranged as follows. In \S1.1 we summarize and discuss
previous observations of SN~2004dj.  In \S2 we present MIR ({\it
Spitzer}) photometric and spectroscopic observations of SN~2004dj,
extending to more than 3~years after the explosion.  This MIR coverage
is one of the most extensive ever achieved for a SN~IIP.  We also
present late-time optical and NIR photometry and spectra of
SN~2004dj. In \S3 we analyze these data. Corrections are derived in
\S3.1 for the effects of the line-of-sight cluster S96, and in \S3.2
the mass of ejected $^{56}$Ni is determined. We compare the data with
blackbodies in \S3.3, in order to assess the likely number and nature
of the contributing sources. In \S3.4 the origins of the IR radiation
are examined in detail. The work is then summarized in \S4.  \\ \\ \\
\subsection{SN~2004dj}
\label{sec:04dj}
SN~2004dj was discovered in the nearby spiral galaxy NGC~2403 on 2004
July 31 by \citet{nak04} and was classified as a normal Type~IIP SN by
\citet{pat04}.  It was the nearest such event in over three decades;
the host galaxy lies within the M81 group. In \citet{kot05} we adopted
a distance to NGC~2403 (and the SN) of $3.13\pm0.15$~Mpc (statistical
errors only), this being the Cepheid-derived, zero metallicity value
reported by \citet{fre01} using the \citet{uda99} period-luminosity
slopes.  We continue to use this distance in the work presented here.
\citet{vin06} have estimated the distance to SN~2004dj using a
combination of the Freedman et al. value plus their own expanding
photosphere method (EPM) and standard candle estimates.  This yields
an average distance of $3.47\pm0.29$~Mpc, implying that the SN
luminosity could be $\sim$20\% larger than the values used herein.\\

The progenitor of SN~2004dj was almost certainly a member of the
compact star cluster Sandage~96 (S96)
\citep{san84,bon04,fil04,mai04,wan05,vin06}. The cluster age is
variously estimated to be $14\pm2$~Myr \citep{mai04}, $\sim$20~Myr
\citep{wan05}, and 10--16~Myr \citep{vin09}.  The main-sequence mass
of the progenitor is estimated at $\sim$15~M$_{\odot}$
\citep{mai04,kot05}, $\sim$12~M$_{\odot}$ \citep{wan05}, and
12--20~M$_{\odot}$ \citep{vin09}.  \citet{mai04} and \citet{kot05}
favor a red supergiant (RSG) progenitor.\\

\citet{gue04} used echelle spectroscopy of Na~I~D absorption lines in
NGC~2403 along the line of sight to SN~2004dj to infer a heliocentric
velocity of $+164.8\pm0.1$~\kms.  This is somewhat larger than the
heliocentric velocity of the nuclear region of NGC~2403 of
$\sim$130~\kms\ given in the SIMBAD and NED databases, but this is not
surprising given the likely dispersion of velocities within the host
galaxy. Indeed, \citet{vin06} point out that H~I mapping of NGC~2403
\citep{fra01} suggests that the true radial velocity of the SN~2004dj
barycenter is about +221~\kms. We adopt this value here.  \\

There is no firm consensus about the value of the reddening to
SN~2004dj. Stellar population fitting for S96 yields total (Galactic +
host) $E(B-V)$ values of $0.17\pm0.02$~mag \citep{mai04},
$0.35\pm0.05$~mag \citep{wan05}, and $0.1\pm0.05$~mag \citep{vin09}.
Direct color comparisons of SN~2004dj with other CCSNe yield $E(B-V)$
values of $\sim$0~mag \citep{zha06}, 0.06~mag \citep{chu06}, and
$0.07\pm0.1$~mag \citep{vin06}.  Perhaps most significantly, the
resolved Na~I~D observations of \citet{gue04} yield a host-only $E(B-V)$ 
value of just $0.026\pm0.002$~mag. The smaller values of $E(B-V)$
obtained using direct measurements toward SN~2004dj suggests that the
SN actually lies near the front of S96.  Based on the
extinction maps of \citet{sch98}, \citet{chu05} find a Galactic
reddening of $E(B-V)=0.062$~mag while \citet{zha06} report
$E(B-V)=0.04$~mag (the same value as obtainable from NED). We
therefore adopt a Galactic $E(B-V)=0.05\pm0.01$~mag.  If we add the
\citet{gue04} host value to the Galactic value, we obtain
$E(B-V)=0.076\pm0.01$~mag, or $A_V=0.24\pm0.03$~mag for a
\citet{car89} extinction law with $R_V = 3.1$.  Given the range of
published values, in the present work we used a total extinction
$A_V=0.31$~mag, the same as that preferred by \citet{vin09}.  Adoption
of even the largest published value of $E(B-V)$ would increase our
shortest MIR wavelength (3.6~$\mu$m) flux by just a few percent. \\

Estimates of the explosion date of SN~2004dj vary by several weeks.
On the basis of an early-time spectrum, \citet{pat04} placed the explosion
at about 2004 July 14.  This is consistent with the date obtained by
\citet{bes05}, who used the radio $L_{peak}$ versus rise-time relation
\citep{wei02} to yield an explosion date between 2004 July 11 and July
31.  On the other hand, based on the EPM method, \citet{vin06} derive
an explosion date as early as 2004 June 30.  On the assumptions that
the light curve of SN~1999gi was typical of SNe~IIP and that
the SN~2004dj plateau was of similar length, \citet{chu05} obtained an
even earlier explosion date: 2004 June 13.  Likewise, assuming similar
evolution between SN~2004dj and SN~1999em, \citet{zha06} find an
explosion date of 2004 June 11.  Nevertheless, partly as a compromise
with the later \citet{pat04} value, \citet{chu06} subsequently adopted
2004 June 28 as the date of the explosion.  \citet{chu07} also used
this explosion date. Given the weight of evidence for a later
explosion date, we reject that preferred by \citet{zha06}. In
\citet{kot05} we adopted an explosion date of 2004 July 10, or
MJD = 53196.0. We use the same explosion date here but recognize that
there is an uncertainty of about $\pm7$~d. All epochs will be with
respect to MJD = 53196.0 (i.e., $t=0$~d). \\

Optical light curves of SN~2004dj are presented by
\citet{kor05,chu05,leo06, vin06,zha06,vin09}. The $VR$ light curves
fell by 10\% and 90\% of the total decline from the plateau to the
start of the radioactive tail at, respectively, epochs $70 \pm 5$~d
and $96 \pm 3$~d.  At the start of the radioactive tail the SN
luminosity was about 15\% of the plateau value.  During the early
nebular phase (up to $\sim$300~d) the $V$~band declined at about 1.1
mag (100~d)$^{-1}$ \citep{vin06}, which is typical for a SN~IIP.  On
the basis of the light curves, the mass of ejected $^{56}$Ni has been
estimated at 0.02--0.03~M$_\odot$ \citep{chu05,kot05,vin06,zha06}.
Notwithstanding, in the present work we argue that these authors have
overestimated the $^{56}$Ni mass and that the true mass is more like
0.01~M$_\odot$. \\

At the end of the plateau phase, SN~2004dj exhibited a remarkable and
rapid change in some of its prominent optical lines, especially
H$\alpha$.  At 89~d the H$\alpha$ profile still had a typical P~Cygni
morphology with a symmetric peak blueshifted by only about
$-270$~\kms\ after correction for the heliocentric velocity of the SN.
Yet, by just $\sim$10~d later the profile had developed a strong
asymmetric profile with a peak at $-1610$~\kms\ \citep{chu05}. As time
went by during the first year, this blueshift gradually decreased and
the asymmetry became less pronounced.  \citet{chu06} interpreted this
unusual behavior as being due to the gradual emergence of an
asymmetric, bipolar jet whose more massive component is travelling
towards the observer.  They propose that the lines are driven by the
radioactive decay of spherical fragments of $^{56}$Co cocooned in
metals and helium, lying within the core.  Using spectropolarimetry
\citet{leo06} also found evidence for departure from spherical
symmetry.  The polarization was observed to increase dramatically at
the end of the plateau phase, implying the presence of significant
asphericity in the inner regions of the ejecta. \\

Early-time evidence of a significant circumstellar medium (CSM) around
the progenitor of SN~2004dj has also been reported.  Radio emission
was detected by \citet{sto04} at 23~d, by \citet{bes05} between 26~d
and 145~d, and by \citet{cha04} at 33~d and 43~d.  The SN was also
detected at X-ray wavelengths \citep{poo04} at 30~d. The X-ray
luminosity was about three times that of the Type~IIP SN 1999em and
nearly fifteen times that of the Type~IIP SN 1999gi at similar epochs.
\citet{bes05} point out that both types of emission arise from a
shocked CSM.  \citet{chu07} have used the presence of a high-velocity
absorption component in the H$\alpha$ line during the late
photospheric phase to deduce the presence of a cool dense shell (CDS),
with a mass of $3.2\times10^{-4}$~M$_\odot$, produced by interaction
of the ejecta with the pre-existing CSM.  To this evidence for a CDS
we add our observation of an early-time IR echo in SN~2004dj (see
\S3.4.2). \\

In \citet{kot05} we presented MIR photometric and spectroscopic
observations of SN~2004dj at epochs 97--137~d after explosion.  Simultaneous
modelling of the fundamental (1--0) and first overtone (2--0) of CO
was carried out.  The results favor a 15~M$_\odot$ RSG progenitor and
indicate post-explosion CO formation in the range 2000--4000~\kms.
\citet{kot05} also noted an underlying NIR continuum. A possible
origin in CSM dust was suggested, but RSGs in S96 were favored as the
more likely cause. Nevertheless, in the present work (\S3.2.2) we find
that the bulk of the early-time NIR and MIR continuum is most
plausibly explained as an IR echo from CDS dust.  The presence of
$1.7\times10^{-4}$~M$_\odot$ of Ni$^+$ in the ejecta was also deduced
by \citet{kot05}. \\

In summary, while the early-time optical light curves and spectra of
SN~2004dj are typical of a Type~IIP event, its early radio, MIR, and
X-ray behavior point to an exceptionally strong ejecta/CSM
interaction. Moreover its earlier nebular-phase spectra imply an
atypically asymmetric core. SN~2004dj is only ``typical'' in some
respects.  \\ 

\section{Observations}
\subsection{Mid-Infrared Photometry}
Imaging at 3.6, 4.5, 5.8, and 8.0\,$\mu$m was obtained with the
Infrared Array Camera (IRAC) \citep{faz04}, at 16 and 22\,$\mu$m with
the Peak-up Array (PUI) of the Infrared Spectrograph (IRS)
\citep{hou04}, and at 24\,$\mu$m with the Multiband Imaging Photometer
for Spitzer (MIPS) \citep{rie04}. Imaging observations spanned epochs
89.1 to 1393.3~d, plus four observations, at 3.6~$\mu$m only, covering
1953.9--2143.5~d.  Aperture photometry was performed on the images
using the Starlink package GAIA \citep{dra02}.  A circular aperture of
radius $\sim3\farcs7$ was used for the photometry.  The background
flux was measured and subtracted by using a concentric sky annulus
having inner and outer radii of 1.5 and 2.2 times the aperture radius,
respectively.  These parameters were chosen as a compromise between
maximizing the sampled fraction of source flux and minimizing the
effects of the bright, complex background.  The aperture radius
corresponds to $\sim$55~pc at the distance of SN~2004dj.  The aperture
was centered according to the SN WCS coordinates.  Aperture
corrections were derived from the IRAC and MIPS point-response
function frames available from the Spitzer Science Center, and ranged
from $\times$1.16 at 3.6\,$\mu$m to $\times$2.79 at 24\,$\mu$m.  A
2$\sigma$ clipped mean sky estimator was used, and the statistical
error was estimated from the variance within the sky annuli.  Fluxing
errors due to uncertainties in the aperture corrections are about
$\pm5\%$. \\

\begin{figure*}
\epsscale{0.87}
\plotone{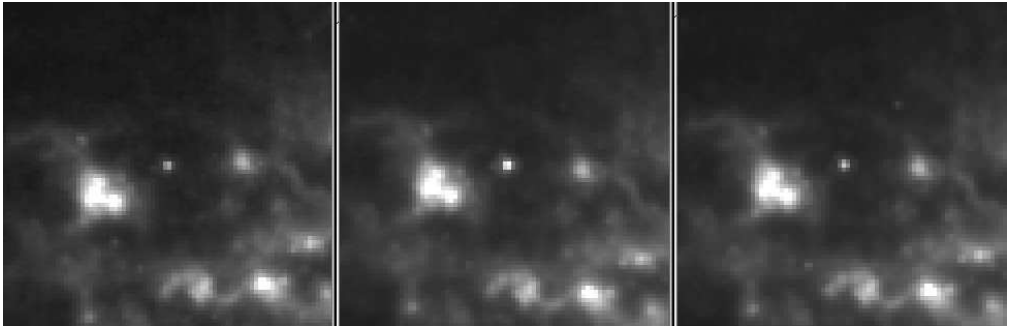}
\caption[]{Sequence of images at 8~$\mu$m at 257~d (LHS), 621~d
(middle) and 996~d (RHS). SN~2004dj is the point source at the center
of each image.  The supernova is clearly brighter on 621~d. This is
due to the epoch being close to the peak of the thermal emission from
the ejecta dust at this wavelength (see Fig.~\ref{fig2}). The fields
are about 2~arcmin across ($\sim1.8$~kpc at the distance of the
supernova). North is $\sim30^{\circ}$ clockwise from the upward
vertical
\label{fig1}
}
\end{figure*}

\begin{figure*}
\epsscale{0.87}
\plotone{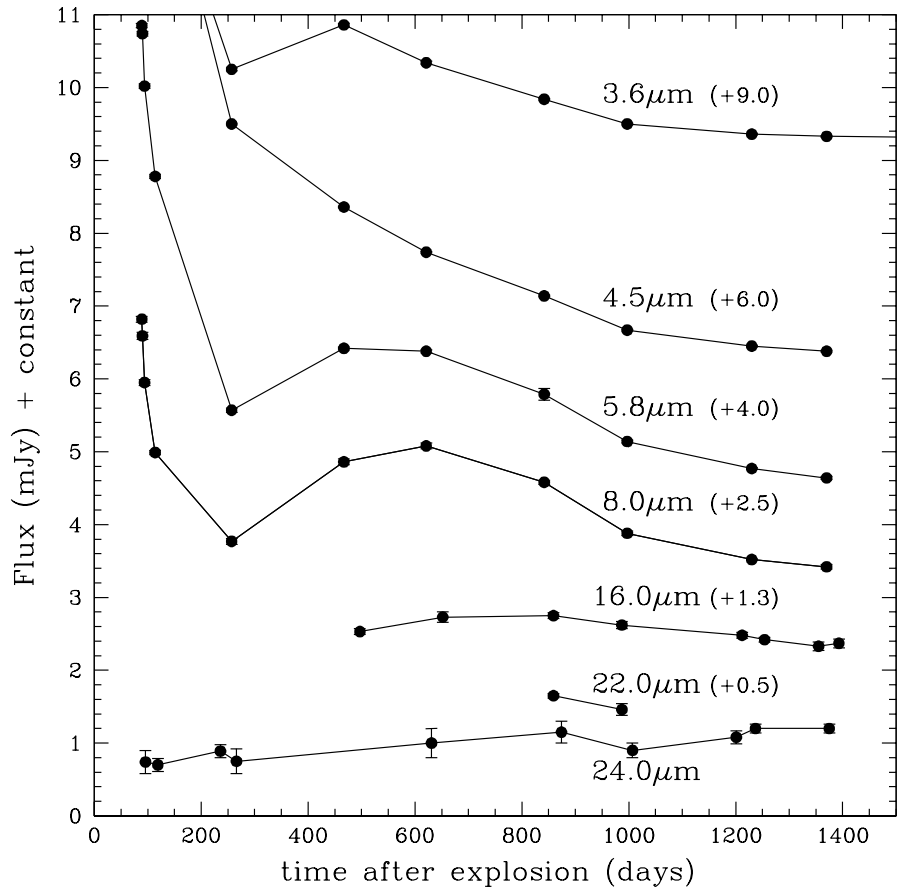}
\caption[]{MIR light curves of SN~2004dj.  They are uncorrected for
reddening.  For clarity, the plots have been shifted vertically by the
amounts shown in brackets (mJy).  Off to the right of the plot there
are four additional 3.6~$\mu$m points spanning 1954~d to 2143~d
(Table~\ref{tab1}).  They show little significant change during this
period, with a mean flux of $0.32\pm0.02$~mJy. This is consistent with
there having been no decline since 1372~d.  The MIR flux of S96 alone
has not been measured and so the light curves are uncorrected for S96.
An estimate of the S96 contribution is given in Table~\ref{tab1} and
is discussed in \S3.1.
\label{fig2}
}
\end{figure*}

\begin{figure*}
\epsscale{0.87}
\plotone{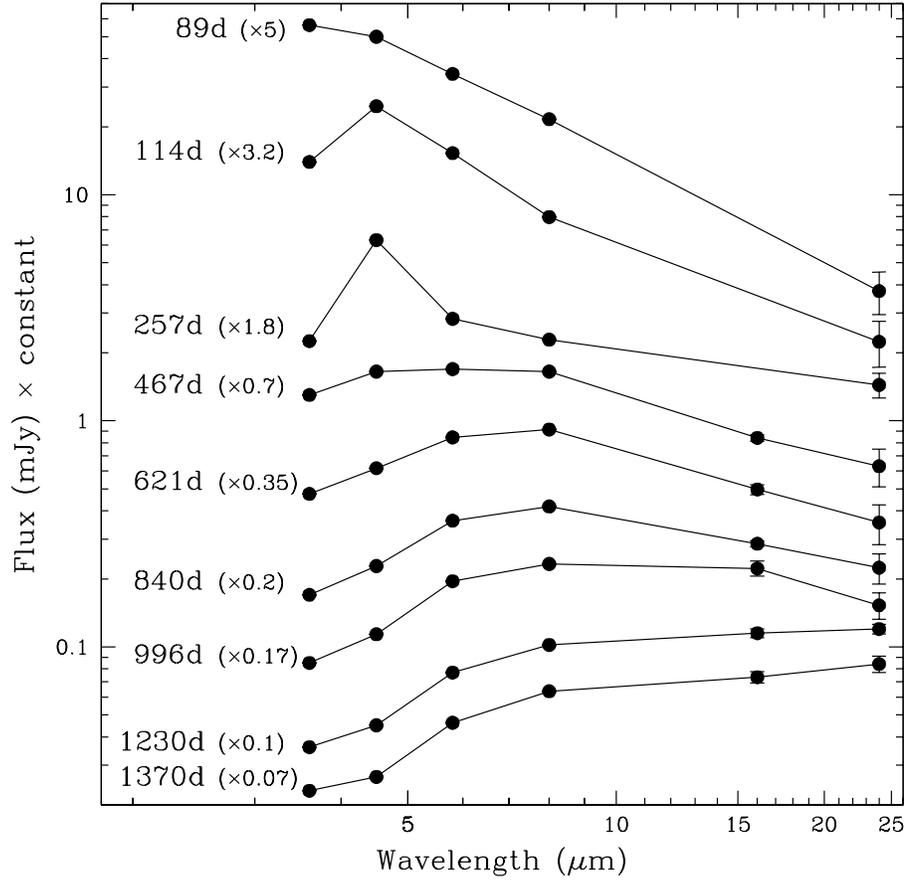}
\caption[]{Evolution of the MIR spectral energy distribution of
 SN~2004dj.  The SEDs are uncorrected for reddening.  For clarity, the
 individual SEDs have been scaled by the amounts shown in brackets.  As
 in Fig.~\ref{fig2}, the fluxes from S96 have not been subtracted. The
 large peak at 4.5~$\mu$m present at 114~d and 257~d is due to the
 dominance of CO fundamental emission in this band during this period.
\label{fig3}
}
\end{figure*}

The MIR photometry is presented in Table~\ref{tab1}.  The {\it
Spitzer} programs from which the imaging data were taken are listed in
the caption.  The pre-explosion MIR flux of S96 has not been measured,
so the tabulated values are uncorrected for S96. An estimate of the
S96 contribution is given at the bottom of the Table.  The effect of
emission from S96 is discussed in \S3.1.  A temporally varying point
source at the SN position is clearly visible in all bands.
Figure~\ref{fig1} shows a sequence of images at 8.0~$\mu$m at 257,
621, and 996~d. The SN is clearly brighter at 621~d (about $\times2$
relative to 257~d). This is due to the epoch being close to the peak
of the thermal emission from the dust (see \S3.4.4.2).  It can also be
seen that the SN lies in a region of relatively bright, complex
background emission.  The MIR photometry is displayed as light curves
and spectral energy distributions (SEDs) in Fig.~\ref{fig2} and
Fig.~\ref{fig3}, respectively. \\

In Fig.~\ref{fig2}, a rapid initial decline is seen at wavelengths
3.6--8.0~$\mu$m.  With the exception of 4.5~$\mu$m, all of the light
curves (with sufficient temporal coverage) exhibit secondary maxima,
with the peak emission occurring at epochs of $\sim$450~d at
3.6~$\mu$m to $\sim$850~d at 24~$\mu$m.  As will be discussed later,
these second maxima constitute strong evidence of dust formation.  
The absence of a delayed peak at 4.5~$\mu$m is due to the earlier
appearance and dominance of CO fundamental emission in this
band. After $\sim$1000~d, the 24~$\mu$m light curve starts to climb
again. The evolution of the 24~$\mu$m flux is complex, as it is a
combination of the detailed behaviors of emission from the ejecta 
dust and from an interstellar (IS) IR echo (see \S3.4.3.1).  In
Fig.~\ref{fig3} we see a steady reddening of the MIR SED with time.
It is argued below (\S3.4.4.2) that this effect also constitutes strong
evidence of dust formation and cooling in the SN ejecta.  The
large peak at 4.5~$\mu$m at 114~d and 257~d is due to the
aforementioned dominance of CO fundamental emission in this band.  As
in Table~\ref{tab1}, neither Fig.~\ref{fig2} nor Fig.~\ref{fig3} have
been corrected for S96.

\subsection{Optical and Near-Infrared Photometry}
\label{sec:nirphot}
NIR imaging of SN~2004dj was obtained using LIRIS (Long-slit
Intermediate Resolution Infrared Spectrograph) on the 4.2~m William
Herschel Telescope (WHT), La Palma, and at an effectively single epoch
(spanning two days) with the OSU-MDM IR Imager/Spectrograph on the
2.4~m Hiltner Telescope of the MDM Observatory, Arizona.  The wavebands
are $Z$ (1.033~$\mu$m), $J$ (1.250~$\mu$m), $H$ (1.635~$\mu$m), and
$K_s$ (2.15~$\mu$m).  The LIRIS data were reduced using standard IRAF
routines.\footnote{IRAF is written and supported by the IRAF
programming group at the National Optical Astronomy Observatories
(NOAO) in Tucson, Arizona, which are operated by the Association of
Universities for Research in Astronomy (AURA), Inc., under cooperative
agreement with the National Science Foundation (NSF).} The jittered
on-source exposures were median-combined to form sky frames. In each
band the sky-subtracted frames were then aligned and
median-combined. \\

\begin{figure*}
\epsscale{0.88}
\plotone{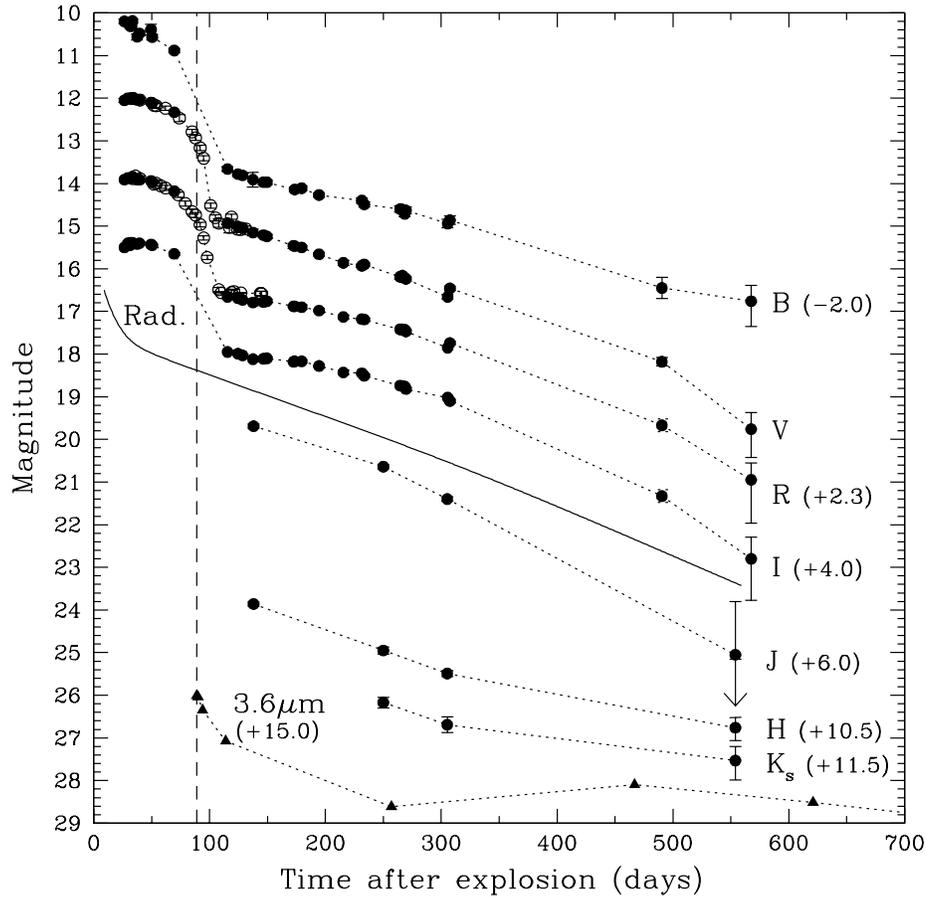}
\caption[]{Optical and NIR light curves of SN~2004dj. For clarity,
they have been displaced vertically by the values shown in brackets.
The optical light curves are from \citet{vin06,vin09} (solid dots). In
addition the $V$ and $R$ points of \citet{zha06} (open circles) around
the end of the plateau have been added to enhance the detail of this
phase.  For all data sets the epoch has been adjusted to our explosion
epoch of MJD=53196.0.  The NIR data are from the present work.  Also
shown is the 3.6~$\mu$m light curve from the present work.  The
vertical dashed line is at the epoch of the earliest 3.6~$\mu$m point
and indicates the corresponding phase in the other light curves.  S96
fluxes at $BVRI$ and at $JHK_s$ have been subtracted from the data.
Also shown for comparison (labelled ``Rad.'') is the temporal
evolution of the radioactive energy deposition as specified by
\citet{li93} \& \citet{tim96} for SN~1987A with the addition of the
early-time contribution of $^{56}$Ni assuming complete absorption.
\label{fig4}
}
\end{figure*}

Aperture photometry was performed on the reduced
images using the Starlink package GAIA \citep{dra02} with the same
aperture and sky annuli as for the MIR photometry.  The aperture was
centered by centroiding on the sources.  The magnitudes at $J$, $H$, and
$K_s$ were obtained by comparison with four field stars lying within
$\sim100^{\prime\prime}$ of SN~2004dj. The field-star magnitudes were
acquired by measurement of 2MASS images \citep{skr97}.  For the single
$Z$-band measurement, the magnitude was obtained by comparison with the
four field stars with their $JHK_s$ SEDs extrapolated to the
$Z$~band. The resulting SN photometric measurements are listed in
Table~\ref{tab2} and plotted in Fig.~\ref{fig4}.
Errors shown include uncertainties in the magnitudes of the four 2MASS
comparison field stars.  Pre-explosion $JHK_s$ fluxes of S96 were
measured from the 2MASS survey (see Table~\ref{tab2}) and used to
correct the $JHK_s$ light curves. Also shown for comparison is the
3.6~$\mu$m light curve from the present work.  In $H$ and $K_s$ the
slopes flatten after $\sim$300~d accompanied by a {\it rise} at
3.6~$\mu$m. This is suggestive of radiation from warm, newly-forming
dust.\\

Optical photometry was taken from \citet{vin06,vin09}.  The optical
light curves are displayed in Fig.~\ref{fig4} and have been corrected
for emission from the S96 cluster using the $BVRI$ magnitudes given by
\citet{vin06}.  In addition, the $V$ and $R$ points of \citet{zha06}
around the end of the plateau were added to enhance the detail of this
phase.  Also shown for comparison (labelled ``Rad'') is the temporal
evolution of the radioactive energy deposition for SN~1987A as
specified by \citet{li93} (0 to 1200~days) and \citet{tim96} (500 to
3500 days) with the addition of the early-time contribution of
$^{56}$Ni decay assuming complete absorption. The radioactive isotopes
include $^{56}$Ni, $^{56}$Co, $^{57}$Co, $^{60}$Co, $^{22}$Na, and
$^{44}$Ti.  (In subsequent use of the \citeauthor{li93} and
\citeauthor{tim96} deposition specifications, our addition of the
early-time contribution of $^{56}$Ni decay is assumed.)  In the
optical light curves, for about 35~d (115--150~d) after the end of the
fall from the plateau the decline rate matches the radioactive
deposition quite closely, indicating that this was powering the
emission during this phase.  The optical decline rates then flatten
during $\sim$150--250~d, indicating the emergence of an additional
power source (see below). After about 250~d, the optical light curves
exhibit a steepening (possibly also present in the $J$~band) which
increases up to the final observations. \\

\subsection{Mid-Infrared Spectroscopy}
\label{sec:spec}
Low-resolution ($R \approx 60$--127) MIR spectroscopy between 5.2 and
14.5~$\mu$m was acquired at nine epochs between 106.3~d and 1393.3~d
with the IRS in low-resolution mode.  Long-Low (LL; 14--38~$\mu$m)
observations were also attempted. Unfortunately, the LL observations
were unusable. The LL slit lies at $90^\circ$ to the SL slit. This
meant that, given the scheduling constraints, the LL slit always lay
across the host galaxy, resulting in heavy contamination.  The MIR
spectroscopic observations were drawn from the MISC programs plus one
epoch at 1207~d from the SEEDS program.  The log of spectroscopic
observations is given in Table~\ref{tab3}.\\

The data were processed through the Spitzer Science Center's pipeline
software, which performs standard tasks such as ramp fitting and 
dark-current subtraction, and produces Basic Calibrated Data (BCD) frames.
Starting with these data, we produced reduced spectra using both the
SPICE and SMART v6.4 software packages. We first cleaned individual
frames of rogue and otherwise ``bad'' pixels using the IRSCLEAN
task. The first and last five pixels, corresponding to regions of
reduced sensitivity on the detector, were then removed. The individual
frames at each nod position were median-combined with equal
weighting on each resolution element.  Sky background was removed from
each combined frame by subtracting the combined frame for the same
order taken with the other nod position. We also experimented with
background removal by subtracting the adjacent order. In general,
nod-nod subtraction was preferred, as the background sampled in this
way is expected to most closely represent the background underlying
the SN.  Any residual background was removed by fitting low-order
polynomials to regions immediately adjacent to the SN position.\\

One-dimensional spectra were then extracted using the optimal
extraction tool within the SPICE software package, with default
parameters. We found that in all cases the source was point-like, with
a full width at half-maximum intensity (FWHM) that was never wider
than the point-spread function (PSF).  This procedure results in
separate spectra for each nod and for each order.  The spectra for
each nod were inspected; features present in only one nod were treated
as artifacts and removed. The two nod positions were subsequently
combined. The nod-combined spectra were then merged to give the final
spectrum for each epoch. Overall, we obtained excellent continuum
matches between different orders. \\

\begin{figure*}
\epsscale{1.15}
\plotone{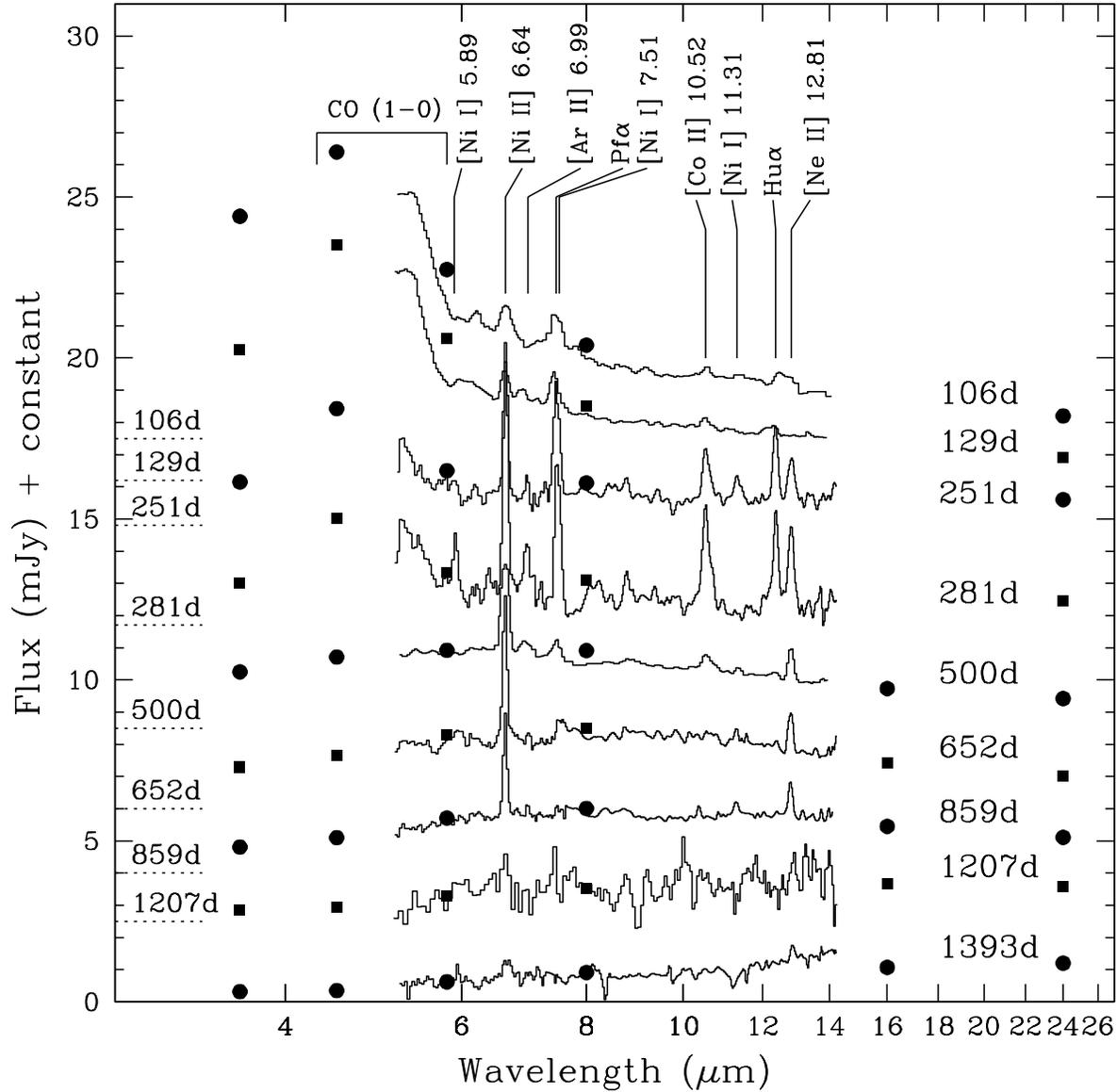}
\caption[]{Evolution of the MIR spectra of SN~2004dj.  The spectra
have not been dereddened and the S96 flux has not been been
subtracted.  No correction has been applied for the heliocentric
velocity of the SN.  The 1207~d spectrum has a small exposure time and
so is of relatively low S/N compared with those of 859~d and 1393~d
(see Table~\ref{tab3}). Also shown are the fluxes obtained through
aperture photometry of the IRAC, PUI and MIPS images, interpolated to
the epochs of the spectra. These photometric data are represented by
round or square points respectively on alternate epochs.  All plots
except the latest have been displaced vertically for clarity, with the
zero flux levels indicated by the horizontal dotted lines on the left.
The stronger features are identified, with the fiducials redshifted to
the SN rest frame. The wavelengths are labelled in microns.
\label{fig5}
}
\end{figure*}

Despite our careful reduction procedure, the fluxes of the IRS spectra
and the IRAC photometry were not completely consistent.  This was due
to (a) differences in the fixed sizes of the spectrograph aperture
slits and the circular apertures used for the image photometry, and (b)
the fact that the spectra were generally taken some days before or
after the imaging data, during which time the SN flux changed.  We
therefore recalibrated the IRS spectra against contemporaneous
photometry in the 8~$\mu$m band obtained by interpolation of the light
curve.  This band was chosen since it was completely spanned by the
short-low (SL) spectrum.  For each epoch, the IRAC 8~$\mu$m
transmission function was multiplied by the MIR spectra and by a model
spectrum of Vega\footnote{The R. Kurucz Stellar Atmospheres Atlas,
1993, ftp://ftp.stsci.edu/cdbs/grid/k93models/standards.}.  The
resulting MIR spectra for the SN and for Vega were integrated over
wavelength.  The total SN spectral flux in the 8~$\mu$m band was then
obtained from the ratio of the two measurements using a zero (Vega)
magnitude of 64.1~Jy (IRAC Data Handbook, Table~5.1).  These were then
compared with the 8~$\mu$m photometry to derive scaling factors by
which the spectra were multiplied.  The spectra are plotted in
Fig.~\ref{fig5}, together with contemporary photometric data.\\

The MIR spectra comprise both continua and emission features. Up to at
least 281~d, strong emission from the CO fundamental was present in
the IRS spectra and IRAC photometry. This had disappeared by
500~d. Strong lines of H~I, [Ni~I], [Ni~II], [Co~II], and [Ne~II] were
also present during the first year, but by 500~d only
[Ni~II]~6.64~$\mu$m, [Ni~I]~7.51~$\mu$m, and [Ne~II]~12.81~$\mu$m were
still relatively strong (see Fig.~\ref{fig5}).  Apart from the CO
region during 106--281~d, the 5--14~$\mu$m region was dominated by
continuum emission.  Moreover, simple extrapolation below and above
the spectral coverage to the limits of the photometric coverage
suggests that the continuum dominated over at least 3.6--24~$\mu$m.

\subsection{Optical and Near-Infrared Spectroscopy}

\begin{figure*}
\epsscale{1.15}
\plotone{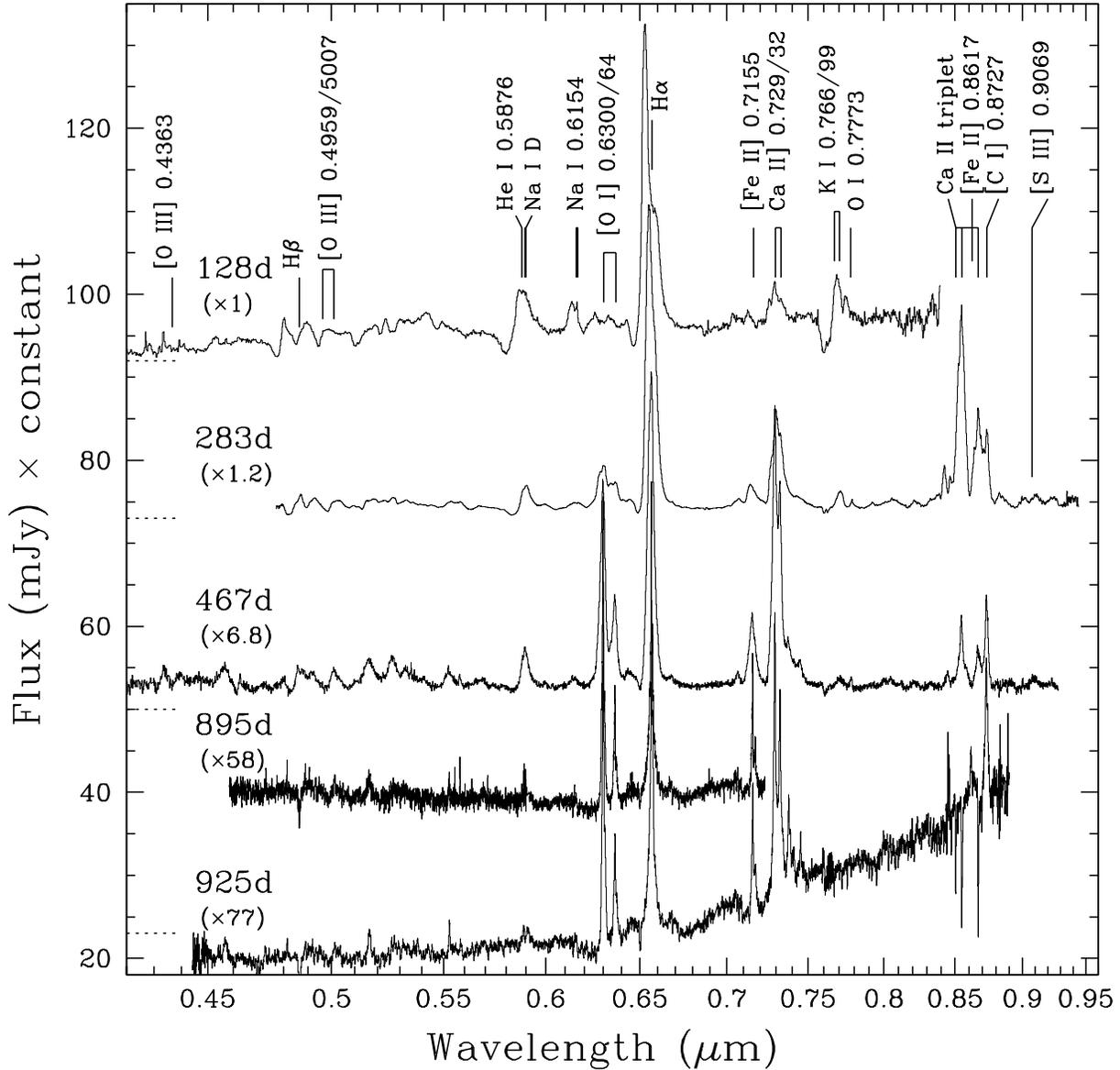}
\caption[]{Evolution of the optical spectra of SN~2004dj.  The 128~d
spectrum is from \citet{vin06} and the 895~d spectrum has already been
presented in \citet{vin09}.  The spectra have not been dereddened.  In
addition, the S96 flux has not been subtracted and this accounts for
the relatively large continuum fluxes at epochs 895~d and 925~d. No
correction has been applied for the heliocentric velocity of the SN.
All plots except the latest have been displaced vertically for
clarity, with the zero flux levels indicated by the horizontal dotted
lines on the left. Note that the zero flux shown at 23~mJy corresponds
to the 895~d spectrum.  Also, the spectra have been flux-scaled by the
amounts shown in brackets.  Locations of spectral lines of interest
are identified, with the fiducials redshifted to the SN rest frame.
The wavelengths are labelled in microns.
\label{fig6}
}
\end{figure*}

\begin{figure*}
\epsscale{1.15}
\plotone{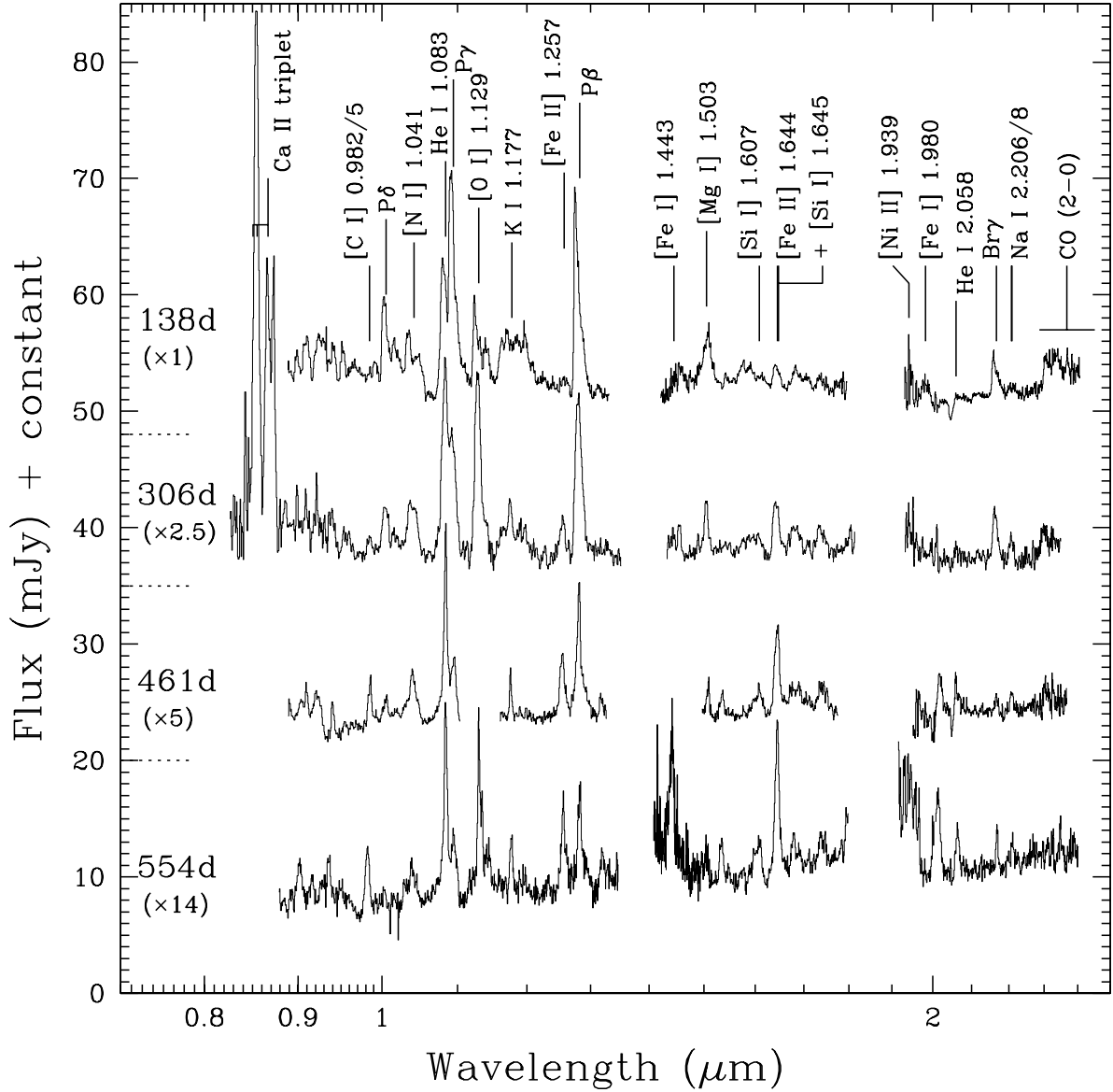}
\caption[]{Evolution of the NIR spectra of SN~2004dj.  The spectra
have not been dereddened and the S96 flux has not been subtracted.  No
correction has been applied for the heliocentric velocity of the SN.
All plots except the latest have been displaced vertically for
clarity, with the zero flux levels indicated by the horizontal dotted
lines on the left. Also, the spectra have been flux-scaled by the
amounts shown in brackets.  The stronger features are identified, with
the fiducials redshifted to the SN rest frame.  The wavelengths are
labelled in microns.
\label{fig7}
}
\end{figure*}

We acquired optical spectra using ISIS on the WHT, La Palma, and
DEIMOS \citep{faber03} on the 10~m Keck~2 telescope, Hawaii.  The
895~d Keck spectrum has already been presented by \citet{vin09}.  We
also made use of earlier post-plateau optical spectra obtained by
\citet{vin06} at 89~d and 128~d, \citet{leo06} at 95~d, and
\citet{chu05} at 100~d.  NIR spectroscopy of SN~2004dj was obtained
using LIRIS on the WHT and with the OSU-MDM IR Imager/Spectrograph on
the 2.4~m Hiltner Telescope of the MDM Observatory, Arizona.  The data
were reduced using standard procedures in Figaro \citep{sho02} and
IRAF.  The observing log for the optical and NIR spectra is given in
Table~\ref{tab4}.  The post-100~d spectra are plotted in
Figs.~\ref{fig6} and \ref{fig7}.\\

\begin{figure*}
\epsscale{0.67}
\plotone{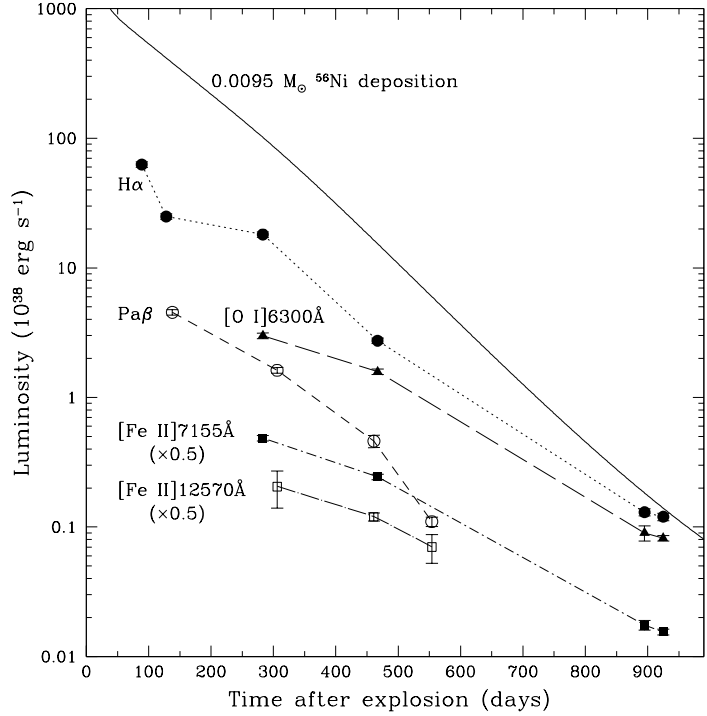}
\caption[]{Evolution of spectral line luminosities of SN~2004dj.  The
luminosities have been corrected for extinction.  Also shown is the
deposition power of 0.0095~M$_{\odot}$ of ${56}$Ni, scaled from the
SN~1987A case described by \citet{li93} \& \citet{tim96}.  The errors
in the fluxes and luminosities are primarily due to uncertainties in
the absolute fluxing and the levels of the underlying continua. Not
included in the luminosity errors are systematic uncertainties in the
distance and extinction (see \S1.1).
\label{fig8}
}
\end{figure*}

The earlier post-plateau spectra still exhibited pronounced P~Cygni
features in H$\alpha$, He~I~5876~\AA\ + Na~I~D, He~I~10830~\AA,
He~I~20581~\AA, and O~I~7771~\AA\ + K~I~7665/99~\AA.  By 461/467~d the
absorption components had largely vanished, with a broad-line emission
spectrum now being observed.  A few lines persisted to as late as the
final optical spectroscopy epoch at 925~d.  We examined in detail the
evolution of the more isolated of these lines, specifically H$\alpha$,
Pa$\beta$, [O~I]~6300~\AA, [Fe~II]~7155~\AA, and [Fe~II]~12567~\AA.
Table~\ref{tab5} lists the line luminosities (dereddened) versus
epoch, together with the radioactive deposition power specified by
\citet{li93} and \citet{tim96} for SN~1987A, but scaled down to an
initial 0.0095~M$_{\odot}$ of $^{56}$Ni (see \S3.2).  The evolution of
the luminosities is plotted in Fig.~\ref{fig8}.
This indicates that from just after the plateau phase to $\sim$460~d,
the H$\alpha$ and Pa$\beta$ luminosities declined at a rate roughly
comparable to that of the radioactive deposition.  In contrast, from
their earliest observation at $\sim$300~d, the [O~I] and [Fe~II] lines
decline significantly more slowly than the radioactive rate.
Moreover, by 895~d the summed luminosity of just the H$\alpha$,
[O~I]~6300~\AA, and [Fe~II]~7155~\AA\ lines exceeds that of the
radioactive input by $\sim$40\%, rising to over 60\% by 925~d.
Thus, as with the optical light curves, we deduce the appearance 
of an additional source of energy, possibly earlier than 300~d.
\\

\begin{figure*}
\epsscale{0.67}
\plotone{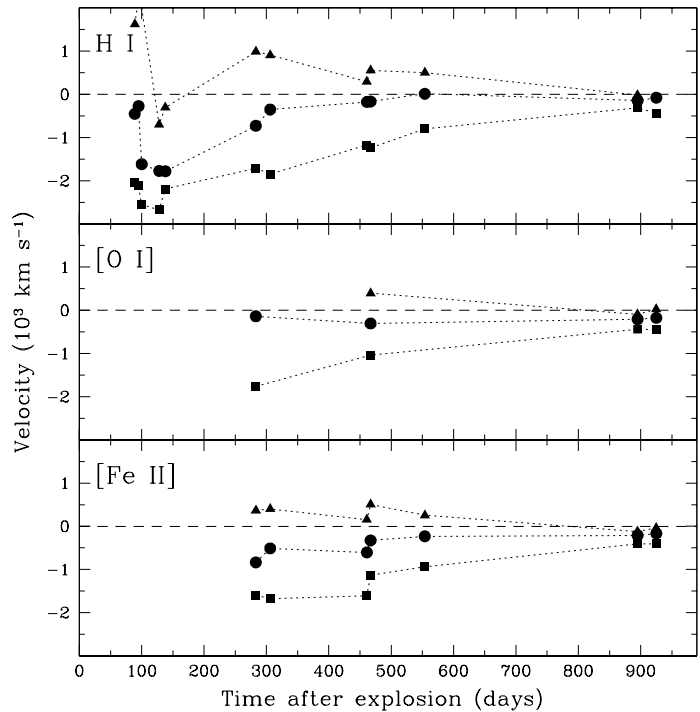}
\caption[]{Evolution of spectral line velocities of SN~2004dj (shifted
to the center-of-mass rest frame of the SN).  The plots are as
follows: triangles=red wings, half-width half-maximum, round dots=peak
intensity, squares=blue wings, half-width half-maximum.  On 895~d and
925~d, the red wing half-width half-maximum refers to the blueshifted
component only. Typical uncertainties in the period 89--554~days, in
\kms, are: $\pm100$ (RHM), $\pm50$ (peak), $\pm50$ (BHM). At 895~d and
925~d the uncertainties are smaller by a factor of $\sim4$.
\label{fig9}
}
\end{figure*}

\begin{figure*}
\epsscale{0.98}
\plotone{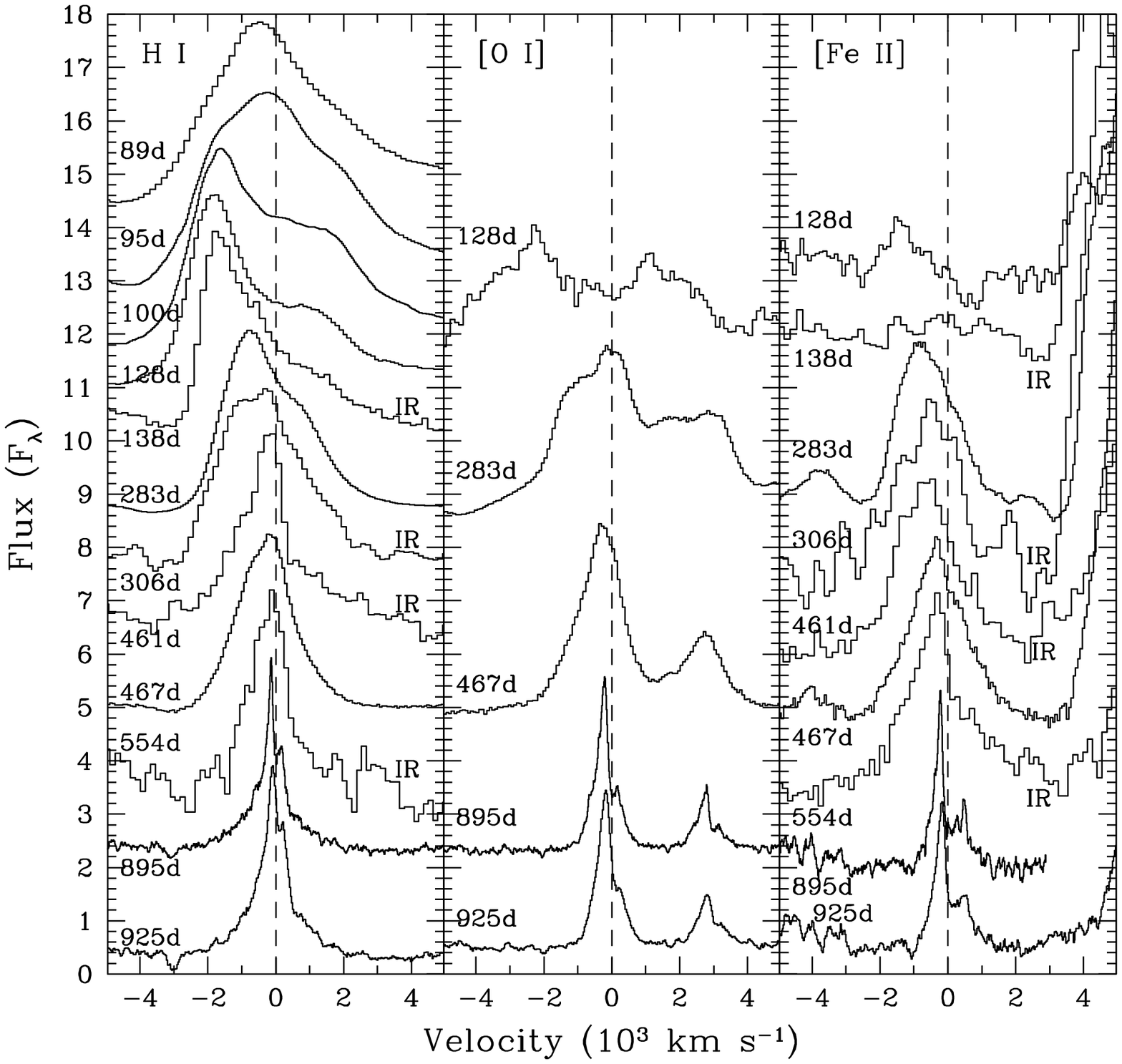}
\caption[]{Evolution of the spectral line profiles of SN~2004dj.  The
89~d and 128~d spectra are from \citet{vin06}, while the 95~d and
100~d spectra are from, respectively, \citet{leo06} and \citet{chu05}.
All other spectra are from the present work.  The horizontal axes are
in terms of equivalent velocity with respect to the SN center of mass,
which has a velocity of +221~\kms\ relative to the Earth.  The LH
panel shows H$\alpha$ and Pa$\beta$ profiles. The Pa$\beta$ profiles
are labelled ``IR''. The middle panel shows [O~I]~6300~\AA\
([O~I]~6364~\AA\ can also be seen lying at $\sim$+3000~\kms in the
[O~I]~6300~\AA\ rest-frame plots).  The RH panel shows
[Fe~II]~7155~\AA\ and [Fe~II]~12567~\AA\ profiles. The
[Fe~II]~12567~\AA\ profiles are labelled ``IR''.  All the profiles
have been flux scaled and shifted vertically for clarity.
\label{fig10}
}
\end{figure*}

Table~\ref{tab6} lists profile parameters expressed as velocities for
the more isolated lines over a range of epochs, shifted to the
center-of-mass rest frame of SN~2004dj.  Also listed (Col. 7) for
461--925~d are the maximum blue-wing velocities derived from profile
model matches (see \S3.4.4.1).  Preliminary inspection indicated that
for lines within a given element (e.g., H$\alpha$ and Pa$\beta$), the
velocities exhibited similar values and evolution. Therefore, in order
to improve the temporal coverage and sampling, H$\alpha$ and Pa$\beta$
were grouped together, as were [Fe~II]~7155~\AA\ and
[Fe~II]~12570~\AA. The evolution of the line velocities (shifted to
the center-of-mass rest frame of the SN) is plotted in
Fig.~\ref{fig9}, and in more detail in Fig.~\ref{fig10}.  These plots
reveal a complex velocity evolution.\\

Up to 138~d, of the three elements considered, only hydrogen lines
could be reliably identified.  At 89~d (corresponding to about
half-way down the plateau-edge), the half width at half-maximum (HWHM)
velocity was $1820\pm60$~\kms, although the red and blue wings
extended to much higher values.  In addition, the peak emission
exhibited a blueshift of $-450\pm50$~\kms. Then, as already described
in \S1, a strong asymmetry rapidly developed, this being attributed to
the emergence of an asymmetric, bipolar core \citep{chu06}.  Rather
than being entirely due to the bulk motion of the ejecta however, some
of the width of the H$\alpha$ and Pa$\beta$ lines may also have been
produced by scattering from thermal electrons as in the cases of
SN~1998S \citep{chu01} and SN~2006gy \citep{smi10}, but this effect is
unlikely to have a significant influence at later epochs.  By the time
of the next observation at 283~d, the asymmetry had diminished, with
the blueshift of the peak now only $-730\pm50$~\kms.  By this time the
lines of [O~I] and [Fe~II] had emerged, also with asymmetric
blueshifted profiles. As the SN continued to evolve, the line widths
narrowed and by 461~d we see the first signs of a sharp suppression of
the red wing. By the time of the next season's observations (895~d,
925~d), this suppression is very pronounced in all three species.
This phenomenon suggests dust formation resulting in the obscuration
of the far side of the ejecta and will be examined in more detail in
\S3.4.4.1.\\

By 895~d and 925~d a second, weaker peak redshifted by $\sim$170~\kms
had also appeared in the H$\alpha$ and [O~I]~6300~\AA\ profiles
(Fig.~\ref{fig10}).  (The redshifted peak can also be seen in the
weaker [O~I]~6364~\AA\ component lying at $\sim$+3000~\kms in the
[O~I]~6300~\AA\ rest-frame plots.)  The previous observations of these
lines were at 467~d when the line luminosities were about a factor of
20 greater (Table~\ref{tab5}) as well as being much wider
(Table~\ref{tab6}).  Therefore an underlying, weaker, redshifted
component could have been present at 467~d or earlier, but was swamped
by the main component of the line.  The 895~d spectrum has a blueshift
in the main peak in H$\alpha$ of $-140\pm10$~\kms\ with the minor peak
showing a redshift of about $+160\pm10$~\kms.  In [O~I]~6300~\AA\ the
corresponding velocities are, respectively, $-210\pm10$~\kms\ and
$+170\pm10$~\kms.  Thus, the main (blueshifted) peak and the weaker
(redshifted) peak lie roughly symmetrically about the local zero
velocity. This suggests that a minor fraction of the line flux
originates in an emission zone centered on the SN and having the
geometry of an expanding ring, jet, or cone.  The [Fe~II]~7155~\AA\
line also shows a secondary peak but at a much larger redshift of
$+480\pm10$~\kms. There may actually be a peak also at around
+170~\kms, similar to those seen in H$\alpha$ and [O~I]~6300~\AA, but
which is swamped by the stronger peak at the larger redshift. The
profiles are analyzed in \S3.4.4.1.\\

We do not consider the MIR line-profile kinematics due to the much
lower resolution ($R_{\rm FWHM} \approx 100$). The formation and evolution
of the MIR lines will be analyzed in a future paper.

\section{Analysis}
\label{sec:analysis}
The evolution of the MIR spectral continuum indicates IR emission from
dust playing a major role in the post-plateau flux distribution of
SN~2004dj. We now make use of the observations described above to
explore the origin, location, distribution, energy source(s), and
nature of these grains. This will be done by comparison of a variety
of simple models with the observations.

\subsection{Correction of the Supernova Flux for S96.}
The position of SN~2004dj coincides with that of the compact star
cluster S96 \citep{san84}, and it seems likely that the progenitor was
a member \citep{wan05}.  Regardless of whether or not this is
the case, it is still important to correct for the contribution of the
cluster to the photometry and spectra, especially at the later epochs,
before we embark on modelling the observed SED.  To do this, in the
optical region we made use of the pre-explosion optical photometry
compiled in table~4 of \citet{vin09}. NIR photometry of S96 was
obtained from 2MASS (see Table~\ref{tab2}).  Unfortunately there are
no pre-explosion MIR images of S96 and so its contribution to the flux
had to be assessed indirectly. \\

\begin{figure*}
\epsscale{0.87}
\plotone{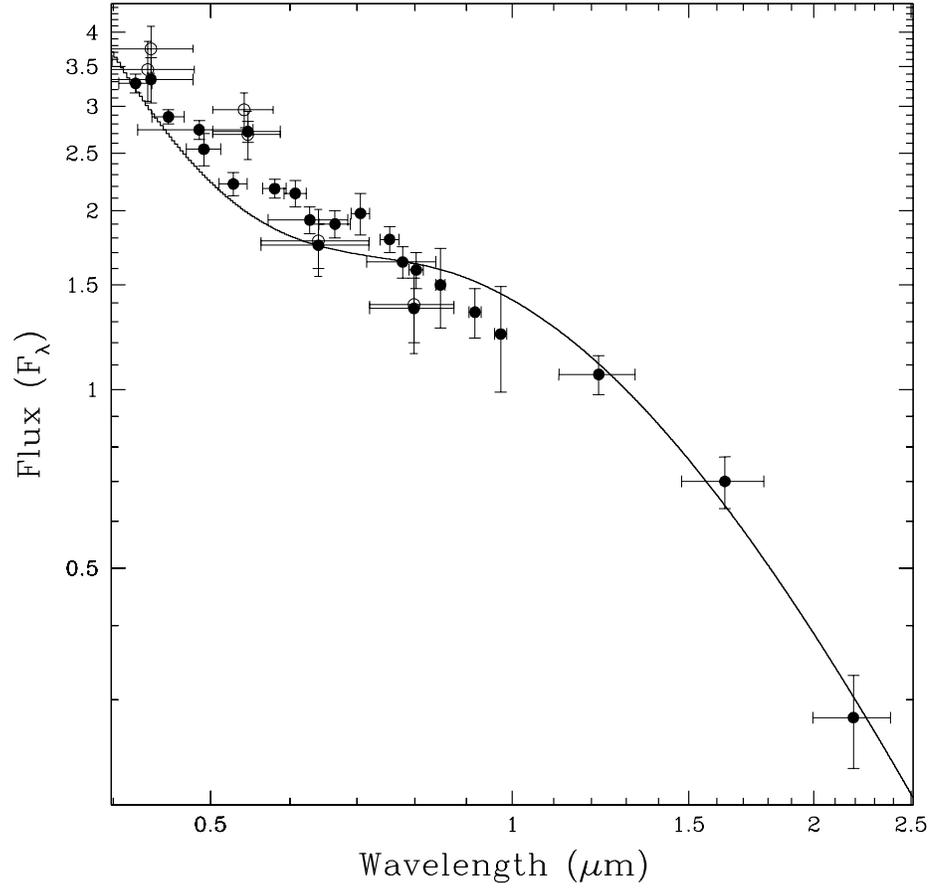}
\caption[]{Two-blackbody representation of the S96 SED. The solid and
open dots represent the pre-explosion and post-explosion fluxes
respectively \citep{vin09}.  The latter data were acquired at 800~d.
In the model, the fluxes longward of $\sim$1~$\mu$m are dominated by
the cold component, with a temperature of 3500~K. The temperature of
the hotter component, 50,000~K, is not intended to have a particular
physical meaning, but simply serves as a means of representing and
extrapolating the optical SED.
\label{fig11}
}
\end{figure*}

We found that longward of $\sim$0.4~$\mu$m the optical/NIR
photometric points could be fairly represented by a combination of two
blackbodies, reddened according to the \citet{car89} law with
$E(B-V)=0.1$~mag and $R_V = 3.1$. This is illustrated in
Fig.~\ref{fig11}.
In this representation, the fluxes longward of $\sim$1~$\mu$m are
dominated by a component with a temperature of 3500~K. At shorter
wavelengths, the hotter component (50,000~K) becomes increasingly
important.  This hot blackbody is not intended as an explanation for
the shortwave radiation, but rather it simply serves as a means of
representing and extrapolating the optical SED.  The contribution of
S96 to the MIR photometric points was then obtained by extrapolation
of the cooler blackbody. (We did not make use of the \citet{vin09}
models to correct for S96 as it was unclear how they should be
extrapolated into the MIR region.) Some support for the effectiveness
of our estimation method, at least for the shorter wavelengths, comes
from four serendipitous 3.6~$\mu$m images of the SN~2004dj field
spanning 1954--2143~d (Table~\ref{tab1}), obtained in {\it Spitzer}
program 61002 (PI W.~Freedman).  These show that, by this period, the
light curve at this wavelength had levelled out at a mean value of
$0.32\pm0.02$~mJy which in good agreement with our estimate for S96 of
$0.28\pm0.05$~mJy (see Fig.~\ref{fig2}).\\

It is possible that the above procedure could underestimate a
contribution from cooler material but it is unlikely that S96 would be
the source of such emission.  \citet{mai04} estimate a cluster age of
13.6~Myr and point out that by this age its parent molecular cloud
should have been dispersed by stellar winds and SN explosions.
\citet{wan05} find an age of $\sim$20~Myr while \citet{vin09} obtain
$\sim10-16$~Myr.  This suggests that the flux contribution from S96 to
longer MIR wavelengths would be small. On the other hand a significant
SN-driven IR echo from the general IS dust of the host
galaxy is quite likely (see \S3.4.3.1). \\

We conclude that the 3500~K blackbody extrapolation provides a
reasonable estimate of the MIR flux contributions from S96.  The
inferred S96 fluxes in the $3.6-24~\mu$m range are shown in
Table~\ref{tab1}.  The values are insensitive to the extinction over
the ranges of $E(B-V)$ ($0.06-0.35$~mag) suggested in the literature.  It
can be seen that the contribution of S96 at 3.6~$\mu$m is significant
as early as $\sim$250~d, and dominates by $\sim$1000~d. As we move to
longer wavelengths, the effect of S96 declines, becoming negligible
for wavelengths longward of $\sim$10~$\mu$m even at the latest epochs.
In the optical-NIR-MIR continuum modelling (see \S3.4.4.2), we use the S96
blackbody representation to correct for its contribution to the SED.

\subsection{Mass of $^{56}$Ni in the Ejecta} 
\begin{figure*}
\epsscale{1.2}
\plotone{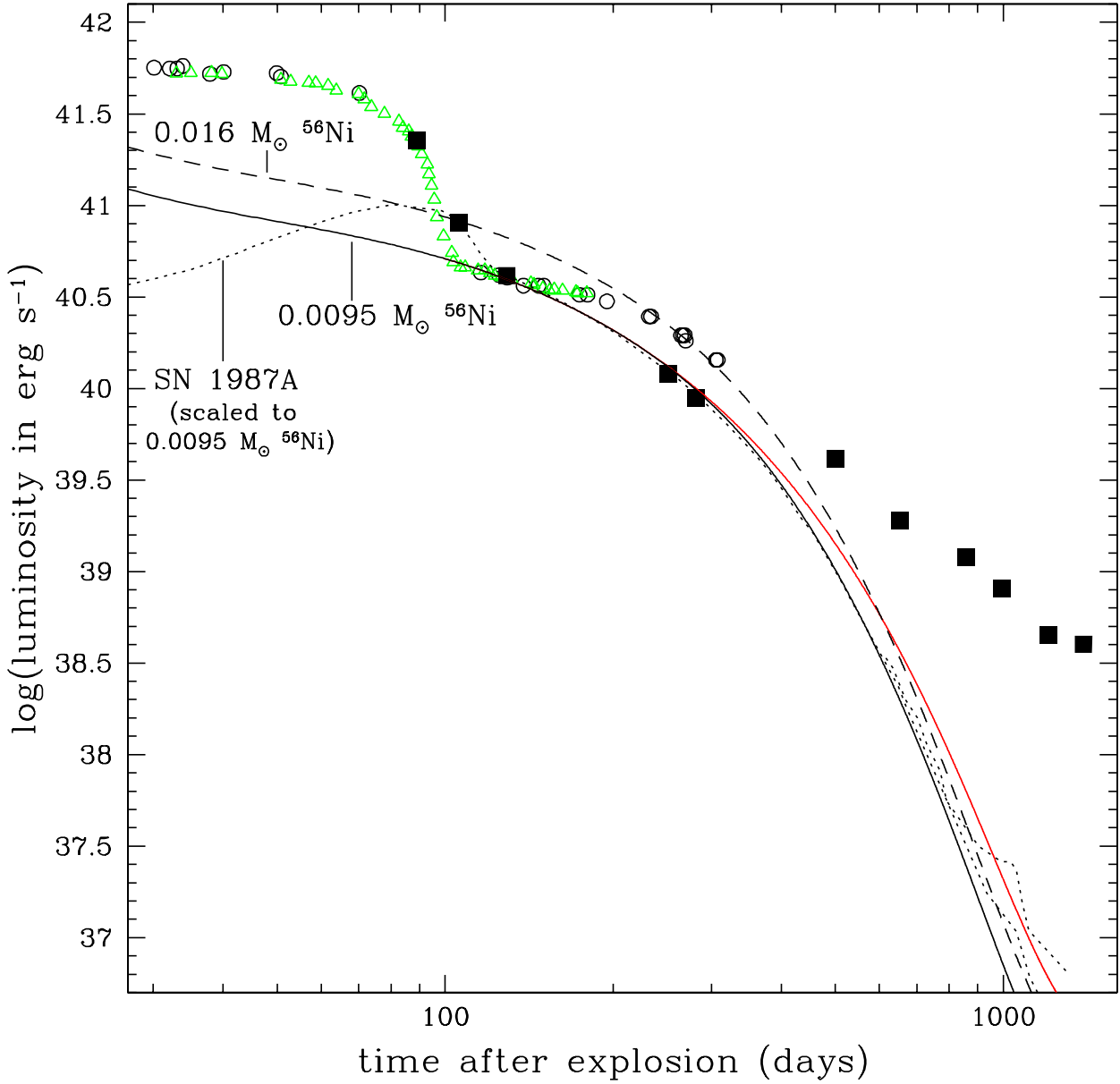}
\caption[]{BLCs of \citet{vin06} (open circles), \citet{zha06} (open
triangles) and the hot+warm luminosities obtained in the present work
via blackbody continuum matching (solid squares) (see \S3.3 and
Table~\ref{tab7}, col.~11.).  The phases of the \citeauthor{vin06} and
\citeauthor{zha06} BLCs have been shifted to an explosion date of
MJD=53196.0, the luminosities scaled to 3.13~Mpc, and the extinction
to $A_V=0.31$~mag.  These BLCs are compared with the radioactive
deposition power for SN~1987A as specified by \citet{li93} \&
\citet{tim96}.  Two scaled cases are shown viz. 0.0095\,M$_{\odot}$
(solid line) and 0.016\,M$_{\odot}$ (dashed line) of $^{56}$Ni.  We
also show the total radioactive luminosity (red) for the
0.0095\,M$_{\odot}$ case.  In addition (dotted line) are the actual
UV-augmented bolometric light curves of SN~1987A \citep{pun95} scaled
to the case of an initial $^{56}$Ni mass of 0.0095\,M$_{\odot}$.  The
divergence between the two SN~1987A datasets at late times is ascribed
to differences in the IR flux measurements between the two
observatories at, respectively, CTIO (lower curve) and ESO (upper
curve).

\label{fig12}
}
\end{figure*}

It is important to establish the mass of $^{56}$Ni in the ejecta of
SN~2004dj since this will allow us to test for the presence of energy
sources, other than radioactive decay, which might be responsible for
the SN luminosity.
In Fig.~\ref{fig12} we show the bolometric light curves (BLCs) of
\citet{vin06} (open circles) and \citet{zha06} (open triangles).  The
phases of these have been shifted to our adopted explosion date of
MJD=53196.0.  This date is nearly one month later than that of
\citeauthor{zha06} reducing their derived $^{56}$Ni mass by about
25\%.  The explosion date of \citeauthor{vin06} is about 10~d earlier
than ours but it is not clear if this would significantly
affect the $^{56}$Ni mass they derived.  In addition to the phase
shifts, the BLCs of these authors have been scaled downward to our
adopted distance of 3.13~Mpc.  This has the effect of reducing the
$^{56}$Ni masses of \citeauthor{vin06} and \citeauthor{zha06} by 19\%
and 10\% respectively.  We scaled the \citet{vin06} BLC by a further
$\times1.1$ in order to allow approximately for the higher total
extinction ($A_V=0.31$~mag) adopted by \citet{vin09} and the present
work.  We scaled the BLC of \citet{zha06} by a further factor of
0.68 to force agreement with our adjusted version of the
\citeauthor{vin06} BLC.  The need for this was due to the much larger
extinction, $A_V=1.02$~mag, adopted by \citeauthor{zha06}, compared
with the $A_V=0.31$~mag adopted in the present work.  It was found
that these adjustments brought our 89~d, 106~d and 129~d blackbody
total luminosities (see \S3.3 and Table~\ref{tab7}, col.~11.)  into
fair coincidence with the other two BLCs (see Fig.~\ref{fig12}).\\

\citet{vin06} constructed their BLC by integrating observed
fluxes in the $BVRI$ bands and then extrapolating linearly from the
$B$ and $I$ fluxes assuming zero flux at 3400~\AA\ and
23,000~\AA. Their BLC extended to 307~d.  \citet{zha06} simply
integrated the observed fluxes in 12 narrow bands between 4000~\AA\
and 10,000~\AA.  Their BLC extended to 154~d, with five additional
points to 180~d obtained by interpolation within a reduced number of
bands. Thus, neither of these BLCs included unobserved excess flux
beyond about 1~$\mu$m. However, for 89--129~d the unobserved MIR flux
made up no more than 10\% of the total luminosity (see
\S3.4.2). Moreover, the optical/NIR region was dominated by continuum
emission at this time.  Consequently, the hot+warm continuum
luminosities obtained from the present work via blackbody matching
(see \S3.3 and Table~\ref{tab7} col.~12.) and plotted in
Fig.~\ref{fig12} (solid squares) agree well with the two adjusted
BLCs. (We exclude the cold component because, as argued in \S3.4.3.1,
it is due to an IS IR echo which was predominantly powered by the
peak luminosity of the SN prior to the earliest epoch of observation.)
By 251/281~d our hot+warm continuum luminosities make up only
$\sim$60\% of the adjusted \citeauthor{vin06} BLC.  This is due to the
relatively strong contribution of line emission to the total
luminosity during this time. Line emission luminosity was not included
in our blackbody matches (see \S3.3).\\

In Fig.~\ref{fig12}, the SN BLCs are compared with the radioactive
deposition power in SN~1987A, as specified by \citet{li93} \&
\citet{tim96}.  These radioactive decay light curves are scaled to,
respectively, 0.0095\,M$_{\odot}$ (solid line) and 0.016\,M$_{\odot}$
(dashed line) of $^{56}$Ni.  Also shown (red) is the total radiocative
luminosity in the case of 0.0095\,M$_{\odot}$.  We also show (dotted
lines) the actual UV-augmented BLCs of SN~1987A \citep{pun95} derived
from observations at ESO and CTIO, scaled to an initial $^{56}$Ni mass
of 0.0095\,M$_{\odot}$.  It can be seen that the 0.0095\,M$_{\odot}$
case provides a good match to the SN~2004dj BLC during $115-150$~d,
just after the end of the plateau phase. After 150~d, unlike SN~1987A,
the SN~2004dj BLC begins to exceed the luminosity of the
0.0095\,M$_{\odot}$ case with the discrepancy growing steadily with
time.  Indeed, even the total radioactive luminosity of the scaled
SN~1987A is exceeded by the SN~2004dj light curve, implying that
an additional source of energy has appeared. \\

Our 0.016\,M$_{\odot}$ deposition plot corresponds approximately to
the 0.02\,M$_{\odot}$ case of \citet{vin06} in their Fig.~18.  We
agree that this case provides a fair match to the BLC during
$\sim260-310$~d.  Nevertheless, viewed within the context of the whole
BLC, it can be seen that this ``match'' is actually due to an
inflection section during the growth of the BLC excess relative to the
true radioactive deposition. Adoption of the 0.016\,M$_{\odot}$ case
would imply an unexplained BLC {\it deficit} during $\sim95-250$~d.
Given the phase of the event and the unexceptional progenitor mass it
is difficult to see how such a discrepancy would come about.
Consequently, we argue that only during the $115-150$~d phase was the
BLC of SN~2004dj actually dominated by radioactive decay.  Beyond this
period, and as deduced also in \S2.2 and \S2.3, an additional
luminosity source emerged. We therefore reject the $^{56}$Ni mass
deduced by \citet{vin06}.\\

We also reject larger $^{56}$Ni masses reported by other authors.
\citet{kot05} used the $V$~band exponential tail method of
\citet{ham03} to derive a $^{56}$Ni mass of $\sim 0.022\,$M$_{\odot}$.
This was based on the $V$ magnitude at 100~d which the subsequently
more complete database shows was not quite yet on the radioactive tail
(see Fig.~\ref{fig4}), thus leading to an overestimate of the
$^{56}$Ni mass.  This method was also one of those used by
\citet{zha06} who obtained $0.025\pm0.010$~M$_{\odot}$ of $^{56}$Ni.
As already indicated, their larger value was due mostly to their much
earlier explosion epoch and much larger extinction, neither of which
we view as likely.  \citet{chu05} obtained $0.020\pm0.006$~M$_{\odot}$
of $^{56}$Ni based on comparison of the $V$ magnitude at 200~d with
that of SN~1987A.  The difficulty here is that by this epoch (as also
in the \citet{vin06} case) an additional power source had appeared in
SN~2004dj, biasing the derived $^{56}$Ni mass to higher values. In
addition, \citet{chu05} used an exceptionally early explosion date,
pushing their result even higher.  Finally we note that both
\citet{chu05} and \citet{zha06} also used the $V$-light curve
``steepness'' method of \citet{elm03}, which is insensitive to
distance and extinction uncertainties.  \citeauthor{chu05} obtained
$0.013\pm0.004$~M$_{\odot}$, consistent with our result.
\citeauthor{zha06} applied the same method to a number of wavebands,
including $V$, but obtained a larger $0.020\pm0.002$~M$_{\odot}$.
Their steepness parameter at just $V$ yields about $0.019$~M$_{\odot}$
suggesting that their use of multiple bands is not the cause of the
apparent disagreement with \citeauthor{chu05} However, the difference
between the \citeauthor{chu05} and \citeauthor{zha06} determinations
is only at the level of $\sim1.5\sigma$ significance.  \\

We conclude that, taking into account the uncertainties in fluxing,
adopted distance, extinction and explosion epoch (see \S1.1) the mass
of $^{56}$Ni ejected by SN~2004dj was
$0.0095\pm0.002$\,M$_{\odot}$. We adopt this value for the rest of the
paper.

\subsection{Comparison of Observed Continua with Blackbody Radiation}
Here we begin to consider the location and energy source of the SN
continuum, especially longward of 2~$\mu$m where thermal emission from
dust would appear.  To take an initially neutral standpoint on the
interpretation, we compared optical, NIR and MIR spectra and
photometry with blackbody continua. This provides us with the minimum
radii of the emitting surfaces.  The epochs were selected primarily as
those for which MIR spectra were available, although the earliest such
epoch, 106~d, was already during the nebular era.  The earliest MIR
photometry was acquired at 89~d when the SN light curve was only about
half way down the fall from the plateau to the nebular level.  Given
the potential interest of this epoch we began our model comparisons at
this epoch despite the lack of an MIR spectrum. In addition, to
compensate for the large gap between 859~d and 1207~d we also
considered the 996~d SED based on MIR photometry only.  Optical
photometry was taken from \citet{vin06,vin09,zha06}. Details about the
sources of the other data are given in Tables~\ref{tab1}--\ref{tab4}.
Apart from 89~d and 996~d, all the optical and NIR data plus the MIR
photometry were flux-scaled by interpolation of the light curves to
the epochs of contemporaneous MIR spectra.  \\

\begin{figure*}
\epsscale{1.3}
\plotone{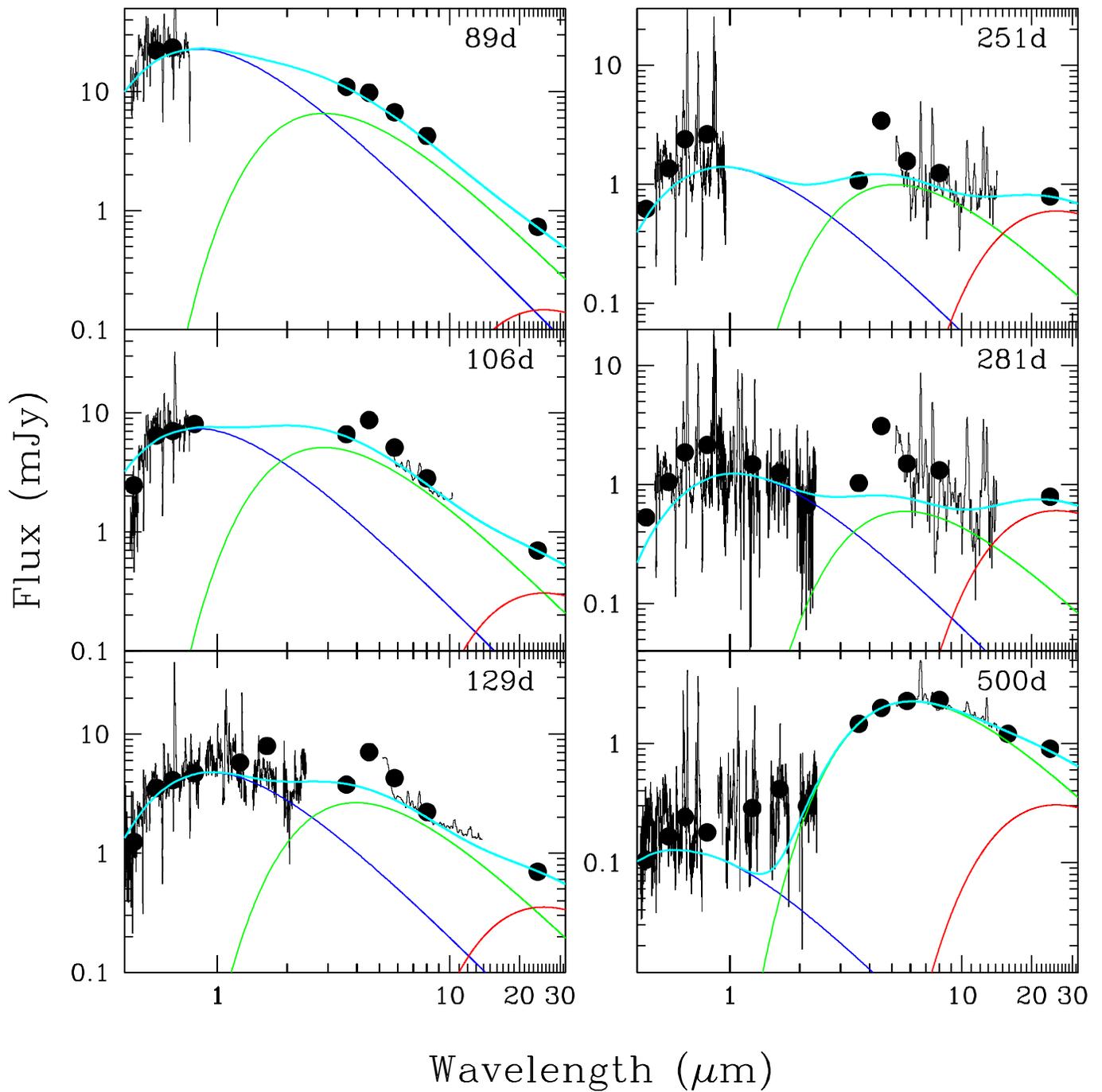}
\caption[]{Comparison of blackbody continua with the optical-NIR-MIR
photometric fluxes and spectra of SN~2004dj at six epochs spanning
$89-500$~d. The solid dots show contemporary photometry.  The blue,
green and red lines show, respectively, the hot, warm and cold
blackbodies. The cyan line shows the total model spectrum.
\label{fig13}
}
\end{figure*}

\begin{figure*}
\epsscale{1.3}
\plotone{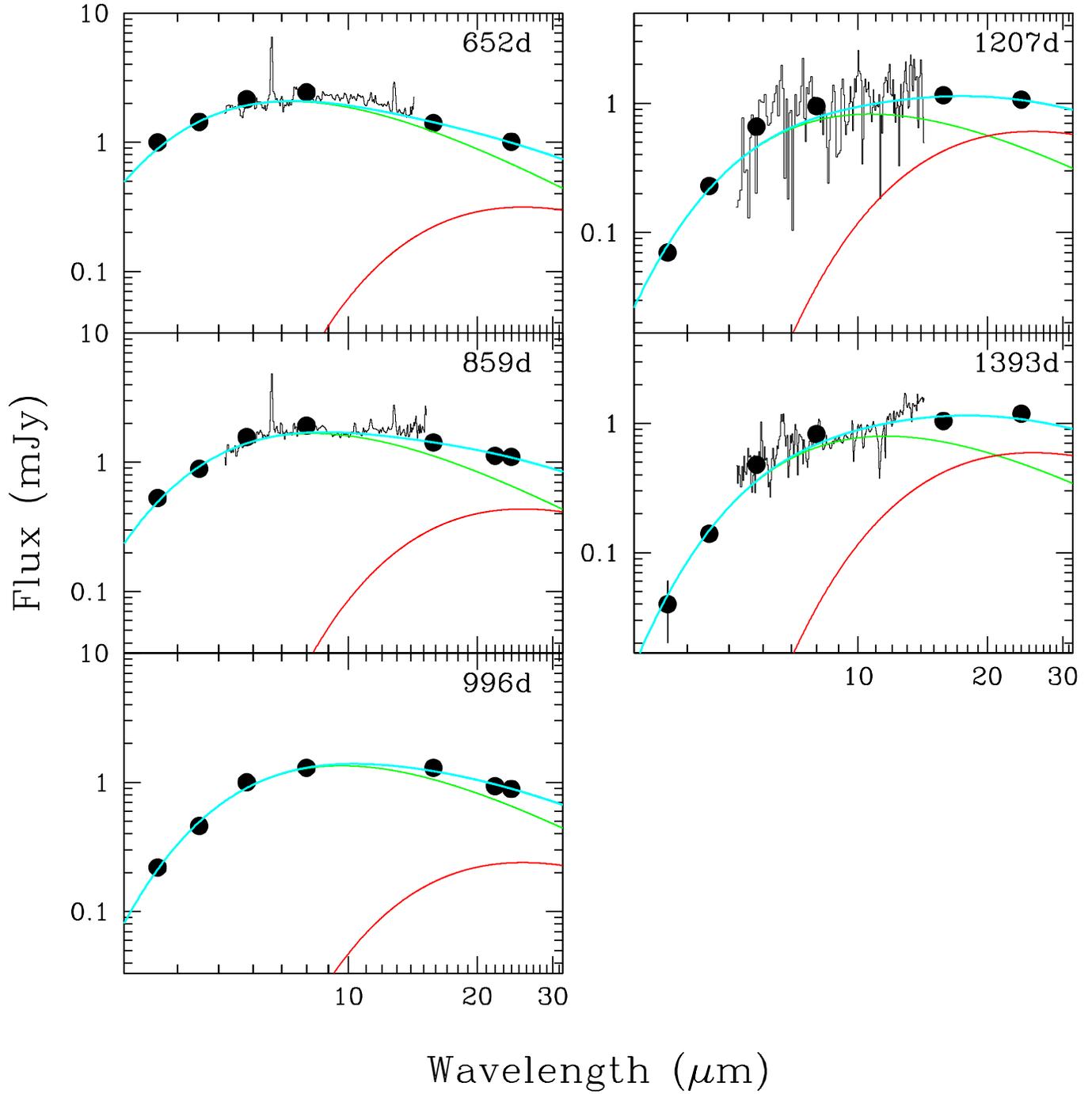}
\caption[]{Comparison of blackbody continua with the MIR photometric
fluxes and spectra of SN~2004dj at five epochs spanning
$652-1393$~d. The solid dots show contemporary photometry.  The green
and red lines show, respectively, the warm and cold blackbodies.  The
cyan line shows the total model spectrum.  A hot component was also
included in the matches but had a negligible effect on the warm-cold
matches to the MIR continua.  Consequently, and in order to show the
MIR behavior in more detail, the optical/NIR region is not shown.
Estimates of the hot component contributions are given in
Table~\ref{tab7}.
\label{fig14}
}
\end{figure*}

The contribution of S96, represented by a 3500~K blackbody of radius
$1.5\times10^{14}$~cm (see \S3.1), was first subtracted from all the
data.  To model the resulting $0.4-24~\mu$m SN continuum it was found
to be necessary to use three blackbodies (``hot'': $5300-10000$~K,
``warm'': $320-1750$~K and ``cold'': $\sim$200~K).  These were
reddened and then matched visually to the continua.  The hot blackbody
was first added and adjusted to match the optical continuum. While the
hot blackbody provides some information about the energy budget of the
shorter wavelength part of the spectrum, the main reason for its
inclusion in this study was to allow correction for its effect on the
net continuum in the NIR where, up to about 500~d, it is comparable in
strength to the warm component.  The warm blackbody was then added and
adjusted to match the $\sim5-10~\mu$m continuum plus the long
wavelength end ($2-2.4~\mu$m) of the NIR continuum.  It was found
that, starting at the earliest epoch, as the SN evolved the hot+warm
blackbody flux longward of 10~$\mu$m increasingly fell below that of
the observations.  Therefore a cold blackbody was added and adjusted
to provide the final match.  For epochs where photometry but no
spectra were available we used the temporally nearest spectral matches
to indicate the likely position of the underlying continuum.  The
expansion velocities of the blackbody surfaces, $v_{hot}$, $v_{warm}$
and $v_{cold}$, and temperatures, $T_{hot}$, $T_{warm}$ and
$T_{cold}$, for these matches are tabulated in Table~\ref{tab7}. The
warm blackbody radii, $R_{warm}$ are also listed. The model matches
are displayed in Figs.~\ref{fig13} ($89-500$~d) and \ref{fig14}
($652-1393$~d).  In Table~\ref{tab7} we also show the luminosities,
$L_{hot}$, $L_{warm}$ and $L_{cold}$, of the three blackbody
components together with the sum of the hot and warm components
($L_{total}$).  We exclude the cold component because, as will be
argued in \S3.4.3.1, it is due to an IS IR echo which was predominantly
powered by the peak luminosity of the SN prior to the earliest epoch
of observation.  In the final column is listed the radioactive
deposition power corresponding to the ejection of 0.0095\,M$_{\odot}$
of $^{56}$Ni, scaled from the SN~1987A case specified by \citet{li93}
\& \citet{tim96}.\\


We stress that the blackbody matches were to the underlying spectral
{\it continua} where this could be reasonably judged, and not to the
photometric points which also contained flux from line emission.
Thus, the models sometimes lie below the average level of the spectra.
This is particularly so for later epochs at wavelengths shortward of
2~$\mu$m where the spectra are dominated by broad, blended emission
lines. This tends to mask most of the underlying thermal continuum,
leading to a possible overestimation of the hot
continuum. Nevertheless, the blackbody luminosities tend to {\it
underestimate} the total luminosity as they do not allow for the total
line emission. This is particularly the case around $200-500$~d when
the relative contribution of nebular line emission to the total
luminosity is at a maximum.  There is less of a problem before this
era when the hot continuum is relatively strong, or afterwards when
the warm/cold continuum increasingly dominates. Also, by 652~d, the
relative weakening of the hot continuum means that it has a negligible
effect at wavelengths longward of 2~$\mu$m.  In the $2-14~\mu$m region
the true continuum level is easier to judge.  By 251~d and later, the
total $2-14~\mu$m flux exceeds that of the continuum model by no more
than 25\%.  \\

The hot continuum declined monotonically and dominated the total SN
continuum luminosity up to about a year post-explosion.  It presumably
arose from hot, optically-thick ejecta gas.  At 89~d most of the hot
continuum was probably still driven by the shock-heated photosphere,
with the remainder being due to radioactive decay (see
Table~\ref{tab7} and Fig.~\ref{fig12}).  After 500~d the hot continuum
became relatively weak and of low S/N. In addition, the NIR spectral
dataset ended on 554~d. For subsequent epochs the strength of the hot
component was estimated by extrapolation.  At 500~d the hot blackbody
match was achieved with a temperature of 10,000~K. Between 251~d and
500~d the blackbody velocity declined exponentially with an e-folding
time of $\sim$130~d and so, for epochs after 500~d the hot blackbody
temperature was fixed at 10,000~K and the velocity obtained by
extrapolation of the earlier exponential behavior (see
Table~\ref{tab7}).  While this is likely to be increasingly inaccurate
with time, it is unlikely to be a serious source of error in
determining the warm continuum; for example, by 500~d the hot
blackbody contributed barely 1\% of the flux at
3.6~$\mu$m. Consequently, and in order to illustrate the MIR behavior
in more detail, the optical/NIR region is not shown in the later plots
(Fig.~\ref{fig14}).\\

The warm component luminosity declined monotonically to 281~d but
then, unlike the hot component, increased by a factor of 3.5
by 500~d.  By that time the hot+warm SN continuum luminosity exceeded
that of the radioactive energy input by a factor of 4 and this
excess continued to increase with time. The cause of the growing
excess was the warm component luminosity.  We also note that
$R_{warm}$ more than doubled between 281~d and 500~d, but remained
roughly constant thereafter. In fact, as we show later (\S3.4.4.1,
\S3.4.4.2), after 500~d slow shrinkage in the size of the warm emission
region occurred.  \\

The cold component is primarily defined by the 24~$\mu$m point, and
can be fairly reproduced using a range of temperatures ($150-300$~K)
and velocities.  Its luminosity remained roughly constant throughout
the observations.  In Table~\ref{tab7} we show the case with the
temperature fixed at 200~K.  \\

It is interesting that, even as early as 89~d, warm and cold
blackbodies had to be included to achieve a fair match to the observed
fluxes in the NIR-MIR region. This will be discussed in \S3.4.2 and
3.4.3.1.  By 106~d, the nebular phase was just beginning and by 129~d
the hot+warm luminosity was driven predominantly by radioactive decay.
By 251~d and 281~d, the radioactive deposition exceeded the sum of the
hot+warm blackbody luminosity by about 15\%. This excess probably went
into powering the line emission not included in the blackbody matches.
As already indicated, the nebular line emission was particularly
strong at this time.  Indeed it was responsible for a $\sim50\%$
excess of the total bolometric luminosity relative to the radioactive
deposition (Fig.~\ref{fig12}), indicating the emergence of an
additional energy source, probably the reverse shock (\S3.4.4.2).  The
appearance of an additional power source has already been indicated by
the flattening of the optical light curves (\S2.2).  Indeed, after
$\sim$150~d the true BLC exhibited a steadily growing excess relative
to the radioactive deposition (see Fig.~\ref{fig12}). \\

The hot and warm blackbody velocities never exceeded 1750~\kms,
indicating continuum emission consistent with an origin in the ejecta
or ejecta/CSM interface.  The cold component exhibits velocities
between 4500~\kms\ and 8500~\kms\ during the earlier phase pointing
to an origin more likely to be outside the ejecta, specifically an IR
echo from pre-existing dust. \\

We conclude that the IR continuum comprised at least two components.
The temperatures and temporal variation of these components point to
thermal emission from dust whose energy source is ultimately the
supernova. The surge in the luminosity of the warm component by 500~d
suggests the emergence of an additional source of radiation --- i.e., 
that the warm component was driven by different energy sources at,
respectively, early and late times. It also raises the possibility
that distinct dust populations were responsible for the early and 
late-time warm components. \\

\subsection{Origin of the Infrared Radiation}
We now explore the origin of the IR continuum radiation from
SN~2004dj, especially the warm component.  To do this we have
constructed a model continuum comprising hot gas, warm local dust, and
cold IS dust. We also made use of spectral line red-wing suppression
in the study of the warm dust.  The continuum model was adjusted to
provide visual matches to the observations.

\subsubsection{The Hot Component}
As in \S3.3, the hot continuum component, presumably due to
optically-thick ejecta gas, was represented using a hot blackbody
having a temperature of 5300--10000~K.  This blackbody radius and
temperature was adjusted to obtain a match to the optical-NIR
continuum.  The warm component model (see \S3.4.2 and \S3.4.4.2) was
then added and adjusted to match the $\sim$5--10~$\mu$m continua plus
the long-wavelength end (2--2.4~$\mu$m) of the NIR continuum.  As with
the pure blackbody matches it was found that, at all epochs, the warm
component model flux longward of 10~$\mu$m increasingly fell below
that of the observations. This excess is attributed to an IS IR echo
(see \S3.4.3.1).

\subsubsection{The Warm Component: Early Phase IR Excess due to a CDS}
As pointed out above, a striking result from the hot-blackbody matches
is that an NIR-MIR excess (relative to the hot component flux) was
present as early as 89~d post-explosion. (For brevity we henceforth
refer to the NIR-MIR continuum excess as the ``IR excess''.)  The MIR
spectra from 106~d onward show that this was primarily due to
continuum emission. We have no reason to suspect that the IR excess in
the 89~d SED was not also due to continuum emission.  Indeed, the CO
peak at 4.5~$\mu$m is noticeably suppressed at this epoch, compared
with 106~d (Fig.~\ref{fig3}).  The obvious interpretation of the IR
excess is that it arose from warm dust heated by the supernova. The
very early appearance of this emission, when the H-recombination front
had not yet reached the He-metal core, argues against an origin in
newly-formed ejecta dust.  In addition, SN-ejecta dust-formation
models \citep[e.g.][]{tod01,noz03} suggest that dust formation in a
CCSN is unlikely to occur until after one year post-explosion. A
second possibility of direct shock-heating of pre-existing
circumstellar dust is also ruled out.  The SN UV flash would evaporate
dust out to $(0.5-1.0)\times10^{17}$~cm ($0.016-0.032$~pc) (see
\S3.4.3.1).  To reach this distance by 89~d would require a shock
velocity of at least 65,000~\kms.  Yet the velocities of the H$\alpha$
trough \citep{kor05,vin06} as well as of the extreme blue-profile edge
(our measurements of the spectra of \citet{kor05,vin06}) during the
period $25-90$~d suggests that the bulk of the ejecta never exceeded
velocities much more than $\sim$15,000~\kms. Indeed \citet{chu07}
adopt a velocity of 13,000~\kms as the boundary velocity in their
treatment of SN~2004dj.  A third possibility, CSM dust heating by
X-rays, is also implausible. The luminosity of the warm continuum
component at 89~d was $1.3\times10^{40}$~\ergs (Table~\ref{tab7}) but
the X-ray luminosity at 30~d was only about 1\% of this \citep{poo04}.
In the case of dust in a CDS, collisional heating is ruled out as the
energy available in the CDS is about a factor of $10^5$ less than that
required to account for the observed IR excess.  The heat capacity 
of the grains is also insufficient to account for required energy. \\

The most likely explanation for the early-phase IR excess is an IR
echo of the SN early-time luminosity from circumstellar dust. There
are two possible scenarios here. In the first of these, the IR echo is
from pre-existing dust in the CSM.  Such early IR echoes have been
suspected before in other CCSNe.  \citet{woo93} found an IR excess in
SN~1987A as early as 260~d and possibly also at just 60~d. They
hypothesised that the origin of the excess was warm, SN-heated dust in
the CSM. \citet{woo97} reiterates that the cause was CSM dust
``echoing the light curve''.  \citet{fas00} reported a strong $K-L'$
excess in the emission from the Type~IIn SN~1998S at 130~d.  They
attributed this to pre-existing CSM dust heated either by the SN
luminosity (a conventional IR echo) or by X-rays from the CSM-shock
interaction. \citet{poz04} argued in favor of the former scenario.\\

In the second IR echo scenario, as the fast moving ejecta collides
with the CSM, a CDS forms between the forward and reverse shocks.  As
noted in \S1, \citet{chu07} have used the H$\alpha$ spectrum to deduce
the existence of a CDS in SN~2004dj.  Within the CDS conditions can
allow new dust to form \citep{poz04}. CDS dust has been invoked as the
origin of the early-time IR excess of SN~2006jc
\citep{smi08,dic08,mat08}.  \citet{mat08} showed that the IR emission
was probably an IR~echo from the CDS dust.  IR excesses in CCSNe at
later times have also been attributed to emission from newly-condensed
CDS dust viz. in the Type~IIn SN~1998S \citep{poz04} and the Type~IIP
SN~2007od \citep{and10}.  \\

We explored the possibility that the early IR excess was due to an IR
echo of the SN luminosity from spherical distributions of either
pre-existing circumstellar dust or newly-formed CDS dust. Details of
the model are given in \citet{mei06}. It fully allows for the effects
of light-travel time across the dust distribution.  Versions of this
model have also been used in \citet{mei07,mat08,bot09,kot09}. The
model assumes a spherically symmetric cloud of grains centered on the
SN, with a concentric dust-free cavity at the center.  The SN is
treated as a point source. For simplicity, a single grain radius, $a$,
is adopted.  For ease of computation, we assumed that the grain
material was amorphous carbon where, for wavelengths longer than $2
\pi a$, the grain absorptivity/emissivity can be well-approximated as
being proportional to $\lambda^{-1.15}$ \citep{rou91}.  For shorter
wavelengths, an absorptivity/emissivity of unity was used.  The
material density is 1.85\,g\,cm$^{-3}$ \citep{rou91}.  Free parameters
are the grain size, grain number density, radial density law and
extent.  The input luminosity is a parametrized description up to
550~d of the BLC shown in Fig.~\ref{fig12} viz.: \\

$L_{\rm bol} = L_0 {\rm exp}(-t/\tau)$, where \\
$L_0=57.0$, $\tau=1000.0$~d for $t \le 0.2$~d,\\
$L_0=1.70$, $\tau=23.4$~d for $0.2 < t \le 23.0$~d,\\
$L_0=0.66$, $\tau=141.9$~d for $23.0 < t \le 76.4$~d,\\
$L_0=50.7$, $\tau=15.9$~d for $76.4 < t \le 112.0$~d.\\
$L_0=0.079$, $\tau=174.8$~d for $112.0 < t \le 550.0$~d \\
$L_0=0$ for $t > 550.0$~d \\

\noindent $L_0$ is in units of $10^{42}$~\ergs.  The brief but
highly luminous first term represents the energy in the UV flash.  The
second term, which covers the pre-discovery phase, was estimated by
using the BLC of SN~1999em \citep{elm03}, adjusted so that the epoch
and luminosity of the beginning of its plateau phase coincided
approximately with the earliest observed point on the SN~2004dj BLC.\\

For the third, fourth and fifth terms, the adjusted BLCs of
\citet{vin06} and \citet{zha06} were used (see \S3.2).  As already
pointed out, these BLCs did not include excess flux beyond about
1~$\mu$m, nor any flux beyond 2.3~$\mu$m; virtually all of the warm
and cold components were excluded. Thus, use of the above parametrized
description avoids ``double-counting'' of a putative CSM/CDS IR echo.
Indeed the BLC description slightly underestimates the true SN
luminosity input as it excludes all IR emission longward of 2.3~$\mu$m,
as well as all UV emission shortward of 0.34~$\mu$m.  Consequently the
input luminosity was scaled by a factor of about 1.1 to allow
for unobserved UV and IR fluxes. \\

For the case of pre-existing CSM dust the model was adjusted to
reproduce the IR-excess SEDs for the three earliest epochs (89, 196
and 129~d).  The outer limit of the circumstellar dust was initially
set at 10 times that of the cavity radius, and a $r^{-2}$ (steady
wind) density profile was assumed.  However, it was impossible to
reproduce the quite rapid temporal decline of the IR~excess without
raising the temperature of the hottest grains to above their
evaporation temperature. While a better match to the SED shape and
evolution was obtained by setting the density profile to steeper than
$r^{-4}$, the best match was achieved with a discrete, thin shell. We
therefore adopted this configuration, setting the shell thickness at
$\times0.1$ the cavity radius.  \\

We investigated a range of cavity radii.  For pre-existing CSM dust
the minimum size of the concentric dust-free cavity is fixed by the
extent to which the dust was evaporated by the initial UV flash from
the supernova. In this scenario, while the bolometric light curve (BLC)
dominates the heating of the surviving dust, the size of the dust-free
cavity is determined by the luminosity peak of the UV flash, with
$r_{evap} \propto L_{peak}^{0.5}$ \citep{dwe83}. For a Type~IIP SN the
flash luminosity is estimated to peak at about $10^{45}$~\ergs
\citep{kle78,tom09} although it has never been observed directly.  A
similar peak luminosity is estimated for the Type~IIpec SN~1987A
\citep{ens92}.  \citet{dwe83} provides an approximate estimate of the
flash-evaporated cavity size for a Type~II supernova. More recently,
in a detailed study \citet{fis02} determined that for SN~1987A the UV
flash would have totally evaporated graphite CSM dust out to a radius
of $(0.5-0.9)\times10^{17}$~cm ($0.016-0.029$~pc).  We therefore adopt
the \citeauthor{fis02} estimates and apply them to the case of
SN~2004dj.  In any case, matches to the data are fairly insensitive to
the cavity radii.  Fair matches to the early-time IR-excess SEDs were
obtainable with cavity radii $(0.5-1.3)\times10^{17}$~cm
($0.016-0.042$~pc) and corresponding grain radii of
$0.1-0.04~\mu$m. For cavity radii exceeding $\sim$0.05~pc the model
continuum slopes were inconsistent with the observations.\\

As an example of a pre-existing CSM dust model match we have: \\

\noindent cavity radius: $0.7\times10^{17}$~cm (0.023~pc), \\
shell thickness: $0.7\times10^{16}$~cm (0.0023~pc),\\ 
grain size: 0.07~$\mu$m, \\
grain number density: $6.0\times10^{-9}$~cm$^{-3}$,\\ 
total grain mass: $0.38\times10^{-5}$~M$_{\odot}$. \\ 

The optical depth through the shell in the optical band is 0.006,
which is easily encompassed within the observed total extinction of
$A_V=0.3$~mag.  For a dust/gas mass ratio of 0.005 (see \S3.4.3.1), the
dust mass corresponds to a total shell mass of
$0.76\times10^{-3}$~M$_{\odot}$.  For the adopted shell thickness and
a typical RSG wind velocity of 20~\kms, this mass would be produced by
a mass loss event about 1100~years ago, lasting for 110~years, with a
mass loss rate rate of $7\times10^{-6}$~M$_{\odot}$~yr$^{-1}$.
However, while this mass loss and loss rate are plausible for RSGs,
such discrete events are not thought to occur in this type of star.
This prompts us to seek a more natural explanation for the thin dust
shell. An obvious candidate is the CDS inferred by \citet{chu07}.\\

For the CDS scenario the size of the cavity is essentially the radius
of the CDS. From \citet{chu07}, the CDS radius is given by 
\begin{equation}
 r_{CDS}=5.2\times10^{15}\times((t+11.5)/64.0)^{6/7}~cm 
\end{equation}
\noindent where $t$ (in days) is with respect to our explosion epoch. Thus,
at the earliest of our epochs, 89~d, the CDS radius was just
$7.7\times10^{15}$~cm, or about one tenth of the pre-existing CSM dust
cavity. This is only 3~light days implying that light travel time
effects are small.  Nevertheless, for convenience and ease of comparison
with the pre-existing dust case, we applied the IR~echo model to the
CDS case. Although the \citeauthor{chu07} study terminates at just
99~d, we assume that the CDS radius continues to increase as described
in equation~(1) until at least 500~d, by which time its
contribution to the IR excess is small. \\

We compared the CDS case of the IR~echo model with the observations,
with the dust lying in a thin shell of radius $r_{CDS}$ as given
above. We found that setting the dust mass at a constant value
produced a poor match to the observations. With a match at 106~d, the
model at 89~d yielded a continuum which matched the observed IR excess
at 8~$\mu$m but exceeded the IR excess by nearly $\times2$ at
3.6~$\mu$m.  On the other hand, at 129~days, while the model continuum
slope was similar to that of the observed IR excess, it significantly
underproduced it with the deficit being as large as $\sim$35\% at
8~$\mu$m. We propose that this problem is due to the unjustified
assumption of a fixed CDS dust mass.  CDS dust could not form until
the supernova flux at the CDS had faded sufficiently for proto-dust
material to cool below the condensation temperature. This occurred at
about 50~d, assuming amorphous carbon dust. Therefore, following
\citet{mat08} we allowed the CDS dust mass to grow as
$M_d=M_0(1-exp(-(t-t_0)/t_d))$ where $t$ is time, $t_0$ is the time at
which dust condensation began (set at 50~d), $t_d$ is the
characteristic grain growth timescale and $M_0$ is the
asymptotically-approached final mass.  (We note that
\citeauthor{mat08} also deduced an epoch of 50~d for the start of the
CDS dust condensation in SN~2006jc.)  No attempt was made to simulate
the growth of individual grains which were assumed to appear
instantaneously at their final size.  Owing to the light travel time
differences across the CDS, the grain condensation is seen to commence
during the epochs ($t_0-(r_{CDS}/c)$) to ($t_0+(r_{CDS}/c)$) days.
Yet, even as late as 500~d, $(r_{CDS}/c)$ was only about 12~light days
and so this effect was ignored. \\

\begin{figure*}
\epsscale{0.9}
\plotone{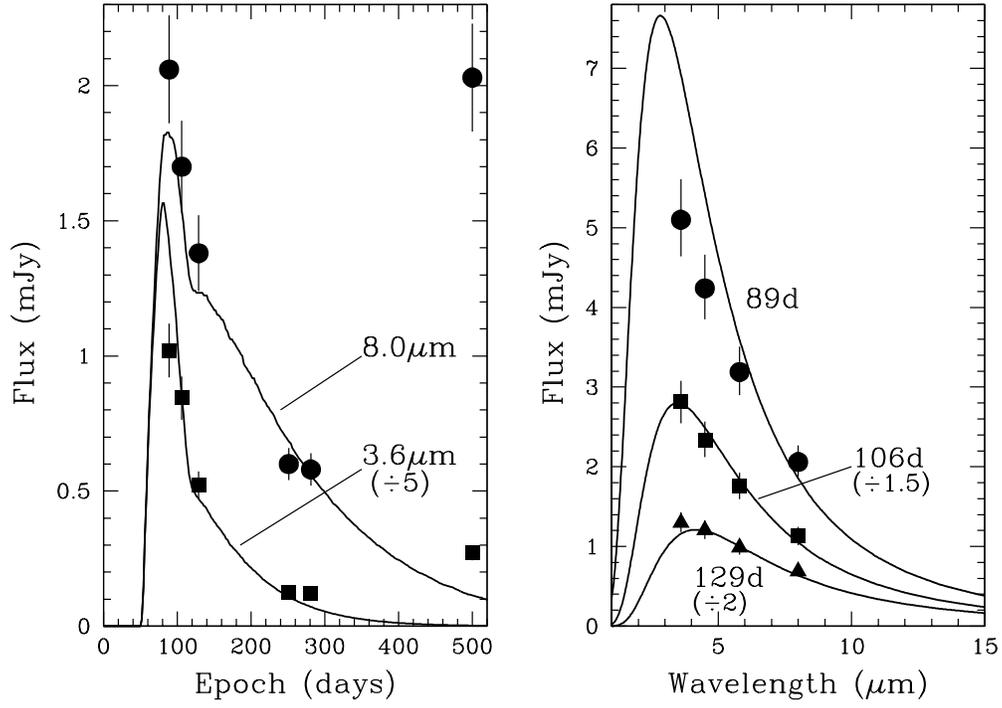}
\caption[]{CDS IR~echo model compared with the 3.6~$\mu$m and
  8.0~$\mu$m {\it MIR excess} light curves of SN 2004dj (LH panel),
  and the 89, 106 and 129~d {\it MIR excess} SEDs (RH panel).  The
  grain growth began at 50~d with a growth timescale also of 50~d.
  Amorphous carbon grains were adopted of radius 0.2~$\mu$m.  The
  final dust mass is $0.33\times10^{-5}$~M$_{\odot}$.
\label{fig15}
}
\end{figure*}

In Fig.~\ref{fig15}, we compare the CDS model light curves and spectra
with, respectively, the observed MIR excess fluxes at 3.6~$\mu$m and
8.0~$\mu$m (LH panel), and with the 89, 106 and 129~d MIR excess SEDs
(RH panel).  The free parameters for the CDS case were the dust mass
scaler, the grain radius and the grain growth timescale.  Satisfactory
reproduction of the IR excess was achieved for all five epochs
spanning 89--281~d with $t_d=50$~d, a grain radius of 0.2~$\mu$m and a
final dust mass of $M_0=0.33\times10^{-5}$~M$_{\odot}$, similar to the
dust mass derived in the pre-existing CSM case above.  The CDS mass is
$3.2\times10^{-4}$~M$_{\odot}$ \citep{chu07} indicating a plausible
final dust/gas mass ratio of 0.01.  The hottest dust ranged from 1330
K at 89 days declining to about 640 K at 281 days.  The optical depth
in the UV/optical range was 0.077 at 89~d falling to 0.011 by 500~d
--- i.e., consistent with the observed total extinction of
$A_V=0.3$~mag.  By 500~d the MIR excess exceeds that of the model by a
significant factor (especially at longer wavelengths)
(Fig.~\ref{fig15}) implying the appearance of an additional energy
source.  The CDS contribution to the SN continua for 89--500~d is
plotted in Fig.~\ref{fig18}.  The luminosity contribution of the CDS
IR~echo, $L_{CDS}$, is listed in Table~9, Col.~3.  For epochs after
500~d the CDS component was negligible and so was ignored.  \\

We conclude that the early-time IR~excess was probably due primarily
to an IR~echo from newly-formed dust lying within the CDS.  The rapid
decline of the IR excess flux, especially between 89~d and
106~d, is due largely to the ongoing fall from the BLC plateau,
tempered by the growth of CDS dust during this period.  Owing to the
small size of the CDS, light travel time effects are small.  In
contrast, in the pre-existing CSM dust scenario, the rapid decline of
the input luminosity is tempered by the larger size of the dusty shell
which produces significant light travel time effects, smoothing the
observed IR excess light curve over longer timescales. Nevertheless,
given the natural explanation by the CDS model of the required thin
shell dust distribution, in the completion of the analysis below we
use the CDS scenario.  \\
\subsubsection{The Cold Component} 
\paragraph{{\it An Interstellar Echo\/}}
The cold component is defined primarily by the 24~$\mu$m data and can
be fairly reproduced using a range of temperatures ($150-300$~K) and
velocities.  As described above (\S3.3), the luminosity of the cold
component remained roughly constant throughout the observations.  In
addition, the cold blackbody velocities were high, arguing against an
origin in ejecta dust. An obvious alternative is an IS IR echo.  The
possibility of detecting the reflection of SN optical light from IS
grains was first suggested by \citet{vdb65}.  In the IR a potentially
much more important phenomenon is the absorption and re-radiation by
the grains of the SN BLC energy (the ``IR echo''). The possibility of
detecting an IS IR echo from a SN was first proposed by
\citet{bod80}.\footnote{\citet{wri80} also considered an IR echo from
a SN but only in the more restricted case of an explosion within a
molecular cloud.}  The occurrence of cold dust IS IR echoes should be
relatively common for SNe occurring in dusty, late-type galaxies.  The
SEDs of such echoes tend to peak in the $20-100~\mu$m region, allowing
echo detection in nearby galaxies by {\it Spitzer} during its cold
mission.  {\it Spitzer}-based evidence of this phenomenon in the Milky
Way Galaxy have been presented for the Cassiopeia~A SN
\citep{kra05,kim08,dwe08}.  \citet{mei07} showed that an IS IR echo
provided a natural explanation for the strength and decline of the
24~$\mu$m flux between $670-681$~d and 1264~d in SN~2003gd in the
SA(s)c galaxy NGC~628 (M74). \citet{kot09} showed that the cold
component of the SN~2004et SED was most likely due to an IS IR echo in
the SAB(rs)cd host galaxy NGC~6946.  The host galaxy of SN~2004dj,
NGC~2403, is of type SAB(s)cd and so there is a good likelihood of a
similar IS IR echo occurring. Therefore, we included an IS IR echo
component in our modelling of the SED. \\

Our IS IR echo model is the same as that used for the
early-time CDS IR echo (\S3.4.2).  Only the dust distribution and
grain radii are different.  We ignore the CDS dust emission derived in
\S3.4.2 since this is already invoked in the CDS echo model and
included in the continuum modelling up to 500~d.  We recognise that
a more extended, lower density CSM may also have existed.  However, we
found that the addition of yet another model component was unnecessary
to provide plausible matches to the continua and so for simplicity the
possible effects of an extended CSM were ignored.\\

As described in \S3.4.2, pre-existing dust surrounding the SN would
have been evaporated by the SN flash out to a distance of about
0.025~pc.  At sufficiently late epochs, enlargement of this cavity can
be produced by the forward shock. Assuming that the shock velocity is
comparable to the highest ejecta velocity viz. $\sim$15,000~\kms\ (see
\S3.4.2), and that there was no deceleration, the edge of the
UV-flash-determined cavity would be reached after about 600~d.  After
this time, the SN shock would evaporate the dust and enlarge the
dust-free cavity.  Therefore, for epochs earlier than 600~d we fixed
the inner limit of the IS dust at 0.025~pc.  For later epochs the
inner limit of the IS dust increases from 0.028~pc (34~light days) at
662~d to 0.058~pc (70~light days) by 1393~d.  It might be objected
that this ignores the possibility of shock deceleration. However, the
IR~echo contribution which the shock-evaporated dust would otherwise
have made, for $600>t>1393$~d, to the cold component is negligible
since the cavity radius was never more than 70~light days i.e. during
this late period the SN peak luminosity would have long since passed
by the dust near the cavity.  \\

The outer limit of the IS dust is much less certain.  \citet{ben10}
have used {\it Spitzer} imaging of NGC~2403 at 70~$\mu$m and
160~$\mu$m to map the dust column density via its thermal emission.
They also used the far-IR images in conjunction with H~I observations
to map the gas/dust ratio. The objection to using such measurements in
the present work is that they only provide total densities through the
disk (or to the mid-plane assuming symmetry), and not directly to
SN~2004dj.  While it was argued above that SN~2004dj actually lies
towards the front of S96, we are still faced with the uncertainty of
the depth of S96 in the galaxy plane.  Our approach, therefore, is to
assume a spherically symmetric dust distribution centered on the SN,
with an outer limit of 100~pc.  While not appropriate in principle for
the outer limits of the dust in the galactic disk, for the early era
being considered spherical symmetry provides a good approximation; at
this epoch, the echo ellipsoid is extremely elongated and so the
region of the spherical model outer surface intercepted by the
ellipsoid is small.  The plane of NGC~2403 appears to be tilted at an
inclination of about $55^{\circ}$ roughly doubling the face-on column
density, and so the adopted outer limit is equivalent to a $\sim$60~pc
scale height for the IS dust with the host galaxy face-on.\\

The free parameters were (i) the grain size, which influenced the dust
temperatures, and (ii) the grain number density, which determined the
luminosity. These parameters were adjusted in conjunction with the
warm dust models (see \S3.4.2 and \S3.4.4.2) to provide a match to the
longwave excess.  It was found that satisfactory matches at all epochs
were obtained with a grain radius of 0.1~$\mu$m and a dust number
density of $2.6\times10^{-13}$~cm$^{-3}$ (i.e., an IS gas number
density of 0.24~cm$^{-3}$ and a dust/gas mass ratio of 0.005 -
\citealt{ben10}.)  This is comparable to the typical value of
$5\times10^{-13}$~cm$^{-3}$ for the Milky Way \citep{all73} and
$7.0\times10^{-13}$~cm$^{-3}$ obtained by \citet{kot09} for NGC~6946.
In the wavelength region ($\lambda >15~\mu$m)
where the cold component makes a significant contribution ($>20\%$) to
the total flux, the IS IR echo model (15--150~$\mu$m) maintained a
near-constant luminosity of $\sim2.0\times10^{38}$~\ergs between 89~d
and 1393~d.  \\

The \citeauthor{ben10} dust column density map of NGC~2403 indicates
0.05~M$_{\odot}$pc$^{-2}$ at the position of SN~2004dj. From the above
dust number density from the echo model and integrated over 100~pc we
obtain just 0.003~M$_{\odot}$pc$^{-2}$. The absorption opacities used
by \citeauthor{ben10} are consistent, to within a factor of $\sim2$,
with the opacity law used in the present work.  Thus, we have a
$\sim\times$8 discrepancy in column density between \citeauthor{ben10}
and the present work.  Part of this discrepancy may be that the
adopted dust outer limit is too small.  The derived IS dust
density is fairly insensitive to the extent of the outer limit.  For
example, with a 75\% increase in the outer limit the model match to
the longwave continuum is retained with a reduction of just $\sim$10\%
in the IS dust density. Thus, at least some of the discrepancy could
be removed by simply increasing the dust outer limit.  An explanation
for the remaining discrepancy is that the SN and S96 actually lie
significantly above the mid-plane of NGC~2403. This is confirmed as
follows.  The optical depth to UV/optical photons yielded by the model
is about 0.026.  We can use this to estimate independently the host
extinction to SN~2004dj. The optical depth translates to an
absorption-only extinction of 0.028 magnitudes. Assuming an albedo of
about 0.4 \citep{dra03}, we obtain a total host extinction of
0.071~mag. This is reasonably consistent with the host-only value of
$A_V=0.081$~mag ($R=3.1$) obtained via the Na~I~D observations of
\citet{gue04}. A similar result is obtainable from our study of
SN~2004et \citep{kot09}. For SN~2004et, the IS IR echo model yields an
absorption-only extinction of 0.081~mag implying a total host
extinction of about $A_V=0.20$~mag.  High resolution spectra of Na~I~D
lines to SN~2004et gave a total (host plus Galaxy) $E(B-V)=0.41$~mag
\citep{zwi04}, or $A_V=1.27$~mag ($R_V=3.1$). Subtracting the
estimated Galactic contribution of $A_V=1.06$~mag \citep{sch98,mis07}
yields a host-only value of $A_V=0.21$~mag, in excellent agreement
with the IS IR echo-based value.  \\

One possible objection to the IS IR echo interpretation of the cold
component of SN~2004dj is that the steady component of the flux around
24~$\mu$m could be due to cool IS dust in S96, insufficiently
corrected for by the 3500~K blackbody extrapolation (\S3.1).  We
regard this as unlikely. As already indicated, \citet{mai04} argue
that the parent molecular cloud of S96 should have been dispersed by
stellar winds and SN explosions.  In addition we have found that (a) a
single set of IS IR echo parameters provided a fair match {\it
throughout the $89-1393$~d} covered and (b) the derived dust column
density is comparable to that obtained from IS Na~I~D spectroscopy.\\

A second possible objection to the IS IR echo interpretation  is that
the cold component is actually due to free-free emission. This is
discussed and dismissed in \S3.4.3.2. \\

We conclude that the cold component of SN~2004dj was due to an IS IR
echo.  As originally suggested by \citet{bod80}, the study of IS IR
echoes from SNe can provide an independent method of measuring the IS
extinction in galaxies. \\
\paragraph{{\it Free-Free and Free-Bound Radiation\/}}
In their MIR study of SN~1987A, \citet{woo93} suggest that during the
$60-415$~d period, a significant proportion of the MIR flux longward
of $\sim$20~$\mu$m was due to free-free (ff) emission, with ff and
free-bound (fb) emission also contributing at shorter wavelengths. For
SN~2004dj we have looked at the possible contribution of fb-ff
radiation.  It is particularly important to consider ff emission as
this rises in flux towards longer wavelengths, thus potentially
reducing or even dismissing an IS IR echo contribution.  For
consistency in this analysis fb emission must also be included.  For
the wavelengths covered this is strongest in the optical/NIR region.
Free-bound radiation has a strong ``sawtooth'' structure and the
extent to which this structure is undetectable in the optical/IR
continuum places a limit on its strength.  For a particular
temperature this, in turn, constrains the strength of the ff emission
($F_{ff}/F_{fb}\propto \sqrt T / \nu^3$).  \\

The fb-ff radiation was assumed to arise in a hydrogen envelope
centered on the supernova.  \citet{koz98} modelled the time-dependent
behavior of the temperature and ionization of SN~1987A, including the
hydrogen envelope, for epochs after 200~d. Therefore, we made use of
their study to model the ff and fb emission from SN~2004dj at epoch
251~d and later.  We adopted a hydrogen envelope of a few solar masses
and used their density profile:
\begin{equation}
\rho=9.1\times10^{-16}(t/(500~d))^{-3}(v/(2000~\kms)^{-2}~g~cm^{-3}. 
\end{equation}
By 251~d, the highest velocity of observable hydrogen in SN~2004dj was
no more than $\sim3100$~\kms\ (Table~\ref{tab8}).  By 925~d the
maximum H velocity observed in SN~2004dj was no more than 1400~\kms.
There were no optical or NIR spectra covering the final three MIR
epochs. We therefore assumed that the maximum hydrogen velocity at
these late epochs remained at 1400~\kms.  The hydrogen inner limit is
less certain; a value of 1000~\kms was adopted for all epochs.  In any
event, the fb-ff luminosity from 500~d onwards was completely
negligible for any plausible hydrogen velocity limits (see below).\\

We estimated the fb-ff emission from the hydrogen within the velocity
limits.  This was done using the escape probability formalism
\citep{ost89}, although in practice this was unnecessary as the
matches to the data always showed that the hydrogen was very optically
thin.  The mass of the fb-ff-emitting hydrogen fell from
$\sim$4~M$_{\odot}$ at 251~d to 0.75~M$_{\odot}$ at 996, 1207 and
1393~d.  Using modest extrapolation of figure~7 in \citet{koz98} to
the velocities observed in SN~2004dj, we deduced that the hydrogen at
the observed velocity limits stayed at a temperature of $T\sim$5000~K
up to 859~d and then declined to $\sim$2000~K by 1393~d.  The
fractional ionization, $\chi_e$, was obtained from figure~9 of
\citeauthor{koz98} by interpolation. This indicated $\chi_e\sim0.02$
at 251~d, falling to $\chi_e\sim0.0005$ by 1393~d. At each epoch, the
$T$ and $\chi_e$ values were assumed to apply to all the hydrogen
within the velocity limits. The \citeauthor{koz98} study did not
extend to epochs as early as the earliest MIR observations (89, 106,
129~d) of SN~2004dj. We therefore adopted 6000~K as a plausible
temperature for these epochs and allowed the ionization to take the
largest value consistent with the overall match to the observed
continuum.  Plausible values were obtained, viz: $\chi_e=0.003$ at
89~d rising to $\chi_e=0.005$ at 129~d. We used Gaunt factors
tabulated by \citet{hum88}. For the fb radiation we used the continuum
recombination coefficients of \citet{erc06}.  \\

The hydrogen maximum velocity, temperature and free electron density
for each epoch are listed in Table~\ref{tab8}.  The fb-ff luminosities
are listed in Table~\ref{tab9}, col.~4.  It can be seen that by
500~d the fb-ff contribution to the total luminosity is small,
becoming increasingly negligible at later epochs The fb-ff components
for 89--281~d are plotted in Fig.\ref{fig18}.  At 500~d the fb-ff flux
is too weak to appear on the 500~d plot which has been scaled to allow
easy comparison with the earlier epochs.  A more comprehensive
estimation of the fb-ff flux, taking into account the temperature and
ionization gradients, is beyond the scope of this paper.  \\

As noted in \S3.4.3.1, in the wavelength region ($\lambda >15~\mu$m)
where the cold component makes a significant contribution ($>20\%$) to
the total flux, the IS IR echo maintained a near-constant luminosity
(15--150~$\mu$m) of $\sim2.0\times10^{38}$~\ergs between 89~d and
1393~d.  Moreover, in setting the IS IR model to reproduce the
longwave excess at the latest epochs, when the fb-ff flux was
negligible, it was found that a satisfactory match to the longwave
($\lambda >15~\mu$m) excess was automatically achieved for {\it all}
epochs. Consequently, the fb-ff flux could make, at most, only a minor
contribution even at the earliest epochs.  We conclude that any
free-free emission was too weak to account for the cold component at
any epoch.  \\\\

\subsubsection{The Warm Component}
\paragraph{{\it Late-Time Optical and NIR Line Profiles\/}}
In \S2.4 we described the evolution of optical and NIR line profiles
in a number of species.  In particular, we noted the development of a
sharp suppression in the red wing, suggesting dust formation causing
obscuration of the far side of the ejecta.  Could this dust also be
responsible for the warm component at later epochs?  In this and the
next sections we explore this possibility. Here we examine the
distributions of dust that might give rise to the late-time
optical/NIR line profiles. In \S3.4.4.2 we shall then test the
hypothesis that the same dust distributions are responsible for the
warm component.  We modelled the line profiles of H$\alpha$,
Pa$\beta$, [O~I]~6300~\AA, [Fe~II]~7155~\AA\ and [Fe~II]~12567~\AA.
The period $461-925$~d was studied since this was mostly
overlapped by the $500-1393$~d during which the blackbody
analysis and more detailed studies of the MIR emission (\S3.4.4.2)
suggest that substantial quantities of dust were present in the
ejecta.  We note (Table~\ref{tab6}) that the H$\alpha$ line widths
towards the end of the bright plateau phase greatly exceeded those
observed on 461~d and later, implying that there was a negligible
contribution of any light echo of the early phase profiles to the
late-time profiles.\\

We did also examine the blue asymmetry of the H$\alpha$ and
[Fe~II]~7155~\AA\ profiles at a much earlier epoch (283~d) but were
unable to achieve a satisfactory match using the model adopted for
later epochs.  In any case, as already demonstrated in \S3.4.2, we
were able to account for the 251/81~d IR excess as thermal emission
from dust formed in the CDS. Moreover, \citet{chu05} and \citet{chu06}
successfully explained the H$\alpha$ profile up to $\sim$300~d by
invoking the emergence of an intrinsic asymmetric, bipolar core
(i.e. no dust involved).  We conclude that there is no evidence for
the formation of new ejecta dust earlier than 461~d. \\

The line-profile red wings exhibit increasingly abrupt declines after
$\sim$1~year (Fig.~\ref{fig10}) suggesting the condensation of
attenuating dust.  In contrast, the extended blue wings exhibit little
sign of developing suppression.  This behavior points to the
attenuating dust lying at low line-of-sight velocities, such as a
face-on disk-like distribution centered on the SN center of mass.
Nevertheless, in our initial considerations of possible dust
configurations we included the case of a spherically symmetric dust
sphere.  Our line-profile model comprises a homologously expanding
sphere of gas responsible for most of the observed line flux, with the
emission being attenuated by a dust zone lying concentrically with the
gas sphere.  The dust zone is also assumed to participate in the
expansion.  The gas emissivity is assumed to have a power-law
dependence with radius.  \\

\begin{figure*}
\epsscale{1.2}
\plotone{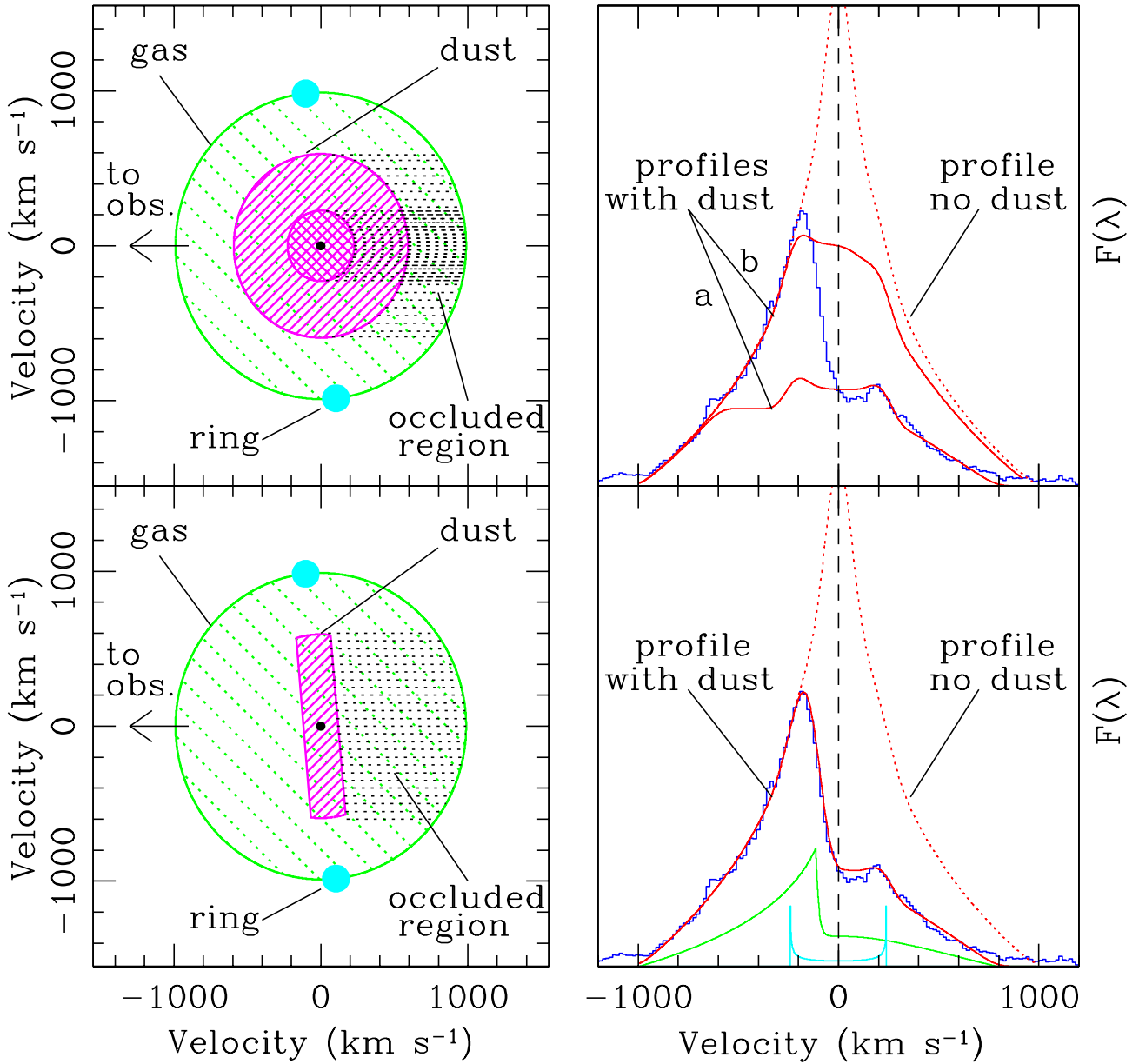}
\caption[]{Illustration of the line-profile models.  The model
profiles are derived from a homologously-expanding, emitting gas
sphere attenuated by an embedded, opaque dust distribution.  A minor
additional contribution to the model is provided by a thin, expanding
concentric ring of emitting gas oriented with the normal to the ring
plane at $6^{\circ}$ to the line-of-sight.  The gas-dust
configurations are illustrated in the LH panels which show sections
through the SN system in the plane defined by the thin ring axis and
the line-of-sight.  The gas sphere is shaded green and the attenuating
dust shaded magenta.  The thin ring, viewed edge-on in the figure, is
represented by two large cyan dots.  The region of line emission
occluded by the dust is shown by the black shading.  The expansion
velocities are indicated by the axes.  The final model profiles are
obtained by combination of the attenuated sphere and unattenuated ring
spectra. The resulting spectrum is then smoothed to the spectral
resolution.  The RH panels show the resulting model line profiles
(solid red), plotted in velocity space, compared with the observed
spectrum (blue). Also shown (dotted red) is the line profile that
would result in the absence of dust attenuation. In this illustration
we compare the model profiles with the observed 895~d [O~I]~6300~\AA\
line. In this case, the gas sphere emissivity declines as
$r^{-1.9}$.\\

The upper panels illustrate the case where the dust is distributed as
an opaque, concentric sphere.  Two sizes of sphere are
considered. Profiles $a$ and $b$ (upper RH panel) correspond,
respectively, to the large and small dust spheres (upper LH panel).
It can be seen that neither reproduce the observed line profile.  The
lower panels illustrate the case where the dust is composed of
amorphous carbon and is distributed as a uniform density, nearly
face-on concentric disk with $\tau=18$ perpendicularly
through the disk plane.  This optical depth is the value required for
the model to simultaneously match the line profile and the
contemporaneous MIR continuum (see Fig.~\ref{fig19}).  While the disk
is shown as being coplanar with the ring, for ease of computation the
disk is taken to be exactly face-on.  Shifting the disk tilt
to match that of the ring would have had only a small effect on the
model profile.  No correction is made for minor edge effects at the
disk's outer limit (radius).  Also shown (lower RH panel) are the
intrinsic line profiles from the attenuated gas sphere (green solid
line) and the thin ring (cyan solid line). The thin ring emission is
not affected by the dust. For clarity, both of these intrinsic profile
plots have been scaled downward in flux by the same amount relative
to the total profile. It can be seen that the final dust-disk model
profile provides an excellent match to the data.
\label{fig16}
}
\end{figure*}

Several arrangements of dust were examined. These were (i) a uniform
opaque sphere (Fig.~\ref{fig16}, upper panels), (ii) a face-on opaque
ring, and (iii) a thin face-on disk (Fig.~\ref{fig16}, lower panels)
whose opacity was uniform or varied radially as an $r^{-\delta}$ power
law along the disk plane but was always uniform normal to the disk
plane.  (For the disks, ``opacity'' refers to the value normal to the
disk plane.)  For the radially-varying version of (iii), a power law
index of $\delta=1.9$ was chosen as typical of the power law indices
determined for the gas emission profiles.  Configurations (i)--(iii)
were initially tested against the 895~d optical profiles since these
were (a) of the highest available resolution, and (b) corresponded to
a time by which the SN spectral continuum was overwhelmingly dominated
by thermal emission from dust.  Configurations (i)--(ii) could not
reproduce the observed line profiles.  This is illustrated in
Fig.~\ref{fig16} (upper panels) where we show two examples of the line
profile produced by an opaque, concentric dust sphere, compared with
the observed [O~I]~6300~\AA\ profile at 895~d.  Only configuration
(iii) - the dust disk - was able to reproduce the observed profiles.
Within this configuration we examined cases where the dust
distribution (a) extended beyond the gas limit, (b) ceased abruptly at
a given velocity within the gas sphere, or (c) extended uniformly to a
given velocity, $v_{duni}$, and then declined as a power law to the
edge of the gas sphere.  Case (a) failed to reproduce the observed
line profiles. Cases (b) and (c) are discussed below. \\

As described in \S2.4, by 895~d and 925~d, a ``secondary'', weaker,
redshifted peak had also appeared in the H$\alpha$ and [O~I]~6300~\AA\
profiles, suggesting that a fraction of the line flux originated in an
emission zone centered on the SN and having the geometry of an
expanding ring, jet or cone.  The second peak had a redshift of only
+160 to +170~\kms. Given the preference for a thin face-on disk of
attenuating dust to account for the asymmetry of the line profiles, a
natural explanation for the low velocity of the second peak is that
its source was actually at a velocity comparable to that of the gas
sphere, but moving in a way which was roughly coplanar with the dust
disk. Consequently the low inclination angle led to the low observed
redshift in the second peak.  The simplest geometry which accounts for
the second peak is a ring-like emission zone whose plane lay at a
small angle to the ``face-on'' plane.  A cone or jet geometry is more
problematic. If we insisted on maintaining axial symmetry then a
cone/jet would have to be normal to the disk but with a mysteriously
low velocity. Alternatively, a cone/jet lying near the disk plane,
while possibly allowing a high intrinsic velocity, would have the
unattractive feature of breaking the axial symmetry.  Therefore, to
account for the second peak, we added a thin, near-coplanar ring of
emitting gas to the line-profile model \citep[cf.][for the case of
SN~1998S]{ger00,fra05}.  The ring was assumed to participate in the
overall homologous expansion of the gas.  \\

We set the intrinsic ring velocity equal to the fastest moving gas
observed viz. hydrogen at 2400~\kms as indicated by the H$\alpha$ blue
wing on 895 and 925~d, although higher velocities could have been
used.  Consequently the ring emission is unattenuated by the dust.
For ease of computation, we retained a face-on disk ($i=0^{\circ}$).
Tilting the disk to become exactly coplanar with the ring would have
had only a small effect on the line profile. \\

In matching the model to the observed profiles the free parameters for
the gas are the maximum (outer) velocity of the gas sphere,
$v_{gmax}$, the gas emissivity scaling factor and power-law index
$\beta$, the inclination, $i_r$, of the ring component assuming an
intrinsic expansion velocity of 2400~\kms, and the ring component
emissivity assumed uniform. In addition, the overall wavelength
positions of the observed profiles were allowed to vary by small
amounts (see below).  For ease of comparison, the dust-disk dimensions
were also expressed as velocities within the homologous expansion.
Thus, for the case (b) dust disk the free parameters are the maximum
radial velocity of the dust disk, $v_{\rm dmax}$, the maximum
velocity, $v_{dth}$, perpendicular to the disk plane and the magnitude
of the dust radial density power-law index, $\delta$.  Multiplication
of $v_{\rm dmax}$ and $2v_{\rm dth}$ by the epoch yields,
respectively, the disk radius and thickness at that time.  For case
(c) $v_{\rm dmax}$ is replaced with $v_{\rm duni}$.  We adopted
amorphous carbon as the grain material (see \S3.4.4.2) and a power law
grain-size distribution, index $m=3.5$ \citep{mat77}, with
$a_{(min)}=0.005~\mu$m and $a_{(max)}=0.05~\mu$m.  The optical depth
normally through the disk was then calculated.  For example, for case
(b) the optical depth for emission from gas lying behind the disk at
wavelength $\lambda$, epoch $t$ and radial velocity $v_d$,
$\tau_{\lambda}(t,v_d)$, is given by:
\begin{equation}
\tau_{\lambda}(t,v_d)=\frac{4}{3}\pi \rho\kappa_{\lambda}k\frac{1}{4-m}
[a^{4-m}_{(max)}-a^{4-m}_{(min)}]v_d^{-\delta}2v_{dth}t^{-2} 
\end{equation}
where $k$ is proportional to the grain number density, at a fiducial
epoch, to the radial velocity and to the grain radius, $\rho$ is the
grain material density and $\kappa_{\lambda}$ is the mass absorption
coefficient of the grain material at wavelength $\lambda$.  For
emission from gas lying within the disk the same formula is used but
replacing $2v_{\rm dth}$ with the velocity equivalent to the path from the
emission point to the near side of the disk. The disk thickness was
always $<20\%$ of the disk diameter and so edge-effects at the disk's
outer limit were assumed to be negligible. \\

The gas emission profiles (i.e., those for the attenuated sphere and
unattenuated ring) were individually convolved with the appropriate
slit PSF.  The PSF FWHM values were 6.5, 4.2, 6.5, 1.25, and 1.75~\AA\
for 461, 467, 554, 895, and 925~d, respectively.  The two components
were then summed to form the final model profile which was compared
with the observed profile. The profile model (case (b) disk) was
adjusted to match the observed profiles as follows.  The parameters
$v_{\rm gmax}$, $\beta$ and the gas sphere emissivity scaling factor
were adjusted to provide a match to the blue wing of the
profile. Then, for a given value of $\delta$ (0 or 1.9), $v_{\rm dth}$
(equivalent to the disk half-thickness) together with the grain number
density scaler, $k$, were adjusted to reproduce the sharp-decline
section of the profile.  The velocity $v_{\rm dmax}$ was varied to
reproduce the suppressed red wing. The gas ring emissivity and
inclination were then adjusted to reproduce the secondary peak.  Some
iteration of all the model parameters was necessary to reach the final
profile match.  For the profile model (case (c) disk) $v_{\rm dmax}$
was replaced with $v_{\rm duni}$, and $\delta$ was allowed to vary to
arbitrarily large steepness. \\

The offset of the sharp-decline section of the profiles from zero
velocity was only about $-100$~\kms\ after correction for the +221~\kms\
of the heliocentric velocity.  Consequently, the match to the
sharp-decline section of the profile (i.e., determination of the disk
half-thickness) could be significantly affected by errors in
individual wavelength measurements or intrinsic variations in the
distributions of different gas species.  Forcing a match in each
case by letting the wavelength position of each observed profile to
vary by small amounts, we were able to estimate the disk
half-thickness and its uncertainty.  For a heliocentric velocity of
+221~\kms\ we obtained a disk half-thickness equivalent to
$v_{\rm dth}=+112\pm18$~\kms. There was no significant variation in
$v_{\rm dth}$ during the period covered by the profile study.  Note that a
change in the heliocentric velocity estimate would yield the same
change in the disk half-thickness velocity. \\

As indicated above, we first carried out model matches to the 895~d
profiles.  It was found that the best case~(b) matches were obtained
when the disk was highly opaque (say, $\tau>5$) at all radial locations.
For case~(c), comparably good matches were achieved by setting the
uniform zone at a high opacity ($\tau>5$) and $\delta$ at values
steeper than $\sim10$ --- i.e., the dust density beyond the uniform zone
had to be extremely steep. Less steep declines tended to suppress the
visibility of the central minimum. We conclude that the dust disk was
highly opaque in the optical region but terminated abruptly at an
approximately fixed radius (see \S3.4.4.2). Given that this
condition was indicated by both cases (b) and (c), we abandoned the
more complicated case (c) model and completed the analysis for all the
461--925~d spectra using only case (b) --- i.e., a dust density which was
uniform or declined radially as $r^{-1.9}$ and which terminated
abruptly at $v_{\rm dmax}$.  \\

The models for 895~d were initially adjusted to determine the minimum
opacity -- that is, the minimum dust mass that was needed to provide a
satisfactory match to the observed profile. 
The dust masses were obtained from the following:
\begin{equation}
M_d= \frac{8}{3}\pi^2\rho\kappa_{\lambda}k\frac{1}{4-m}
[a^{4-m}_{(max)}-a^{4-m}_{(min)}]\frac{1}{2-\delta}v_{\rm dmax}^{2-\delta}
2v_{\rm dth}.
\end{equation}
At 895~d, with a uniform disk ($\delta=0$), the dust extended to
$v_{\rm dmax}=590\pm25~\kms$ i.e. $r_{\rm
  dmax}=(4.6\pm0.2)\times10^{15}$~cm.  The disk half-thickness was
19\% of this.  Simultaneous matches to all three profiles (H$\alpha$,
[O~I]~6300~\AA, [Fe~II]~7155~\AA) required a minimum $\tau$ of 13 at
6300~\AA, corresponding to a minimum dust mass of
$0.27\times10^{-4}$~M$_{\odot}$.  Note that the high optical depth was
demanded by the very sharp decline, unresolved even at the
1.25~\AA\ resolution of 895~d.  Other parameters are:
$r_{\rm gmax}=7.65\times10^{15}$~cm, $\beta=1.9$, (ring flux)/(total flux)
= 0.07 and ring tilt = $6^{\circ}$. \\

\begin{figure*}
\epsscale{1.2}
\plotone{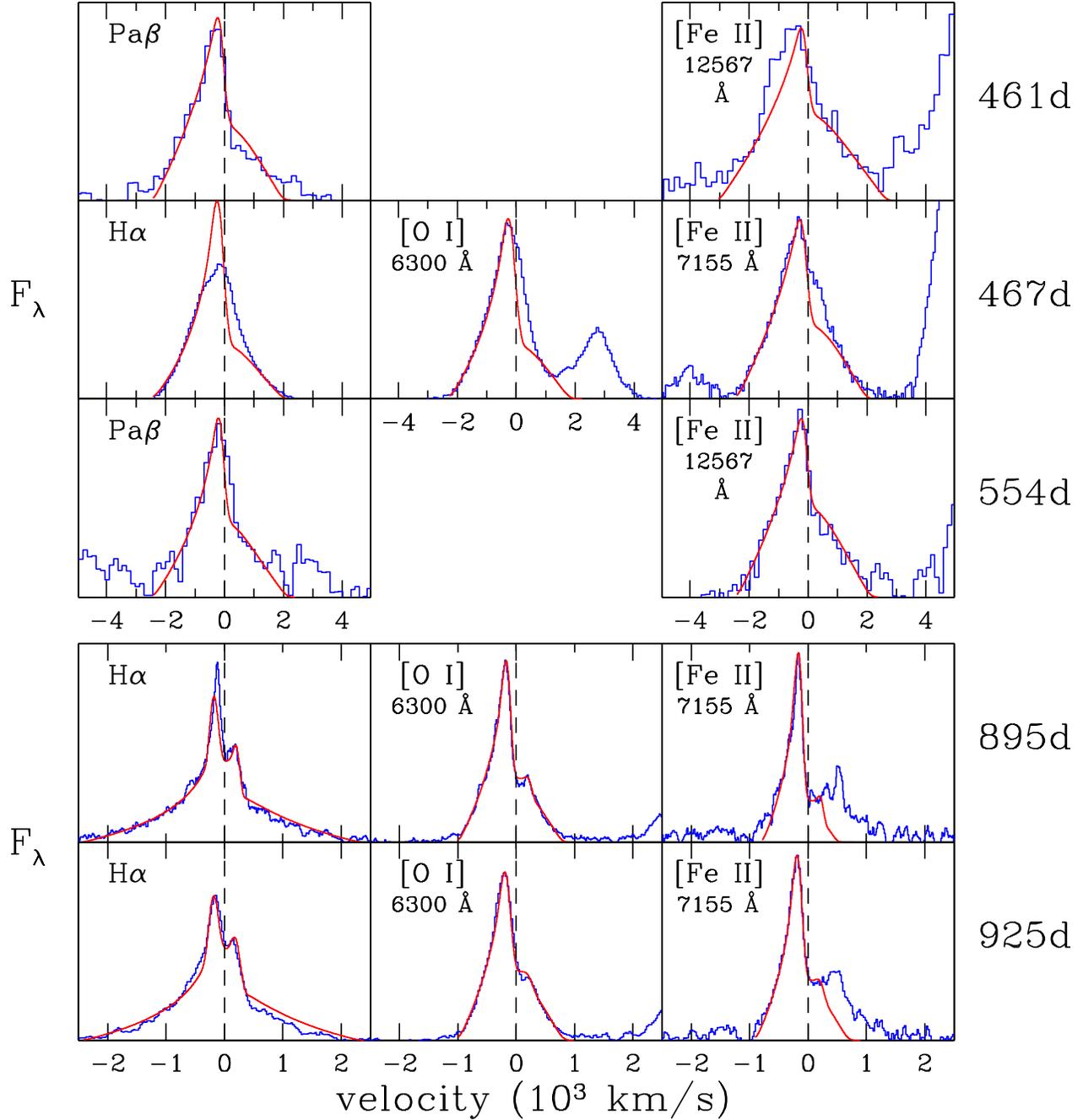}
\caption[]{ Individual dust-disk model matches (red) to observed line
profiles (blue) of SN~2004dj for the case where $\delta=0$ (uniform
dust density).  The velocities of the observed spectra have been
shifted by --221~\kms\ to match the center-of-mass rest frame of the
SN.  The latest two epochs also include a small contribution from a
ring of emission.  Also, the velocity scale of these two epochs has
been expanded to provide more detail.  The matches incorporate dust
disk masses obtained by interpolating to the profile epochs the values
obtained from the MIR continuum modelling.  Between 554~d and 925~d
the dust mass increased from $0.25\times10^{-4}$~M$_{\odot}$ to
$0.4\times10^{-4}$~M$_{\odot}$.  The radius of the dust disk {\it
decreased} slightly from $5.3\times10^{15}$~cm to
$4.6\times10^{15}$~cm. The disk half-thickness expanded from
$0.45\times 10^{15}$~cm to $0.90\times 10^{15}$~cm.  For each profile
the optical depth had a value in the range $3<\tau<19$ (i.e. the disk
was always of high optical depth).
\label{fig17}
}
\end{figure*}

In using the dust-disk model to reproduce the contemporary MIR
continua (\S3.4.4.2) it was found that the opacities and dust masses for
all lines and epochs had to be somewhat higher than the minimum values
required to match the observed optical and NIR line profiles. For
example, for [O~I]~6300~\AA\ at 895~d and a uniform disk it was
necessary to set $\tau \ge 18\pm3$, corresponding to a minimum dust
mass of $(0.38\pm0.06)\times10^{-4}$~M$_{\odot}$.  In the rest of this
subsection, therefore, we present and consider results for profile
matching incorporating dust-disk masses obtained by interpolating to
the profile epochs the values obtained from the MIR continuum
modelling.  A sketch of the case (b) [O~I]~6300~\AA\ uniform
($\delta=0$) disk model at 895~d, is shown in Fig.~\ref{fig16} (lower
LH panel) together with an illustration (lower RH panel) of the model
match (solid red line) to the observed profile (blue).  Also shown are
the intrinsic line-profile contributions from the attenuated gas
sphere (green) and unattenuated gas ring (cyan) components, as well as
the final profile (dotted red line) which would result if the
attenuating dust were removed.  In Fig.~\ref{fig17} we show all the
individual matches to the line profiles for 461--925~d. In general,
good matches were achieved for epochs 554~d, 895~d, and 925~d. Poorer
matches were obtained at 461~d and 467~d. \\

On 895~d and 925~d the model did not reproduce the extended red wing
seen in the [Fe~II]~7155~\AA\ profile.  This suggests an additional,
faster moving component of iron, and may be indicative of $^{56}$Ni
asymmetry or ``bullets'' in the initial explosion
\citep[e.g.][]{bur95}.  At the earliest two epochs (461~d and 467~d)
the model matches are also poorer. Specifically, if the model was
adjusted to match the full suppression of the extreme red wing, then
it also over-attenuated the less redshifted portion of the red wing;
in other words, the steep red decline in the observed profile is
generally less pronounced at these earliest epochs. Alternatively,
matching to the less redshifted portion of the red wing meant that the
full suppression of the extreme red wing was not reproduced. These
points suggest that, during the 461--467~d period, dust formation was
less complete and was not yet fully opaque over the whole disk. We
therefore adjusted the models to reproduce the suppression of the
extreme red wing, but recognise that the derived dust mass lower
limits for 461--467~d may well be overestimated. \\

Between 554~d and 925~d the minimum dust mass increased from
$0.25\times10^{-4}$~M$_{\odot}$ to $0.4\times10^{-4}$~M$_{\odot}$.
During this time the radius of the dust disk appeared to actually {\it
decrease} slightly from $5.3\times10^{15}$~cm to
$4.6\times10^{15}$~cm. We explored the possibility that this was due
to a declining optical depth as the disk expanded --- i.e., that while
the dust disk continued to expand, the optically-thick/thin boundary
declined in radius.  We found that the requirement of a steep, sharp
cutoff at the disk limit (see above) ruled out this explanation
implying that the decrease in the dust radius was real. This issue
will be considered further in \S3.4.4.2. \\

The above analysis was repeated with $\delta=1.9$. Similar disk radii
were obtained, but the minimum dust masses were $\sim\times$20 larger
than for $\delta=0$.  This is as one would expect.  To maintain the
line-profile matches with $\delta=1.9$ required the maintenance of a
high optical depth at the disk edge in the optical/NIR wavelength
range. If we demand that the optical depths be the same at $r_{\rm
dmax}$ for both $\delta=0$ and $\delta>0$ then it may be shown that
$M_d(\delta)/M_d(0)=2/(2-\delta)$, where $M_d(\delta)$ is the dust
mass for a given value of $\delta$ and $M_d(0)$ is the mass for
$\delta=0$.  Thus, for $\delta=1.9$, $M_d(\delta)/M_d(0)=20$.  We note
that this assumes that the power law continues to the center, which is
unlikely to be the case. Indeed for $\delta=2$ or more, the mass would
be infinite.  A more plausible behavior would be that the density law
flattens toward the center. For example, if we assume that within
$0.05r_{\rm dmax}$ the dust density became uniform, with $\delta=1.9$
outside $0.05r_{\rm dmax}$, then the total minimum dust mass would
only be about a factor of 3 larger than for the totally uniform disk
case --- that is, about $10^{-4}$~M$_{\odot}$ \\

The model gas sphere expansion velocity remained at around 2400~\kms\
in hydrogen at all epochs. Similar velocities were obtained in
[Fe~II]~12567~\AA\ at 461~d and 554~d, and in [O~I]~6300~\AA\ and
[Fe~II]~7155~\AA\ at 467~d.  By 895~d and 925~d the expansion
velocities in [O~I]~6300~\AA\ and [Fe~II]~7155~\AA\ had fallen to less
than 1000~\kms, perhaps suggesting some stratification by element.  At
554~d and earlier epochs, the ring flux was negligible. At the later
epochs, the ring inclination is $6^{\circ}$ (i.e., close to face-on).
The fractional contribution of the ring to the total observed flux
never exceeded 20\%.  Indeed, were it not for the dust disk, it is
unlikely that the relatively weak ring emission would have been
detected since it would have been swamped by the emission from the gas
sphere (see Fig.~\ref{fig16}). We therefore regard the ring component
as an interesting but minor effect. \\

We conclude from the line-profile analysis that, in the 461--925~d
period, the mass of the dust disk and its radial extent can be small
enough to be consistent with an origin as newly-condensed ejecta dust.
In \S3.4.4.2 we shall show that the same dust disk can account for the
MIR continua.\\
\paragraph{{\it Later Phase IR Excess and Ejecta Dust Condensation\/}}
We have argued that, up to at least 281~d, the warm
component could be fully accounted for in terms of an IR echo from a
CDS.  By 500~d the warm component luminosity exceeded the CDS
contribution by a factor of $\sim20$.  This ``late IR excess'',
together with the steady shift of the peak emission of the SED to
longer wavelengths in the period $\sim250-1200$~d (Figs.~\ref{fig3},
\ref{fig5}) suggests the appearance of an additional population of
warm, but cooling dust.  What is the location, distribution and
heating mechanism of this dust?  Such IR emission can arise from (i)
heating, by a number of possible mechanisms (see below), of new dust
formed either in the ejecta, or in a late surge of dust growth within
the CDS or (ii) heating of pre-existing circumstellar dust by the
early-time SN luminosity (a conventional IR echo), or possibly the
forward shock travelling into undisturbed CSM beyond the CDS.
We can dismiss immediately hypothesis~(ii):\\

(a) An IR echo does not naturally produce the long delay (at least
$\sim300$~d) before the commencement of the warm component excess flux
rise.  To account for this within an IR echo scenario, we would have
to invoke an {\it ad hoc} asymmetry in the CSM. \\

(b) IR-echo or forward-shock heating would not account for the
observed late-time red wing suppression in the spectral line profiles
discussed in \S3.4.4.1.  In contrast, red-wing attenuation can easily be
produced by dust formation in the ejecta. New CDS dust could
conceivably also have produced such an effect provided that the dust
did not lie completely outside the optical/IR emission zone.\\

(c) Pre-existing CSM dust, whether heated by the SN luminosity or a
forward shock, would not produce steepening of the optical-NIR decline
rate simultaneously with the appearance of the late IR excess.  In
contrast, new dust in the ejecta or CDS can be indicated by
optical-NIR steepening.  In Fig.~\ref{fig4}, after $\sim$470~d the
$VRIJ$ light curves show weak evidence of steepening, providing a
minor additional argument against a CSM IR echo origin for the later
warm component, although the errors on the latest points are large.
We also note that an alternative explanation for such steepening could
be a faster-than-expected decrease in the $\gamma$-ray absorption
relative to that invoked by \citet{li93} for the radioactive energy
deposition in SN~1987A (Fig.~\ref{fig4}). \\

The above points, especially (a) and (b), leaves us with
hypothesis~(i) viz. that the late IR excess is due to emission from
new dust formed in either the ejecta or CDS.  The later-era appearance
of the red wing suppression coincides roughly with the emergence of
the late IR excess, suggesting that the same dust could have been
responsible for both the optical/NIR attenuation effects and the rise
of the MIR emission.  That this dust formed in the ejecta rather than
the CDS is indicated by the fact that the extent of the attenuating
dust derived from late-time line-profile analysis was comparable to
that of the MIR-emitting dust as derived from blackbody analysis; for
example, at 859~d the blackbody radius is $3.3\times10^{15}$~cm compared
with a profile-derived radius of $(4.4\pm0.2)\times10^{15}$~cm at
895~d.  Moreover, the line-profile analysis demonstrates that the dust
was of high opacity in the optical/NIR, and distributed as a near
face-on disk lying concentrically with the SN center of
mass. Attenuation by a CDS could not have produced the late-time line
profile behavior. \\

In view of the above points, we shall now give detailed consideration
to the scenario where the warm component emission originated in
newly-formed ejecta dust. Such dust could be heated by a number of
mechanisms including radioactive decay, ionization freeze-out effect,
embedded pulsar or reverse-shock radiation.  In the case of SN~2004dj
we can rule out radioactivity as the principal heating mechanism of
any putative new ejecta dust.  Inspection of Table~\ref{tab7} shows
that, as early as 500~d, the warm component flux exceeded the total
deposited radioactive luminosity by a factor of 2 growing to a factor
of $\sim450$ by 1393~d.  Indeed, by the latest epoch, the warm
component luminosity was still as high as $\sim4\times10^{38}$~\ergs.
We also rule out the freeze-out effect as the main energy source
\citep{cla92,fra93} since, at $\sim1400$~d, its contribution would
only be $\sim5\times10^{36}$~ergs.\\

Power input from an embedded pulsar, via a pulsar wind nebula, is
another possibility \citep{woo89,che92}.  \citeauthor{che92} predicted
that a distinctive feature of such a source would be the presence of
certain high ionization lines. The earliest epoch they studied was at
1500~d. The nearest of our optical/NIR spectra to this epoch are those
at 895~d and 925~d. Of the high ionization lines for which
\citeauthor{che92} make luminosity predictions, only
[O~III]~4959/5007~\AA\ was covered by these spectra. We co-added the
spectra at the two epochs and searched for the stronger
[O~III]~5007~\AA\ component.  A broad, low S/N emission feature was
detected with a dereddened luminosity of
$(0.8\pm0.2)\times10^{36}$~\ergs.  This is about half the luminosity
predicted by \citeauthor{che92}.  A serious difficulty is that the
redshift-corrected position of the feature lies about +400~\kms\ from
where it would be if due to an [O~III]~5007~\AA\ line subject to the
same dust attenuation deduced in other lines (\S3.4.4.1).  Even without
dust attenuation, the feature would still be +200~\kms\ too far to the
red.  This, together with the low S/N, leads us to conclude that there
is no persuasive evidence for the presence of pulsar-driven
high-ionization features in the latest spectra of SN~2004dj.  This
leaves us with reverse-shock heating of the ejecta following the
ejecta-CSM collision as the most promising mechanism for the late-time
energy source \\

To test further the hypothesis that the same dust was responsible for
the late-time line-profile red-wing suppression and the MIR emission
we modelled the IR emission over a range of late-time epochs using the
same dust-disk configuration as was employed in the line-profile
analysis.  The emission was derived following a similar procedure to
that used in the isothermal dust model by \citet{mei07} and
\citet{kot09}.  The resulting flux was then added to the other
continuum components and the net model compared with the observed
continua.\\

\begin{figure*}
\epsscale{1.3}
\plotone{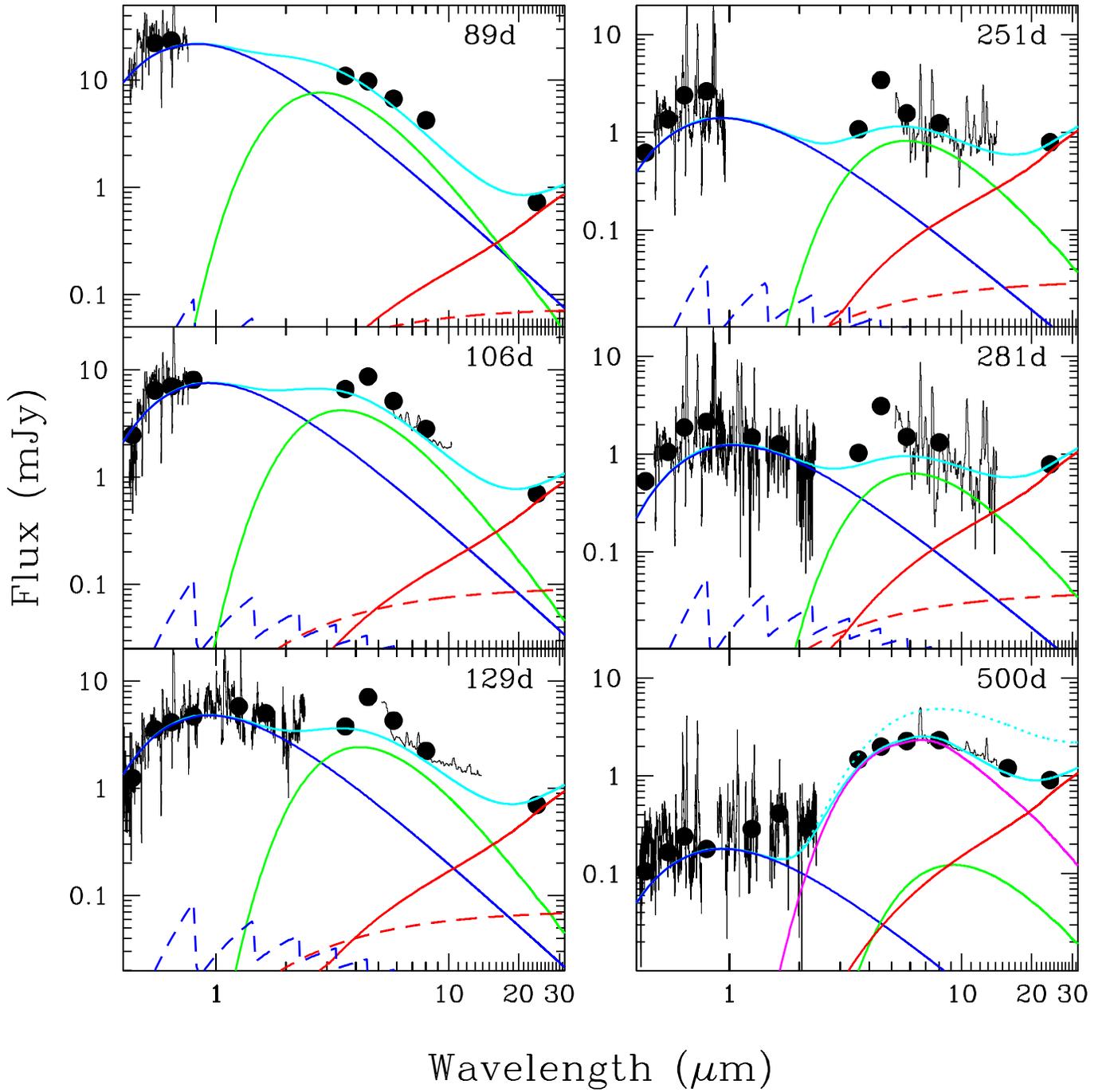}
\caption[]{89--281~d: Model continua comprising hot blackbody (solid
blue), free-bound (dashed blue), free-free (dashed red), CDS IR echo
(green), IS IR echo (solid red) and total flux (solid cyan).  The
observed spectra and contemporary photometry (dots) of SN~2004dj are
plotted in black.\\ 
500~d: As for earlier epochs but with the addition of IR emission from
a warm disk of dust (magenta). The dotted cyan line represents the
model continuum when the IDDM is replaced with a blackbody adjusted to
match the shortwave region of the MIR continuum. At this epoch the ff
and fb fluxes are too weak to appear on the plot.
\label{fig18}
}
\end{figure*}

\begin{figure*}
\epsscale{1.3}
\plotone{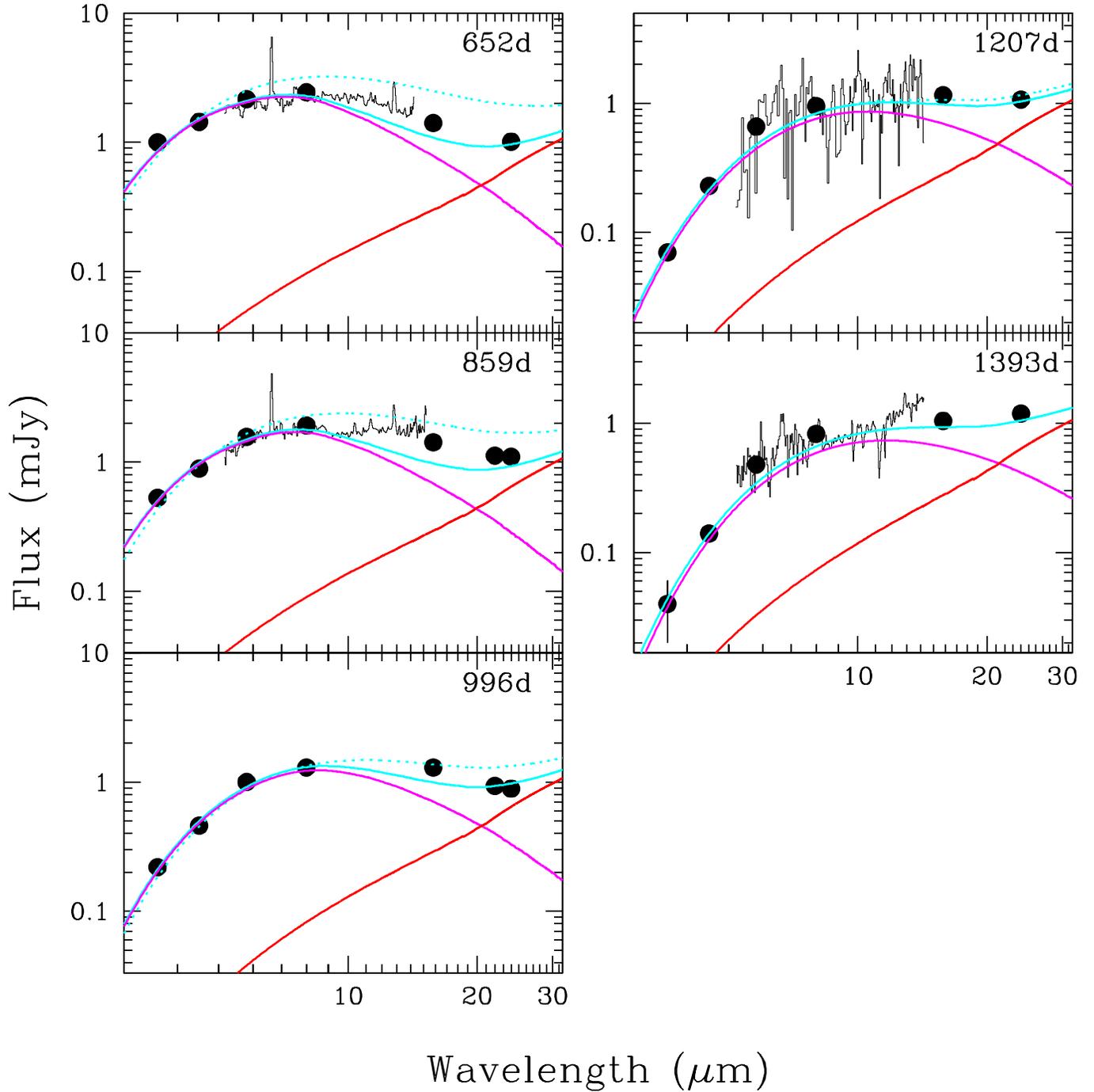}
\caption[]{Model continua at epochs $652-1393$~d, comprising IR flux
from a warm disk of dust (magenta), IS IR echo (red) and total flux
(solid cyan).  The dotted cyan line represents the model continuum
when the IDDM is replaced with a blackbody adjusted to match the
shortwave region of the MIR continuum. By 1393~d the blackbody has
fully merged with the IDDM; i.e., the model is effectively opaque over
the whole observed wavelength range.  The observed spectra and
contemporary photometry (dots) of SN~2004dj are plotted in black.  A
hot blackbody plus free-bound and free-free continua were also
included in the matches but these had a negligible effect on the MIR
continua.  Consequently, and in order to show the MIR behavior in
more detail, the optical/NIR region is not shown.  Estimates of the
hot and fb-ff component contributions are given in Tables~\ref{tab8}
and \ref{tab9}.
\label{fig19}
}
\end{figure*}

We considered the thermal radiation from a warm, isothermal, face-on
disk of dust located symmetrically about the SN center of mass. We
shall refer to this as the IDDM (isothermal dust-disk model).  The
disk radius is $r_{\rm dmax}$ and, as in \S3.4.4.1, the dust
number density declines as $r_d^{-\delta}$ where $\delta$ is set as 0
or 1.9.  As mentioned above the disk thickness was $<20\%$ of the disk
diameter and so we judged that a thin-disk treatment would provide an
adequate means of estimating the flux from the disk (i.e., edge effects
are ignored).  The observed IDDM flux, $dF_{\lambda}(t,r_d)$, at
wavelength $\lambda$ and time $t$ from an elemental ring lying between
radii $r_d$ and $r_d+dr_d$ is: 
\begin{equation}
dF_{\lambda}(t,r_d)= 2 \pi r_d dr_d D^{-2} B_{\lambda}(t)
(1-exp(-\tau_{\lambda}(t,r_d)))
\end{equation}
where $D$ is the distance of the SN, $B_{\lambda}(t)$ is the Planck
function, and $\tau_{\lambda}(t,r_d)$ is the optical depth
perpendicularly through the disk. The total flux is then found by
integrating from $r_d=0$ to $r_d=r_{\rm dmax}$. The dust mass is
obtained from equation~(4).  Amorphous carbon grains were assumed
(silicate grains are discussed below).  As explained before, during
the $89-281$~d period ejecta dust formation was unlikely and in any
case the early IR excess was explainable as emission from CDS dust.
Therefore the IDDM was introduced at 500~d and used at all subsequent
epochs.  By 652~d the CDS IR emission was negligible and so was not
included in the modelling of this or subsequent phases.  We first
describe the results with $\delta=0$ in the IDDM.  The matches
together with the observations are shown in Figs.~\ref{fig18} and
\ref{fig19}.  The overall model parameters are listed in
Table~\ref{tab8}.\\

The three MIR observation epochs 500~d, 652~d and 859~d lay within the
timespan of the line-profile analysis.  We therefore imposed the
constraint that the disk radius had to be consistent with those values
derived from the line profiles -- that is, $r_{\rm dmax} \approx
5\times10^{15}$~cm and $r_{\rm dth} = v_{\rm dth}\times t$ where
$v_{\rm dth}=+112$~\kms, with $k$ (and therefore $M_d$) being
constrained by the demand that the disk have a high optical depth in
the optical/NIR region (\S3.4.4.1).  In practice, matching the IDDM to
the MIR continuum demanded optical depths which were higher than the
minimum values obtained from the line-profile analysis.  The disk
parameters for the specific MIR epochs were set by linear
interpolation of the line-profile-derived values for $r_{\rm dmax}$
and $r_{\rm dth}$.  The $k$ parameter could take values at or above
those set in the line-profile analysis.  Only $T_d$ was a completely
free IDDM parameter.  By 500~d the MIR continuum up to $\sim$20~$\mu$m
was overwhelmingly due to the warm dust disk, with the CDS component
yielding no more than $\sim$8\% of the total flux at any wavelength.
For $\lambda\gtrsim20~\mu$m, the IS IR echo dominated.  Also from
500~d onwards the hot continuum and fb-ff contributions to the MIR
were negligible. In other words, from 500~d onwards, only the dusty
disk and IS IR echo made significant contributions to the total MIR
continuum. Consequently, and in order to show the MIR behavior in more
detail, the optical/NIR region is not shown in the plots for
652--1393~d (Fig.~\ref{fig19}). \\

In spite of the line-profile constraints on $r_{\rm dmax}$ and $r_{\rm
dth}$, for epochs 500~d, 652~d, and 859~d we were nevertheless able to
obtain fair matches to the observed continuum (Figs.~\ref{fig18},
\ref{fig19}).  It was found that the matches required the dust to be
optically thick ($\tau>1$) in the optical/NIR region, but optically
thin in the MIR region. A blackbody spectrum matched to the shorter
MIR wavelengths overproduced the flux at longer wavelengths.  This is
illustrated in Fig.~\ref{fig18} (500~d) and Fig.~\ref{fig19} as dotted
cyan lines.  This restriction, together with the fixed values for
$r_{\rm dmax}$ and $r_{\rm dth}$ allowed us to obtain specific values,
{\it not} limits, for $T_d$, $\tau$ and $M_d$.  At 500~d the ejecta
dust mass was $(0.22\pm0.02)\times10^{-4}$~M$_{\odot}$, with
$T_d=650\pm15$~K. The optical depths through the disk were
$\sim$10 in the $V$ band and $0.45\pm0.05$ at 10~$\mu$m. By 859~d the
dust mass had increased to $(0.33\pm0.05)\times10^{-4}$~M$_{\odot}$,
while $T_d$ fell to $570\pm15$~K.  A particular value of this
500--859~d study is that it demonstrates that the same dust-disk
parameters can account for the line profiles {\it and} the MIR
continuum.  This adds considerable weight to our contention that the
two disks are one and the same.  Moreover specific values, not limits,
for the dust masses and temperatures were determined for the
500--859~d period.  \\

As the SN aged between 500~d and 996~d it was found that the IDDM
steadily approached the blackbody case.  Moreover, there were no line
profiles available after 925~d.  Linear extrapolation to 996~d of the
line-profile-derived values for $r_{\rm dmax}$ gave $r_{\rm
dmax}\sim(4\pm1)\times10^{15}$~cm.  The model match yielded dust with
a high optical depth in the optical/NIR but with $\tau\sim1$ in the
MIR.  The dust mass was $M_d \approx 0.5\times10^{-4}$~M$_{\odot}$ and
$T_d = 520\pm30$~K.  \\

By epochs 1207~d and 1393~d we found that the MIR continuum was best
reproduced by allowing a continuing increase in the IDDM optical depth
such that the disk was optically thick at all observed wavelengths
(see Table~\ref{tab8}).  Indeed, given the uncertainties in the
observed fluxes, we cannot rule out a totally opaque disk at all
wavelengths covered.  Consequently, the IDDM could provide only dust
mass lower limits of $1.0\times10^{-4}$~M$_{\odot}$ at 1207~d and
$1.5\times10^{-4}$~M$_{\odot}$ at 1393~d. The 1207~d and 1393~d plots
shown in Fig.~\ref{fig19} are with $r_{\rm dmax}$ and $T_d$ set at the
limiting values (Table~\ref{tab8}). \\

In Table~\ref{tab9} we show the luminosities of the IDDM ($\delta=0$)
components compared with the radioactive deposition power of
0.0095\,M$_{\odot}$ of $^{56}$Ni, including $^{56}$Ni decay.  The IS
IR echo component is excluded since it is powered primarily by the SN
peak luminosity.  The luminosity of the thin disk, $L_{IDDM}$, is
approximated by:
\begin{equation}
L_{IDDM} \approx 2 \pi R_d^2 \pi B_{\nu} (1-exp(-2\tau_{\nu})).
\end{equation}
As already indicated by the BLC analysis (\S3.2) the post-30~d
evolution of SN~2004dj falls into three phases. At 89~d, (about
half-way down the plateau-edge) the luminosity is still dominated by
the shock-ionized ejecta, with radioactive decay contributing a small
proportion of the total.  There is then a short period just after the
end of the plateau when radioactive decay deposition dominated the
luminosity. This is supported by the fact that at 129~d, the
radioactive luminosity, $L_{rad}$, is highly similar to the total
continuum model luminosity, $L_{total}$.  At the nebular phases of
251~d and 281~d, the small excess in $L_{rad}$, relative to
$L_{total}$, presumably went into powering the line emission which is
not included in the model.  By 500~d, in spite of the exclusion of
much of the line luminosity from the model, $L_{total}$ is $\times 2.5
L_{rad}$, rising to $\times 200 L_{rad}$ by 1393~d.  As argued above,
the most plausible additional power source available at this stage is
a reverse shock. \\

Model matches were also carried out with $\delta=1.9$ in the IDDM
component. Similar dust parameters were found to those obtained for
$\delta=0$.  The only significant difference was that the dust masses
were $\times3$ larger, assuming a uniform density distribution within
the inner $0.05r_{\rm dmax}$ (cf. \S3.4.4.1). These larger values are
due to the growing proportion of dust mass concentrated in the
optically-thick region of the disk.  \\

We also considered a dusty disk of warm silicate grains, with the
radii and thickness determined by the line-profile analysis as before.
The demand that the disk should be of optical depth in the optical/NIR
up to its edge meant that the replacement of amorphous carbon dust
with silicate dust had no effect on these dimensions.  The problem
with silicate dust continuum matching is the absence of the
$8-14~\mu$m silicate feature in the observed continua during the
period when the dust was optically thin in the MIR, up to 996~d.
Attempts to suppress the feature in the IDDM by increasing the dust
mass and hence the optical depth yielded a continuum that was too
bright. Only for 1207~d and 1393~d was silicate dust able to reproduce
the observed continuum. This is not surprising since, by these two
epochs, the warm ejecta dust was close to being opaque over the MIR
range observed.  Between 500~d and 996~d the proportion of silicate
grains by mass could have been no more than 20\%, and was usually
significantly less than this.  Indeed, the data were always consistent
with there being no silicate grains at all.  We also note the absence
of the SiO feature at $7.5-9.3$~$\mu$m.  SiO formation is a necessary
step on the way to silicate grains \citep{tod01,noz03}.\\

The points made in the previous paragraph argue against a substantial
amount of silicate dust in the ejecta of SN~2004dj.  The absence of
silicate grains is consistent with the presence of strong CO
fundamental and first overtone emission in the period $\sim$100~d to
$300-500$~d.  A high C/O ratio in the ejecta could result in most of
the oxygen being absorbed as CO leaving behind an excess of carbon to
provide carbon grains, but little oxygen to provide silicate
grains. However, in the SN environment, the net grain population can
be affected by factors in addition to the C/O ratio.  These include
molecule destruction by high-energy electrons, charge transfer
reactions and ejecta density \citep{liu96,noz03,den06}.  Nevertheless,
we conclude that the dust grains in SN~2004dj were predominantly
composed of non-silicate material.\\

The success in reproducing the MIR continua using the same dust disk
as was invoked to explain contemporary line profiles tends to support
the curious result from \S3.4.4.1 that, rather than expanding, the
radius of the dust disk actually shrunk by a small amount. Indeed, if
we include the disk radii for 1207~d and 1393~d derived from the IDDM
matches (Table~\ref{tab8}, col.~7) we find a shrinkage of 27\% since
500~d.  In SN~2004et \citep{kot09} it was found that the dust radius
remained roughly constant.  The explanation offered was that the dust
was contained within an optically-thin cloud of optically-thick,
pressure-confined clumps.  But the availability of late-time
optical/NIR spectra for SN~2004dj and the profile modelling presented
here shows that a clumping explanation for the non-expanding dust-disk
radius would not work.  We note that, during the 500--859~d period in
SN~2004dj, the product of the blue-wing, half-maximum (BHM) velocities
and epoch for the line profiles, $R_{BHM}$, yields a roughly constant
value of $\sim 4\times10^{15}$~cm (see Table \ref{tab6}) - similar to
that obtained for the dust-disk radius.  This coincidence may imply
that the extent of both the dust and the bulk of the ejecta gas
emission was physically constrained within a radius of $\sim
5\times10^{15}$~cm.\\

We suggest that the apparent shrinkage of the dust-disk radius may
have been due to the presence and inward motion of the reverse shock
such that the shock position defined the disk radius.  Dust formation
could have continued within the disk with the dust taking part in the
overall ejecta outflow, but as the ejecta passed through the reverse
shock it would have been destroyed.  There would have been a net
increase in dust mass with time, as observed, if the dust growth rate
within the disk exceeded the destruction rate.  A possible difficulty
with this scenario is that the dust mass appeared to continue to grow
right up to the final epoch at 1393~d. This is rather later than dust
formation studies suggest. \citet{tod01} find that all grain
condensation would be complete by 800~days. Moreover, both
\citet{tod01} and \citet{noz03} find carbon dust condensation is
complete in not much more than one year.  An alternative explanation,
therefore, might be that the dust mass in SN~2004dj did not increase
after $\sim$400~d, but rather that the density gradient of the outer
region of the disk was actually steeper than $r^{-2}$.  Thus, as the
ejecta expanded, the dust density at a fixed location in the outer
region would have grown, yielding an apparently higher total dust mass
--- i.e., more dust emerged from the optically-thick inner regions.  \\

Radiation from the reverse shock could have been primarily responsible
for heating the dust. As more and more ejecta passed through, the
shock and its radiation would have weakened causing the dust to cool,
as observed.  The reverse shock may also have been responsible for the
approximate constancy of the radius of the line emitting gas sphere.
Further examination of the reverse-shocked dust-disk hypothesis is
beyond the scope of this paper.  We note that in a recent
observational study of Cassiopeia~A \citet{del10} deduce a flattened
ejecta distribution or ``thick disk'' containing all the ejecta
structures. They also deduce the occurrence of a roughly spherical
reverse shock.  \\

Intrinsic axial asymmetry in the form of a bipolar jet biased towards
the observer has been invoked by \citet{chu05} to account for the
H$\alpha$ profile in SN~2004dj up to about 1~year.  We note that the
jet makes an angle of only $\sim15^\circ$ to the normal to the dust
disk plane derived from our line-profile analysis, perhaps indicating
a physical connection.  However, line-profile asymmetry in the earlier
(pre-$\sim$1~year) nebular spectra of other CCSNe also tend to be
blue-biased. This implies that, in general, an intrinsic bias towards
the observer cannot be the explanation for such line blueshifts
\citep[cf.][]{mil10}.  At later (post-$\sim$1~year) nebular epochs,
line blueshifts are also often seen
\citep{luc89,spy90,tur93,fes99,ger00,leo00,fas02,elm03,poz04}
and these are usually attributed to attenuation by newly-formed dust
in the ejecta or CDS.  Our dust-disk model invokes this scenario.
This has the advantage of explaining the red-wing suppression without
invoking an intrinsic observer-biased axial asymmetry in the SN.
Moreover, our model uses the same dust disk to simultaneously account
for the line-profile attenuation {\it and} the MIR emission.  We
consider it unlikely that the \citeauthor{chu05} model can provide a
superior alternative explanation for the line profiles presented and
analysed in \S3.4.4.1.  Further examination of the relationship between
our model and that of \citeauthor{chu05} is beyond the scope of this
paper.  \\

We conclude that dust formation in the ejecta of SN~2004dj had
commenced, and may even have been completed, by 500~d with a
near-face-on, disk-like distribution. {\it This dust was responsible
both for the late-time line-profile red-wing suppression and the bulk
of the MIR luminosity up to $\sim20 \mu$m.}  The main source of dust
heating was probably reverse-shock radiation.  Assuming $\delta=0$,
the dust mass was at least $10^{-4}$~M$_{\odot}$.  For $\delta=1.9$,
the lower limit rises to about $3\times 10^{-4}$~M$_{\odot}$.  The
value of $\delta$ is poorly constrained.  We reject silicates as the
grain material.  \\

\section{Conclusions}
We have presented optical, NIR, and MIR observations of the Type~IIP
SN~2004dj. The combination of wavelength and temporal coverage
achieved makes this SN one of the most closely studied of such
events. In the present work we have analyzed the SN continuum over a
period spanning 89--1393~d, augmented by a line-profile analysis over
461--925~d. Our conclusions are as follows. \\

(1) A mass of $0.0095\pm0.002$~M$_\odot$ of $^{56}$Ni was ejected
from SN~2004dj, which is less than that reported by other authors. The
period during which the radioactive tail dominated the bolometric
light curve lasted for an unusually short period of only
$\sim$35~d. Subsequently, a different energy source dominated;
we suggest reverse-shock heating. \\

(2) At early times the optical/NIR (``hot'') part of the continuum
provided most of the SN luminosity. This emission is attributed to hot,
optically-thick ejecta gas. At later epochs the optical/NIR luminosity
was increasingly due to nebular emission, and formed only a minor,
ultimately negligible, proportion of the BLC. \\

(3) At both early and late times, the long-wave portion of the MIR
continuum (the ``cold'' component) was primarily due to an IS IR echo.
Free-free radiation made only a minor contribution.  As originally
suggested by \citet{bod80}, the analysis of SN IR echoes may provide a
useful way of studying IS dust in nearby galaxies. \\

\noindent 
Subsequent to the submission of this paper, \citet{sza11} reported
their findings on SN 2004dj, making use of some of the data presented
in this work. They dismiss the IS IR echo, and argue against
pre-existing dust since the estimated extinction of SN 2004dj was
lower than that of other SNe. We do not concur.  We have shown that an
extinction of only Av=0.02 is all that is required to account for an
adequate IR echo.  \citet{sza11} also argue that the early UV/X-ray
flash would create a dust-free cavity of up to $10^{17}$ cm and that
OB-stars would have expelled the ISM/dust from the cluster.  In fact
most of the IS IR echo longward of $\sim10~\mu$m comes from dust lying
at considerably more than 1~parsec.  We therefore strongly favour a
scenario whereby the long-wave component was primarily due to an IS IR
echo.\\

(4) The early-time NIR/MIR (``warm'') component was probably due to
thermal emission from non-silicate dust formed in the CDS.  The CDS
dust growth began at about 50~d, reached 90\% of maximum by 165~d, and
approached a maximum of $0.33\times10^{-5}$~M$_{\odot}$; the dust mass
produced in this way was small.  Heating of the dust by the
contemporary optical-NIR BLC completely accounts for the strength and
evolution of the early-time warm component.  \citet{sza11} did not
present contemporary optical or NIR data and consequently did not
identify the early warm component. \\

(5) The late-time warm component was dominated by the luminosity of
newly-formed, non-silicate dust in the ejecta. The dust growth
commenced at some time between 281~d and 461~d.  The same dust was
responsible for the late-time red-wing attenuation of optical and NIR
spectral line profiles. The dust was distributed as a nearly face-on
disk, within a spherical cloud of emitting gas. During 500--996~d the
disk was effectively opaque in the optical/NIR region, but was
optically thin at longer wavelengths. The dust mass appeared to grow
during this period, attaining $(0.5\pm0.1)\times10^{-4}$~M$_{\odot}$
by 996~d for a uniform density disk, or a few times more than this for
an $r^{-1.9}$ gradient outside $0.05r_{\rm dmax}$ and uniform within.
However, it may be that the dust mass ``growth'' was really due to the
emergence of previously formed (i.e., pre-500~d) dust from
optically-thick regions of a disk with an even steeper density
gradient.  For the latest two epochs (1207~d and 1393~d) the dust was
optically thick at all wavelengths and only lower limits could be
obtained for a given gradient --- for example, $>10^{-4}$~M$_{\odot}$
for a flat gradient and a factor of 3 higher limit for an $r^{-1.9}$
gradient outside $0.05r_{\rm dmax}$ and uniform within.  This is
broadly in agreement with \citet{sza11}. For a smooth distribution,
they found a dust mass of $\sim10^{-5}-10^{-4}$~M$_{\odot}$ though,
with clumping, up to $\sim10^{-3}$~M$_{\odot}$ would also be
possible. \\

(6) Rather than expanding, the dust-disk radius appeared to
slowly shrink.  This may have been due to the dust extent being
confined by the reverse shock, which also heated the grains
radiatively. \\

(7) While the latest epochs provide only lower limits to the
mass of dust produced by SN~2004dj, these limits are at least a factor
of 100 below the 0.1~M$_{\odot}$ of grains per SN required to account
for the dust observed at high redshifts. Moreover, measurements as late
as 996~d yield actual dust masses of only $\sim10^{-4}$~M$_{\odot}$.
While not completely ruling out the possibility that typical CCSN
ejecta are major contributors to cosmic dust production, this study
does suggest that, at least for SN~2004dj, the dust-mass
production was small.

\acknowledgements We thank J. Vink\'o for providing us with digitized
versions of his optical spectra.  The work presented here is based on
observations made with the {\it Spitzer Space Telescope}, the
W. M. Keck Observatory, the William Herschel Telescope (WHT), and the
2.4~m Hiltner telescope of the MDM Observatory.  The {\it Spitzer
Space Telescope} is operated by the Jet Propulsion Laboratory,
California Institute of Technology, under a contract with NASA.  The
W. M. Keck Observatory is operated as a scientific partnership among
the California Institute of Technology, the University of California,
and NASA; it was made possible by the generous financial support of
the W. M. Keck Foundation.  We wish to extend special gratitude to
those of Hawaiian ancestry on whose sacred mountain we are privileged
to be guests.  The WHT is operated on the island of La Palma by the
Isaac Newton Group in the Spanish Observatorio del Roque de los
Muchachos of the Instituto de Astrof\'isica de Canarias.  Financial
support for this research was provided by NASA through an award issued
by JPL/Caltech (specifically grant number 1322321 in the case of
A.V.F.).  A.V.F. gratefully acknowledges additional support from NSF
grant AST-0908886 and the TABASGO Foundation.  P.A.H. was supported by
NSF grants AST-1008962 and 0708855.  S.M. acknowledges support from
the Academy of Finland (project 8120503).  J.S. is a Royal Swedish
Academy of Sciences Research Fellow supported by a grant from the Knut
and Alice Wallenberg Foundation.  J.C.W gratefully acknowledges
support from NSF grant AST-0707769.  The Dark Cosmology Centre is
funded by the Danish National Research Foundation.




\clearpage



\newpage

\begin{table}
\caption{Mid-IR Photometry of SN~2004dj.}
\begin{center}
\begin{tabular}{lccclcccccc}
\hline
        &  & & & \multicolumn{7}{c}{Flux (mJy)$^{**}$} \\
\cline{5-11}
& & Epoch$^*$ & t$_{\mathrm{exp}}$ & \multicolumn{4}{c}{IRAC} & \multicolumn{2}{c}{PUI} & MIPS \\
Date    & MJD & (d)    & (s) & 3.6\,$\mu$m  & 4.5\,$\mu$m & 5.8\,$\mu$m  & 8.0\,$\mu$m &  16\,$\mu$m & 22\,$\mu$m & 24\,$\mu$m\\
\hline
 2004 Oct 07$^a$& 53285.1 &  89.1& 134 & 11.25(2) &10.00(2) & 6.85(3) & 4.32(4) & --      & --      & --      \\
 2004 Oct 08$^b$ & 53286.4 &  90.4& 107 & 10.94(2) &10.01(2) & 6.74(3) & 4.09(5) & --      & --      & --      \\
 2004 Oct 12$^b$ & 53290.1 &  94.1& 107 &  8.30(2) & 9.71(2) & 6.02(3) & 3.45(4) & --      & --      & --      \\
 2004 Oct 14$^a$ & 53292.5 &  96.5& 160 &   --     &  --     &  --     &  --     & --      & --      & 0.74(16) \\
 2004 Nov 01$^a$ & 53310.1 & 114.1& 134 &  4.37(2) & 7.70(2) & 4.78(3) & 2.49(3) & --      & --      & --      \\
 2004 Nov 06$^a$ & 53315.7 & 119.7& 160 &   --     &  --     &  --     &  --     & --      & --      & 0.70(9) \\
 2005 Mar 03$^a$ & 53432.3 & 236.3& 160 &   --     &  --     &  --     &  --     & --      & --      & 0.89(9) \\
 2005 Mar 24$^a$ & 53453.9 & 257.9& 134 &  1.25(2) & 3.50(1) & 1.57(3) & 1.27(4) & --      & --      & --      \\
 2005 Apr 02$^a$ & 53462.0 & 266.0& 160 &   --     &  --     &  --     &  --     & --      & --      & 0.75(17)\\
 2005 Oct 20$^c$ & 53663.7 & 467.7& 536 &  1.86(2) & 2.36(1) & 2.42(2) & 2.36(4) & --      & --      & --      \\
 2005 Nov 22$^c$ & 53693.9 & 497.9& 630 &   --     &  --     &  --     &  --     & 1.23(4) & --      & --      \\ 
 2006 Mar 23$^c$ & 53817.9 & 621.9& 536 &  1.34(2) & 1.74(1) & 2.38(2) & 2.58(4) & --      & --      & --      \\
 2006 Apr 02$^d$ & 53827.6 & 631.6&  10 &   --     &  --     &  --     &  --     & --      & --      & 1.0(2)$^\dag$\\
 2006 Apr 23$^c$ & 53848.4 & 652.4& 630 &   --     &  --     &  --     &  --     & 1.43(7) & --      & --      \\  
 2006 Oct 28$^e$ & 54036.2 & 840.2& 536 &  0.85(2) & 1.14(1) & 1.81(2) & 2.09(4) & --      & --      & --      \\
 2006 Oct 31$^f$ & 54039.1 & 843.1&  14 &  0.83(2) & 1.13(2) & 1.50(7) & 2.07(4) & --      & --      & --      \\
 2006 Nov 16$^f$ & 54054.8 & 859.0&  57 &   --     &  --     &  --     &  --     & 1.38(7) & --      & --      \\
 2006 Nov 16$^e$ & 54055.0 & 859.0& 315 &   --     &  --     &  --     &  --     & 1.47(4) & 1.15(3) & --      \\ 
 2006 Dec 01$^f$ & 54070.3 & 874.3& 160 &   --     &  --     &  --     &  --     & --      & --      & 1.15(15)\\
 2007 Mar 24$^e$ & 54183.0 & 987.0&1258 &   --     &  --     &  --     &  --     & 1.28(9)$^{\dag\dag}$ & 0.96(8)$^{\dag\dag}$ & --\\  
 2007 Mar 24$^f$ & 54183.0 & 987.0& 132 &   --     &   --    &  --     &  --     & 1.37(10)$^{\dag\dag}$&  --     & --\\
 2007 Apr 02$^e$ & 54192.8 & 996.8& 536 &  0.50(2) & 0.67(1) & 1.15(2) & 1.37(4) & --      & --      & --      \\
 2007 Apr 02$^f$ & 54192.8 & 996.8& 125 &  0.49(2) & 0.67(1) & 1.13(3) & 1.38(4) & --      & --      & --      \\
 2007 Apr 13$^f$ & 54203.1 &1007.1& 160 &   --     &  --     &  --     &  --     & --      & --      & 0.90(10)\\
 2007 Oct 24$^g$ & 54397.3 &1201.3& 494 &   --     &  --     &  --     &  --     & --      & --      & 1.09(13)\\
 2007 Oct 24$^h$ & 54397.3 &1201.3&1360 &   --     &  --     &  --     &  --     & --      & --      & 1.07(13)\\ 
 2007 Nov 04$^g$ & 54408.3 &1212.3& 283 &   --     &  --     &  --     &  --     & 1.18(4) & --      & --      \\
 2007 Nov 19$^f$ & 54423.3 &1227.3& 322 &  0.36(2) & 0.45(1) & 0.76(2) & 1.01(4) & --      & --      & --      \\
 2007 Nov 23$^g$ & 54427.1 &1231.1& 322 &  0.36(2) & 0.44(1) & 0.76(2) & 1.03(3) & --      & --      & --      \\
 2007 Nov 24$^h$ & 54428.0 &1232.0&3485 &  0.36(1) & 0.45(1) & 0.80(2) & 1.02(3) & --      & --      & --      \\
 2007 Nov 29$^f$ & 54433.4 &1237.4& 494 &   --     &  --     &  --     &  --     & --      & --      & 1.20(6) \\
 2007 Dec 06$^f$ & 54440.0 &1244.0& 283 &   --     &  --     &  --     &  --     & 1.80(35)$^{\dag\dag}$& -- & --      \\
 2007 Dec 15$^h$ & 54450.0 &1254.0&1260 &   --     &  --     &  --     &  --     & 1.12(3) & --      & --      \\
 2008 Mar 25$^h$ & 54551.0 &1355.0&1260 &   --     &  --     &  --     &  --     & 1.03(6)    & --      & --      \\
 2008 Apr 07$^h$ & 54563.8 &1367.8&3485 &  0.33(2) & 0.38(1) & 0.67(3) & 0.90(4) & --      & --      & --      \\
 2008 Apr 12$^g$ & 54568.2 &1372.2& 322 &  0.33(2) & 0.38(1) & 0.63(2) & 0.95(4) & --      & --      & --      \\
 2008 Apr 14$^g$ & 54570.4 &1374.4& 494 &   --     &  --     &  --     &  --     & --      & --      & 1.24(9) \\
 2008 Apr 15$^h$ & 54571.0 &1375.0&1360 &   --     &  --     &  --     &  --     & --      & --      & 1.11(14)\\
 2008 May 03$^g$ & 54589.3 &1393.3& 283 &   --     &  --     &  --     &  --     & 1.07(6) & --      & --      \\
 2009 Nov 14$^i$ & 55149.9 &1953.9&3276 &  0.31(1) &  --     &  --     &  --     & --      & --      & --      \\ 
 2009 Dec 03$^i$ & 55168.7 &1972.7&3276 &  0.30(1) &  --     &  --     &  --     & --      & --      & --      \\ 
 2010 Apr 16$^i$ & 55302.7 &2106.7&3276 &  0.35(1) &  --     &  --     &  --     & --      & --      & --      \\ 
 2010 May 23$^i$ & 55339.5 &2143.5&3276 &  0.31(1) &  --     &  --     &  --     & --      & --      & --      \\ 
\hline	      	      							
 Sandage 96      &         &      &     &  0.28(5) & 0.21(4) & 0.14(3) & 0.08(2) & 0.024(7)& 0.013(4)& 0.011(3)\\ 
\hline
\end{tabular}
\end{center}
\tablecomments{
$^*$ We assume an explosion date of 2004 Jul. 10.0 (MJD=53196.0).\\
$^{**}$ All measurements were made in a $\sim3\farcs7$ radius circular
aperture, with annular sky measured between $\times1.5$ and
$\times2.2$ the aperture radius. The aperture was centered on the WCS
co-ordinates of the SN.  Statistical uncertainties in the last one or
two significant figures are shown in brackets.  The fluxes have not been
corrected for reddening, nor for the contribution of the S96 cluster.
The last line shows the estimated contribution to the total flux by
S96 (see \S3.1). The exposure times in Column~4 are per band. It can
be seen that some same-wavelength observation epochs have negligible
temporal spacing as well as differing exposure times (viz. 840/3~d,
859~d, 987~d, 996~d, 1201~d, 1227/31/32, 1244/54, 1367/72,
1374/5). This was due to the impact of Spitzer scheduling constraints
on the MISC and SEEDS programs.\\
$^a$PID. 00226 Van Dyk (MISC); \\
$^b$PID. 00159 Kennicutt \etal (SINGS);\\
$^c$PID. 20256 Meikle \etal (MISC);\\
$^d$PID. 20321 Zaritsky; \\
$^e$PID. 30292 Meikle \etal (MISC);\\
$^f$PID. 30494 Sugerman \etal (SEEDS);\\
$^g$PID. 40010 Meixner \etal (SEEDS);\\
$^h$PID. 40619 Kotak \etal (MISC);\\
$^i$PID. 61002 Freedman \etal \\
$^\dag$ Target $\sim$50\% off edge of field. Total flux estimated by
extrapolation. \\
$^{\dag\dag}$ Contamination to north-east.} 
\label{tab1} 
\end{table}

\begin{table}
\caption{ Near-IR Photometry of SN~2004dj} 
\begin{center}
\begin{tabular}{lccccccc}
&&&&&&& \\ \hline
 Date         &MJD    & Epoch$^*$ & t$_{\mathrm{exp}}$ & \multicolumn{4}{c}{Magnitudes$^{**}$} \\ 
 (UT)         &       & (d)   &         (s)            &  $Z$      &  $J$     &   $H$    & $K_s$  \\ \hline
 2004 Nov 25  &53334.0& 138.0 & J(70) H(135)           & --        & 13.56(4) & 13.24(4) &   --     \\      
 2005 Mar 17  &53446.0& 250.0 & Z(60) J(60) H(60) K(50)&15.4(1)$^1$& 14.35(4) & 14.15(6) & 14.13(5) \\ 
 2005 May 11  &53501.2& 305.2 & J(300) H(225) Ks(180)   & --        & 14.88(1) & 14.54(2) & 14.42(4) \\ 
 2005 May 12  &53502.1& 306.1 & J(300) H(225)           & --        & 14.88(1) & 14.52(2) &   --     \\ 
 2006 Jan 14  &53749.9& 553.9 & J(675) H(1025) Ks(1215) & --        & 15.87(5) & 15.19(5) & 14.75(5) \\ 
\hline
Sandage 96    &       &       &                         & --        & 15.93(11)& 15.70(12)& 15.15(14)\\
\hline
\end{tabular}
\end{center}
\tablecomments{The 2005 May 11,12 images are from the TIFKAM IR
camera on the 2.4m Hiltner Telescope of the MDM Observatory. All the
other images are from the LIRIS IR imager/spectrograph on the
4.2m William Herschel Telescope of the Observatorio del Roque de los
Muchachos, La Palma,\\
$^*$ We assume an explosion date of 2004 Jul. 10.0 (MJD=53196.0).\\
$^{**}$ All measurements were made in a 3.7'' radius circular
aperture, with annular sky measured between $\times1.5$ and
$\times2.2$ the aperture radius. The aperture was centered on the
centroid of the SN. The error on the last one or two figures is given
in parentheses. Errors shown include uncertainties in the magnitudes
of the four 2MASS comparison field stars. The fluxes have not been
corrected for reddening, nor for the contribution of the S96 cluster
(see \S3.1).  The last line shows magnitudes of S96 measured from
2MASS images \citep{skr97}.\\
$^1$ Estimated by extrapolation of $JHK_s$ SEDs of calibration field
stars to $Z$ band.}
\label{tab2}
\end{table}

\begin{table}
\caption{ Mid-IR Spectroscopy Log of SN~2004dj.}
\begin{center}
 \begin{tabular}{lcccl}
&&&& \\ \hline
 Date        &   MJD    & Epoch$^*$ &t$_{\mathrm{exp}}$ & Program        \\ 
 (UT)        &          & (d)       & (s)   &                                    \\ \hline
 2004 Oct 24 & 53302.32 & 106.3     &  610  & PID. 00226 Van Dyk (MISC)          \\      
 2004 Nov 16 & 53325.77 & 129.8     &  610  & PID. 00226 Van Dyk (MISC)          \\ 
 2005 Mar 18 & 53447.65 & 251.6     &  610  & PID. 00226 Van Dyk (MISC)          \\ 
 2005 Apr 17 & 53477.92 & 281.9     &  610  & PID. 00226 Van Dyk (MISC)          \\ 
 2005 Nov 22 & 53696.83 & 500.8     & 1828  & PID. 20256 Meikle \etal (MISC)    \\ 
 2006 Apr 23 & 53848.47 & 652.5     & 1828  & PID. 20256 Meikle \etal (MISC)    \\ 
 2006 Nov 16 & 54055.08 & 859.1     & 3657  & PID. 30292 Meikle \etal (MISC)    \\  
 2007 Oct 30 & 54403.45 &1207.4     &  117  & PID. 30494 Sugerman \etal (SEEDS) \\ 
 2008 May 03 & 54589.35 &1393.3     & 3657  & PID. 40619 Kotak \etal (MISC)     \\
\hline
\end{tabular}
\end{center}
\tablecomments{Spectral ranges at each epoch were Short-Low (SL) first
and second orders: $7.4-14.5$ and $5.2-8.7~\mu$m, respectively. \\
t$_{\mathrm{exp}}$ gives the exposure time per order, in seconds.\\
$^*$ We assume an explosion date of 2004 Jul. 10.0 (MJD=53196.0).}
\label{tab3}
\end{table}


\begin{table}
\caption{ Optical and Near-Infrared Spectroscopy Log of SN~2004dj.}
\begin{center}
 \begin{tabular}{lcccl}
&&&& \\ \hline
 Date             &MJD  & Epoch$^*$& Spectral range & Telescope \\ 
 (UT)             &     &  (d)     &    ($\mu$m)    & \\ 
\hline						   				      
 2004 Oct 05 &53284.9  &  88.4 &  0.420--0.766  &  DDO 1.88m \\
 2004 Oct 13 &53291    &  95   &  0.480--0.910  &  Lick 3m  \\
 2004 Oct 18 &53296    & 100   &  0.370--0.750  &  LAT Scorpio \\
 2004 Nov 15 &53324.21 & 128.2 &  0.402--0.839  &  DDO 1.88m \\
 2004 Nov 25 &53334.06 & 138.1 &  0.884--2.406  &  WHT 4.2m  \\ 
 2005 Apr 19 &53479.86 & 283.9 &  0.477--0.944  &  WHT 4.2m  \\
 2005 May 12 &53502.15 & 306.1 &  0.827--2.400  &  Hiltner 2.4m\\ 
 2005 Oct 14 &53657.08 & 461.1 &  0.889--2.453  &  WHT 4.2m  \\ 
 2005 Oct 20 &53663.21 & 467.2 &  0.417--0.928  &  WHT 4.2m  \\
 2006 Jan 15 &53750.03 & 554.0 &  0.890--2.343  &  WHT 4.2m  \\ 
 2006 Dec 22 &54091.5  & 895.5 &  0.458--0.723  &  Keck 10m  \\
 2006 Jan 21 &54121.5  & 925.5 &  0.444--0.890  &  Keck 10m  \\
\hline
\end{tabular}
\end{center}
\tablecomments{\\
$^*$ We adopt an explosion date of 2004 Jul. 10.0 (MJD=53196.0).\\
Column~5 details:\\
DDO 1.88m: 1.88m telescope of the David Dunlap Observatory \citep{vin06} \\
Lick 3m: Shane 3m telescope at Lick Observatory \citep{leo06}\\
LAT Scorpio: Scorpio on the 6m large azimuthal telescope \citep{chu05,chu06} \\
WHT 4.2m: 4.2m William Herschel Telescope of the Observatorio del Roque de los
Muchachos. Slit width = 1''.\\
Hiltner 2.4m: 2.4m Hiltner Telescope of the MDM Observatory.\\
Keck 10m: 10m telescope of the W. M. Keck Observatory.}
\label{tab4}
\end{table}

\begin{table}
\caption{Optical and Near-Infrared Line Fluxes, $F$, and Luminosities, $L$.} 
\begin{center}
\begin{tabular}{c|cc|cc|cc|cc|cc|c|}
\hline
     &  \multicolumn{2}{c|}{H$\alpha$} &   \multicolumn{2}{c|}{Pa$\beta$} & \multicolumn{2}{c|}{[O~I]~6300~\AA} & \multicolumn{2}{c|}{[Fe~II]~7155~\AA} &\multicolumn{2}{c|}{[Fe~II]~12570~\AA} &     \\
Epoch (d)&     $F$      & $L$             & $F$       & $L$               &  $F$& $L$                          &$F$& $L$                              &$F$& $L$                              &$L_{\mathrm{rad}}$ \\\hline
  89 & $5300^*$(260)& 63(3)    &  --    &   --    &   --    &   --     &   --     &  --     &  --    &  --      & 586.9 \\ 
 128 & $2120^*$(100)& 25(1)    &  --    &   --    &   --    &   --     &   --     &  --     &  --    &  --      & 417.2 \\
 138 &    --        &   --     & 380(20) &  4.5(2)&   --    &   --     &   --     &  --     &  --    &  --      & 381.4 \\
 283 &   1535(80)   & 18.1(9)  &  --    &   --    &  250(10)&  3.0(2)  & 82(14)   & 1.0(2)  &  --    &  --      & 101.7 \\
 306 &    --        &   --     & 137(7) &  1.62(8)&   --    &  --      &   --     &  --     & 35(10) & 0.4(1)   & 81.5  \\
 461 &              &   --     &  39(4) &  0.46(5)&   --    &  --      &   --     &  --     & 20(2)  & 0.24(2)  & 16.4  \\
 467 &    230(10)   &  2.75(15)&  --    &   --    &  134(7) &  1.58(8) & 41(7)    & 0.49(8) &  --    &  --      & 15.35 \\
 554 &    --        &   --     &  9.7(8)&  0.11(1)&   --    &   --     &   --     &  --     & 12(3)  & 0.14(3)  &  6.0  \\
 895 &    10.7(7)   &  0.13(1) &  --    &   --    & 7.6(9)  & 0.09(1)  &  3.0(2)  & 0.035(3)&  --    &  --      &  0.18 \\
 925 &     9.5(7)   &  0.12(1) &  --    &   --    & 6.9(3)  & 0.082(4) &  2.52(8) & 0.031(1)&  --    &  --      &  0.14 \\ \hline
\end{tabular}
\tablecomments{ Line fluxes, $F$ are in units of $10^{-15}$ erg
cm$^{-2}$s$^{-1}$ while dereddened line luminosities, $L$, are in
units of $10^{38}$~\ergs.  Figures in brackets give the error on the
last one or more significant figures.  The errors in the fluxes and
luminosities are primarily due to uncertainties in the absolute
fluxing and the levels of the underlying continua. Not included in the
luminosity errors are systematic uncertainties in the distance and
extinction (see \S1.1). \\
$^*$Spectra from \citet{vin06}. \\
In the final column is shown $L_{\mathrm {rad}}$, the radioactive
deposition power corresponding to the ejection of 0.0095\,M$_{\odot}$
of $^{56}$Ni, scaled from the SN~1987A case specified by \citet{li93}
\& \citet{tim96}.}
\label{tab5}
\end{center}
\end{table}

\begin{table}
\caption{Optical and Near-Infrared Line Profile Parameters} 
\begin{center}
\begin{tabular}{lcccccc}
\hline
Epoch   &     \multicolumn{4}{c}{Velocity (\kms)} & $R_{\mathrm{BHM}}$ & Model \\ 
(d) &$v_{\mathrm{BHM}}$& $v_{\mathrm{peak}}$&$v_{\mathrm{RHM}}$&$v_{\mathrm{HWHM}}$& $(10^{15}$~cm) & (\kms) \\ \hline
\multicolumn{7}{c}{H$\alpha$ and Pa$\beta$}\\ \hline
  89$^*$      & -2010 &  -450      & 1630    & 1820     & 1.5 & --    \\ 
  95$^{\dag}$ & -2120 &  -270      & 2210    & 2160     & 1.7 & --    \\
 100$^{**}$   & -2550 & -1620      & 2000    & 2270     & 2.2 & --    \\
 128$^*$      & -2670 & -1770      & -700    &  980     & 2.9 & --    \\
 138$(NIR)$   & -2190 & -1780      & -310    &  940     & 2.6 & --    \\
 283          & -1710 &  -730      &  990    & 1350     & 4.2 & --    \\
 306$(NIR)$   & -1840 &  -350      &  910    & 1370     & 4.9 & --    \\
 461$(NIR)$   & -1170 &  -180      &  300    &  730     & 4.7 & -2400 \\
 467          & -1240 &  -160      &  550    &  890     & 5.0 & -2400 \\
 554$(NIR)$   &  -800 &    15      &  500    &  650     & 3.8 & -2400 \\
 895          &  -310 &  -140, 160 &-10$^a$  &  150$^a$ & 2.4 & -2400 \\
 925          &  -430 &  - 80, 210 & --$^b$  &   --     & 3.4 & -2400 \\ \hline
\multicolumn{7}{c}{[O~I]~6300~\AA}\\ \hline		   	      	 	    	    
 283           & -1760 &  -140      &   --    &   --     & 4.3 & --   \\
 467           & -1040 &  -300      &  400    &  720     & 4.5 & -2250\\
 895           &  -440 &  -210, 170 &-90$^a$  &  170$^a$ & 3.4 & -990 \\
 925           &  -440 &  -170, 210 & 20$^a$  &  230$^a$ & 3.5 & -990 \\ \hline
\multicolumn{7}{c}{[Fe~II]~7155~\AA\ \& 12570~\AA}\\ \hline 	       
 283           & -1600 &  -830      &  370    &  980     & 3.9 & --   \\
 306$(NIR)$    & -1670 &  -510      &  400    & 1030     & 4.4 & --   \\
 461$(NIR)$    & -1610 &  -600      &  160    &  880     & 6.4 & -3000\\
 467           & -1130 &  -330      &  510    &  820     & 4.6 & -2250\\
 554$(NIR)$    &  -940 &  -230      &  260    &  600     & 4.5 & -2250\\
 895           &  -400 &  -210, 490 &-120$^a$ &  140$^a$ & 3.1 & -780 \\
 925           &  -410 &  -170, 480 & -30$^a$ &  190$^a$ & 3.3 & -900 \\ \hline
\end{tabular}
\tablecomments{All the velocities are with respect to the center-of-mass 
rest frame of SN~2004dj.\\
Velocity subscripts: \\
\indent BHM: Blue wing, half-maximum \\
\indent RHM: Red wing, half-maximum. \\ 
\indent HWHM: Half-width, half-maximum.\\
$R_{\mathrm{BHM}}$ is the product of $v_{\mathrm{BHM}}$ and the elapsed time (epoch).\\
$^*$: \citet{vin06}, $^{\dag}$: \citet{leo06}, $^{**}$ \citet{chu05}\\
$(NIR)$: Indicates NIR lines Pa$\beta$ and [Fe~II]~12570~\AA.\\
In col.~3, 895~d and 925~d, the velocities of the stronger, blueshifted
peak and the weaker, redshifted peak are given (see \S2.4 and \S3.4.4.1).\\
$^a$: Based on blueshifted component only.\\
$^b$: Blueshifted component insufficiently resolved.\\
Typical uncertainties in the period 89--554~days, in \kms, are:
$\pm50$ (BHM), $\pm50$ (peak), $\pm100$ (RHM), $\pm60$ (HWHM). At
895~d and 925~d the uncertainties are smaller by a factor of
$\sim4$.\\
\indent The last column, ``Model'', gives the maximum velocity derived
from profile (blue wing) matches ($\delta=0$) - see \S3.4.4.1.  }
\label{tab6}
\end{center}
\end{table}

\begin{table}
\caption{Parameters for Triple-Blackbody Matches to SN 2004dj Continua} 
\begin{center}
\begin{tabular}{ccccccccccccc}
\hline
Epoch &$v_{\mathrm{hot}}$ &$T_{\mathrm{hot}}$& $v_{\mathrm{warm}}$&$R_{\mathrm{warm}}$ &$T_{\mathrm{warm}}$&$v_{\mathrm{cold}}$&$T_{\mathrm{cold}}$&  $L_{\mathrm{hot}}^*$&$L_{\mathrm{warm}}^*$&  $L_{\mathrm{cold}}^*$ &  $L_{\mathrm{total}}^{\dag}$&$L_{\mathrm{rad}}$\\
  (d)& (\kms) &  (K)       &(\kms)& ($10^{15}$~cm)&(K)   &(\kms)& (K) & ($10^{38}$ &($10^{38}$&($10^{38}$& ($10^{38}$& ($10^{38}$)\\
     &        &            &      & &      &      &     & \ergs)            &\ergs)&\ergs)&\ergs) & \ergs      \\\hline  
  89 & 450    & 7000       & 1750 & 1.3& 1800 & 7000 & 200 &2141                 &135  &0.33 & 2276  & 592  \\ 
 106 & 220    & 7000       & 1300 & 1.2& 1800 & 8500 & 200 & 695                 &106  &0.69 &  801  & 507  \\
 129 & 180    & 6000       & 1250 & 1.4& 1300 & 7500 & 200 & 372                 &39.5 &0.80 &  412  & 411  \\
 251 &  50    & 6000       &  580 & 1.3& 1000 & 5000 & 200 & 109                 &11.3 &1.3  &  120  & 137  \\
 281 &  50    & 5300       &  470 & 1.1&  900 & 4500 & 200 & 82.9                &6.1  &1.4  &   89  & 103  \\
 500 & 3.8    &10000       &  570 & 2.5&  840 & 1800 & 200 & 19.2                &21.5 &0.69 &   41  & 10.7 \\
 652 &(1.0)   &10000$^{**}$&  550 & 3.1&  700 & 1400 & 200 &  2.3                &16.4 &0.71 &   19  &  2.1 \\
 859 &(0.16)  &10000$^{**}$&  450 & 3.3&  620 & 1250 & 200 &  0.10               &11.7 &0.98 &   12  & 0.26 \\
 996 &(0.05)  &10000$^{**}$&  440 & 3.8&  530 &  800 & 200 &  0.013              &8.1  &0.5  &  8.1  & 0.077\\
1207 &(0.01)  &10000$^{**}$&  330 & 3.4&  480 & 1050 & 200 &  $7.8\times10^{-4}$ &4.5  &1.4  &  4.5  & 0.019\\
1393 &(0.0025)&10000$^{**}$&  320 & 3.8&  440 &  900 & 200 &  $6.5\times10^{-5}$ &4.0  &1.3  &  4.0  & 0.009\\ \hline
\end{tabular}			   
\tablecomments{ $^*$ The continuum luminosities tend to underestimate
the total SN luminosity as they do not include line emission. This is
particularly the case around $200-500$~d when the relative
contribution of nebular line emission is at a maximum.  There
is less of a problem before this era when the hot continuum dominates,
or afterwards when the warm/cold continuum increasingly dominates.\\
$^{\dag}$: $L_{\mathrm{total}}$ excludes $L_{\mathrm{cold}}$. This is
because the cold component is due to an IS IR echo (see \S3.4.3.1)
which was predominantly powered by the peak luminosity of the SN prior
to the earliest epoch of observation.\\
$^{**}$: Hot blackbody velocities and temperatures estimated by fixing
the temperature at the 500~d value and extrapolating the earlier
velocity evolution.\\
In col.~13, $L_{\mathrm {rad}}$ is the radioactive deposition
corresponding to the ejection of 0.0095\,M$_{\odot}$ of $^{56}$Ni,
scaled from the SN~1987A case specified by \citet{li93} \&
\citet{tim96}.  \\
The 89~d and 996~d warm and cold parameters are based on photometry only.}
\label{tab7}
\end{center}
\end{table}

\begin{table}
\caption{Hot Blackbody, Free-Free/Free-Bound and Warm Dust Disk ($\delta=0$) Parameters
for Matches to SN 2004dj SEDs.} 
\begin{center}
\begin{tabular}{cccccccccccc}
\hline
Epoch &$v_{\mathrm{hotbb}}$ &$T_{\mathrm{hot}}$&$v_{\mathrm{fb-ff}}$ &$T_{\mathrm{fb-ff}}$&$n_{\mathrm{e(fb-ff)}}$ &$r_{\mathrm{dmax}}^*$&$T_{\mathrm{d}}^*$& $\tau_{\mathrm{0.55~\mu m}}$& $\tau_{\mathrm{10~\mu m}}$& $\tau_{\mathrm{24~\mu m}}$& $M_{\mathrm{dust}}$\\
  (d)& (\kms) &  (K)   & (\kms)& (K) & ($10^6$~cm$^{-3}$)  &($10^{15}$~cm)  &(K)    &    &        &      &  ($10^{-4}M_{\mathrm{\odot}}$)   \\
            &        &        &      &      &      &        &         &      &       \\\hline  
  89        & 450    & 7000   & 3420 &(6000)&100   & --     & --      & --   & --      & --     & --     \\
 106        & 275    & 6000   & 3420 &(6000)& 86   & --     & --      & --   & --      & --     & --     \\
 129        & 180    & 6000   & 3420 &(6000)& 56   & --     & --      & --   & --      & --     & --     \\
 251        &  50    & 6000   & 3080 & 5000 & 36   & --     & --      & --   & --      & --     & --     \\
 281        &  50    & 5300   & 2980 & 5000 & 24   & --     & --      & --   & --      & --     & --     \\
 500        & 9.0    & 6000   & 2470 & 5000 & 1.8  & 5.2(5) &  650(15)& 9.6  & 0.45(5) & 0.16   & 0.22(2)\\
 652        &(1.0)   &(10000) & 2090 & 5000 & 0.6  & 4.9(5) &  610(15)&15.3  & 0.73(10)& 0.26   & 0.32(5)\\
 859        &(0.16)  &(10000) & 1590 & 5000 & 0.25 & 4.6(5) &  570(15)&17.4  & 0.82(10)& 0.29   & 0.33(5)\\
 996        &(0.05)  &(10000) & 1400 & 2500 & 0.14 & 4(1)   &  520(30)&36.7  & 1.7(5)  & 0.62   & 0.5(1) \\
1207        &(0.01)  &(10000) & 1400 & 2500 & 0.05 & 3.8(6) &  460(30)&$>$80 &$>$4     & $>$1.5 &$>$1.0  \\
1393        &(0.0025)&(10000) & 1400 & 2000 & 0.024& 3.8(6) &  430(20)&$>$120&$>$6     & $>$2   &$>$1.5  \\ \hline
\end{tabular}

\tablecomments{$^*$ $r_{\mathrm{dmax}}$ and $T_{\mathrm{d}}$ are,
respectively, the radius and temperature of the warm, dusty disk
model. Uncertainties in the last one or two figures are shown in
brackets.  Likewise for the estimated disk dust mass,
$M_{\mathrm{dust}}$ (last col.). An example of uncertainties in the
optical depths is shown in col.~10 for 10~$\mu$m. \\ For the hot
blackbody at 652~d and later, the temperature was fixed at 10,000~K
and the velocity obtained by extrapolation, and so these parameters
are shown in brackets. \\ For the fb-ff modelling, the hydrogen
temperature and electron density (cols.~5 \& 6) for 251--1393~d were
derived from the observed hydrogen velocities for SN~2004dj (col.~4)
in conjunction with the late-time SN~1987A study of \citet{koz98} (see
\S3.4.3.2). For 89--129~d, the hydrogen temperature is shown in brackets
as it is a rough estimate (see \S3.4.3.2).  \\ Cols. 9--11 give the
optical depths perpendicularly through the dust disk at the
wavelengths indicated.\\ The matches also included a CDS IR echo at
epochs $89-500$~d and an IS IR echo at all epochs (see \S3.4.2 and
\S3.4.3.1). }
\label{tab8}
\end{center}
\end{table}

\begin{table}
\caption{Hot Blackbody, fb+ff and Warm ($\delta=0$) Dust-disk Luminosities compared with Radioactive Input.\\
Fixed disk radius}
\begin{center}
\begin{tabular}{cccccccc}
\hline
Epoch&  $L_{\mathrm{hot}}$   &$L_{\mathrm{CDS}}$&$L_{\mathrm{ff+fb}}$&$L_{\mathrm{IDDM}}^*$ & $L_{\mathrm{total}}$ &$L_{\mathrm{rad}}$\\
  (d) & ($10^{38}$           &($10^{38}$          &($10^{38}$          &($10^{38}$         & ($10^{38}$           & ($10^{38}$)  \\
      &\ergs)                & \ergs)             & \ergs)             &\ergs)             & \ergs)               & \ergs        \\\hline  
  89   & 2050                &128                 & 8.0                &--                 &                2186  &     592      \\
 106   & 586                 & 55.7               &10.0                &--                 &                 652  &     507      \\
 129   & 372                 & 26.8               & 7.7                &--                 &                 407  &     411      \\
 251   & 109                 & 6.6                & 3.3                &--                 &                 119  &     137      \\
 281   & 82.9                & 4.8                & 4.2                &--                 &                 91.9  &     103      \\
 500   & 14.0                & 0.6                &0.20                &12.4               &                 27.2 &     10.7     \\
 652   & (2.3)               & --                 &0.03                &9.9                &                 12.2 &      2.1     \\
 859   & (0.10)              & --                 &0.005               &6.9                &                  7.0 &     0.26     \\
 996   & (0.032)             & --                 &0.0025              &3.9                &                  3.9 &     0.077    \\
1207   & ($7.7\times10^{-4}$)& --                 &0.0006              &2.3                &                  2.3 &     0.019    \\
1393   & ($6.5\times10^{-5}$)& --                 &0.0002              &1.8                &                  1.8 &     0.009    \\ \hline
\end{tabular}
\tablecomments{$^*$ $L_{\mathrm{IDDM}}$ is the luminosity of the warm,
dusty disk model. \\ For the hot blackbody (col.~2) at 652~d and
later, the temperature was fixed at 10,000~K and the velocity obtained
by extrapolation, and so the luminosities derived from these
parameters are shown in brackets. The post-652~d contribution of the
hot blackbody to the total luminosity is negligible.\\ In col.~7,
$L_{\mathrm {rad}}$ is the radioactive deposition corresponding to the
ejection of 0.0095\,M$_{\odot}$ of $^{56}$Ni, scaled from the SN~1987A
case described by \citet{li93} \& \citet{tim96}.  \\ No IS IR echo
luminosities are shown since these were predominantly powered by the
peak luminosity of the SN prior to the earliest epoch of observation.
In the wavelength region ($\lambda >15~\mu$m) where the cold component
makes a significant contribution ($>20\%$) to the total flux, the IS
IR echo model described in the text maintained a near-constant
luminosity (15--150~$\mu$m) of $\sim2.0\times10^{38}$~\ergs between
89~d and 1393~d.}
\label{tab9}
\end{center}
\end{table}

\clearpage

\section{}



\end{document}